\documentclass[review,3p]{elsarticle}
 
\usepackage{setspace}
\usepackage{graphicx}
\usepackage{subequation}
\usepackage{bm}
\usepackage{natbib}
\usepackage{psfrag}
\usepackage{amssymb}
\usepackage{subfigure}
\usepackage{color}

\begin{document}

\title{Accelerated boundary integral method for multiphase flow in non-periodic geometries}
\author{Amit Kumar}
\author{Michael D. Graham \corref{cor1}}
\ead{graham@engr.wisc.edu}
\cortext[cor1]{Corresponding author}
\address{Department of Chemical and Biological Engineering, University of Wisconsin-Madison, Madison,
Wisconsin 53706, USA}

\begin{abstract}
An accelerated boundary integral method for Stokes flow 
of a suspension of deformable particles is presented for an arbitrary
domain and implemented for the important case of a planar slit geometry.
The computational complexity of the algorithm scales as $O(N)$ or $O(N\log N$),
where $N$ is proportional to the product of number of particles
and the number of elements employed to discretize the particle.
This technique
is enabled by the use of an alternative boundary integral formulation in which
the velocity field is expressed in terms of a single layer
integral alone, even in problems with non-matched viscosities.
The density of the single layer integral is obtained from a
Fredholm integral equation of the second kind involving the double layer integral.
Acceleration in this implementation is provided by
the use of General Geometry Ewald-like method (GGEM)
for computing the velocity and stress fields driven by a set of point forces 
in the geometry of interest. For the particular case of the slit geometry, a Fourier-Chebyshev spectral
discretization of GGEM is developed. Efficient implementations
employing the GGEM methodology are presented for
the resulting single and the double layer integrals.  
The implementation is validated with test problems on 
the velocity of rigid particles and drops between
parallel walls in pressure driven flow, the Taylor deformation
parameter of capsules in simple shear flow and the particle trajectory
in  pair collisions of capsules in simple shear flow. The
computational complexity of the algorithm is verified with results from several large
scale multiparticle simulations.
\end{abstract}
\maketitle

\section{Introduction} 
Multiphase flow in confined geometries is ubiquitous in
nature and technological applications. A very common 
example is  blood flow in the microcirculation. Recall that
blood is primarily a suspension of red blood cells (RBCs) in plasma,
with the volume fraction $\phi$ of RBCs (hematocrit) typically ranging
between $\phi \sim 0.1-0.3$ in the capillaries and reaching as
high as $\phi \approx 0.5$ in large arteries \citep{fung96}. The
diameter of the blood vessels in the microcirculation, which includes the capillaries,
arterioles and venules, is typically in the range $10-125 \,\mu m$ \citep{fung96},
such that a discoidal RBC with a typical diameter and thickness of
$8 \, \mu m$ and $2 \, \mu m$ respectively can be strongly to moderately
confined. Therefore, any realistic computational study of blood flow in the
capillaries must account for confinement.
Other examples of technological interest where confinement
effects are usually significant include multiphase flows
in microfluidic devices \citep{stone04}. Again, any
realistic model must account for confinement in such problems.
Given the importance of multiphase flows under confinement, or more generally
speaking in non-periodic geometries, it is imperative
to develop efficient and accurate computational
techniques which faithfully represent the system
under study, including the aspect of system size (meaning number of particles here).
The algorithm presented herein has been motivated by our goal to
study the class of problems described above. We next discuss some related previous
efforts on the computational studies of multiphase flows under confinement.

Boundary integral  based methods have emerged as a powerful tool for
studying the flow behavior of multiphase systems in the limit of negligible
Reynolds number, i.e. under Stokes flow conditions. Such methods
have been employed in the past to study the flow behavior of a variety
of particle types including drops, capsules, RBCs, and vesicles
among others. Most of these prior implementations scale as $O(N^2)$,
where $N$ is proportional to the number of degrees
of freedom in the system. For a system with $N_p$ particles, each of which have
been discretized into $N_{\Delta}$ elements,  the number of degrees of freedom
in the system scales as $N \sim N_p N_{\Delta}$. The $O(N^2)$ scaling
above assumes an iterative solution of the discretized system of equations, where
the number of iterations is independent of $N$; a direct solution will result 
in a scaling of $O(N^3)$, while a system size dependent number
of iterations with an iterative solution results in a scaling higher than $O(N^2)$;
the worst case scaling being $O(N^3)$. The $O(N^2)$ scaling is usually
prohibitive, such that it precludes a numerical study of large system sizes.
It is therefore not surprising that many of the past studies have been
limited to an $O(1)$ number of particles. 

To overcome these limitations, there have been several efforts to develop 
\textit{accelerated} techniques, where an accelerated technique is assumed
to give a scaling closer to the ideal $O(N)$, while being sufficiently
accurate at the same time. These accelerated techniques employ
either some variant of the particle-particle-particle-mesh ($\textnormal{P}^3$M)
method \citep{deserno98}, or the Fast Multipole Method (FMM) \citep{greengard87}.
One of the earliest Stokes flow boundary integral implementation with acceleration
was perhaps presented by \citet{greengard96}, who employed the FMM for acceleration in
complex domains. Using the particle-mesh-Ewald (PME) method, \citet{metsi2000} 
developed an accelerated implementation of the Stokes flow boundary integral method 
for her studies on two dimensional periodic suspensions of emulsions and foams.
\citeauthor{zinchenko2000} \citep{zinchenko2000,zinchenko2002} employed multipole expansion accelerated
boundary integral method to study large number of drops in a periodic geometry under
shear. \citet{freund07} used the smooth particle-mesh-Ewald
method to study the motion of periodic suspensions of  RBCs and leukocytes
in two dimensions; this was later extended to three dimensions by
\citet{freund10}. In the latter study \citep{freund10}, the effect of confinement
was incorporated by explicitly discretizing the walls, which generally has unknown
tractions and known no-slip velocity conditions. This explicit discretization
is necessary because the periodic Green's function does not inherently satisfy the
no-slip condition on the walls. Additionally, the previous authors employed a
staggered time integrator, such that the wall tractions and the particle surface
velocities were not determined simultaneously; this is 
due to the large cost associated with their simultaneous solution.
Another potential drawback with the periodic
Green's function is that it has a zero mean flow and a non-zero mean pressure gradient associated
with it \citep{hasimoto59}. As a consequence, the pressure drop in the
system is not directly a specified quantity, and must be solved for by varying
the mean flow, which is a specified quantity \citep{freund07,freund10}. Note
that many experiments on pressure driven flow have a specified pressure
drop, and it is therefore desired  to  specify the pressure drop directly
in numerical simulations without incurring additional computational costs. We also remark that the
specified mean flow includes the flow outside the walls,
as that is a part of the simulation box \citep{freund10}; consequently,
in such a method, neither the mean flow between the walls nor
the pressure drop is a directly controlled quantity. Other recent 
work of possible interest is \citet{biros10}, where a FMM accelerated boundary
integral method is presented. While this  implementation was developed for an 
arbitrary domain, its applicability is restricted to two dimensional systems. In a 
subsequent article \citep{biros11}, the previous authors generalized their implementation
to three dimensions, though only an unbounded domain was considered. 
At this point, it must be emphasized that all the prior accelerated implementations 
of the boundary integral method are based on either the free space Green's
function or the periodic Green's function; in such a case, the boundaries
of the confined domain are required to be explicitly discretized.

We next discuss previous boundary integral implementations employing the
Green's function for the geometry of interest. Such a Green's function
satisfies the appropriate boundary conditions at the domain boundaries;
consequently, the unknowns at the domain boundaries, e.g.
hydrodynamic tractions, do not enter the boundary integral equation.
A popular geometry for which several boundary integral implementations have
been developed is a slit -- the region between two parallel
walls. The Green's function for this geometry has been
provided by \citet{liron76}. A boundary integral implementation based on this
Green's function was developed by \citet{zinchenko03} for rigid
particles. This was later extended by  \citet{zinchenko07} for
studies on drops in the same geometry. In a related work, \citet{pieter07}
also implemented a boundary integral method for drops
between two parallel walls, though that was restricted to matched
continuous and dispersed phase viscosities. This was later extended
by \citet{janssen10} to include non-matched viscosity problems.
It is important to emphasize that none of these implementations
are accelerated and have a computational cost of at least $O(N^2)$.
Consequently, it is not surprising that all of the studies
described above were limited to a few particles and are thus not suitable for studying suspension dynamics. 

We now briefly discuss examples of other numerical techniques
employed in the literature for studies on the flow behavior
of particles under confinement.
One such simulation technique is the immersed boundary method,
which has been employed, e.g. by \citet{bagchi09} and \citet{pratik10},
for studies on capsules in a slit. Another popular technique
is the lattice-Boltzmann method. As an example, \citet{aidun09} developed a
coupled lattice-Boltzmann and finite element method to
study deformable particles, which included studies
under confinement. A somewhat related algorithm is the multiparticle
collision dynamics, which has been used by, e.g., \citet{gompper05}
to study RBCs and vesicles in capillaries. The ideal computational cost 
of all the above numerical techniques is $O(N)$. Finally, \citet{swan:2011fh} have developed an accelerated Stokesian dynamics method for rigid spherical particles in a slit that uses ideas related to those presented here and scales as $O(N\log N)$.

In the present study, we develop an accelerated boundary integral
method for multiphase flow in an arbitrary geometry and implement it for a slit geometry as shown in Fig.~\ref{fig:geom}. The computational complexity of this algorithm scales as $O(N)$ or $O(N\log N)$
depending on the specific numerical scheme employed.
The latter scaling of $O(N\log N)$ is associated with the use of fast Fourier
transforms (FFTs) if one or more directions have periodic
boundary conditions, though that is not a requirement of our
method. In the present effort, we provide a detailed description 
using the example of the slit geometry; its extension to other geometries
is straightforward.  The acceleration in our method is provided
by the use of General Geometry Ewald-Like Method (GGEM) \citep{ortiz07}. The choice of GGEM 
as the acceleration technique necessitates the use of an alternative boundary
integral formulation in which the velocity field is expressed solely
in terms of a single layer integral \citep{pozrikidis92}; its unknown density is
obtained from a second kind integral equation involving the double
layer integral. The resulting single and double layer integrals are computed efficiently
employing the GGEM methodology, which results in the aforementioned favorable scaling of $O(N)$
or $O(N\log N)$.

The organization of this article is as follows. In Sec. (\ref{sec:overview}), 
we provide a brief overview of the GGEM accelerated boundary integral method and 
also discuss some of the limitations of the current implementation.
Following this, in Sec. (\ref{sec:form}), we present the boundary integral formulation
and discuss its numerical implementation using GGEM. In Sec. (\ref{sec:memb}),
we present the procedure to compute the hydrodynamic traction jump at a particle
surface using the example of a capsule with a neo-Hookean membrane.
The solution procedure for the discretized boundary integral equation
is presented next in Sec. (\ref{sec:sol}). An extensive validation of our method
is presented in Sec. (\ref{sec:val}). Lastly, in Sec. (\ref{sec:mult_part}), we present results from
several large scale multiparticle simulations and verify the computational complexity of our algorithm.

\section{Overview of the current work} \label{sec:overview}
We summarize here some of the key aspects of the present work. Each
of these points are discussed further later in the article.
\begin{itemize}

\item The general geometry Ewald like method (GGEM) employed in the current work for acceleration
essentially yields the \textit{geometry-dependent} Green's function and other associated
quantities. This work is a first instance of an accelerated boundary integral method 
based on the geometry-dependent Green's function. Prior implementations have employed either the free-space
Green's function (in case of FMM accelerated methods) or the periodic Green's function (in case of PME
accelerated methods).

\item The GGEM methodology decomposes the overall problem into a \textit{local} problem and a \textit{global} problem, essentially by splitting the Green's function into local (singular but exponentially-decaying) and global (smooth but long-ranged) parts. 
The implementation of the local problem is similar to that of the traditional boundary integral
method. However, since the local Green's function decays exponentially with distance from the source of the singularity, distant elements are not coupled and the local solution can be obtained in $O(N)$ operations.

\item The global problem involves solving a \textit{single} phase Stokes equation in the domain
of interest with known boundary conditions and with a known smooth distribution of force densities. It is in this problem that the coupling between distant elements appears.
In solving the global problem, one is not concerned with the particle interfaces,  or 
the different viscosity fluids present inside and outside the particle (if that is the case in the 
	original problem). This major simplification allows the use of a wide variety of
	fast and accurate numerical techniques present in the literature for the solution of
	 Navier-Stokes equation in an arbitrary domain. All these methods, including those
	based on finite difference, finite volume, finite element, and spectral methods are suitable here \citep{canuto06,canuto07,esw05,ferziger02}. In addition, various fast and efficient implementations of Navier-Stokes solvers on GPUs and distributed memory systems are readily available \citep{henniger10,thibault09}.

\item We present a spectral $O(N \log N)$ Stokes flow solver for the global problem in a slit geometry, which, as indicated 
in the introduction, is one of the most widely-studied confined geometries studied in the literature. 
This solver employs a Fourier-Chebyshev Galerkin method in conjunction with the influence matrix approach \citep{canuto07}. 
The unknown coefficients of the Fourier-Chebyshev series expansion are computed with a direct $O(N)$ algorithm --
no iterations are necessary here, as would be the case with FMM or PME accelerated methods in a slit. 
\end{itemize}
	
The implementation of the methodology presented in this paper also has some limitations.
An important limitation of the current implementation concerns the evaluation of the near singular integrals --
these integrals arise when the gaps between the particles become 
very small, and, if not treated appropriately, may cause the simulations to diverge.
In the present work this is alleviated by requiring that the minimum interparticle gap in the system
be always maintained above a specified value; this is achieved by the use of an overlap correction procedure
in an auxiliary step. There are several other minor limitations of the current implementation.
For example, we currently use linear elements to discretize the particle surface. 
It may be beneficial to employ higher order discretizations, like a spectral discretization,
which could be particularly helpful for the accurate evaluation of the near singular integrals. 
In this paper results are reported only for a slit geometry. It will be appropriate 
to develop efficient implementations of the our methodology for other geometries like a cylinder. 
It must be emphasized that none of the above limitations are inherent to our methodology and
we hope to address these in future efforts.

\section{Problem Formulation and Implementation}\label{sec:form}

\begin{figure}[!t]
\centering
\includegraphics[width=0.5\textwidth]{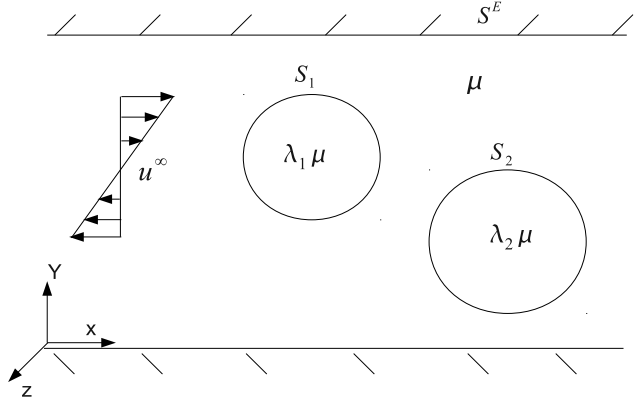}
\caption{Schematic of the problems considered here:  a dispersed phase with viscosity $\mu$ inside the
domain boundaries denoted by $S^E$, containing (for example) two particles
with internal viscosities $\lambda_1 \mu$ and $\lambda_2 \mu$
respectively; their surfaces are denoted by $S_1$ and $S_2$.
The undisturbed flow is denoted by $\mathbf{u}^{\infty}$.}\label{fig:geom}
\end{figure}

\subsection{Boundary integral equation for fluid motion}\label{sec:BI_eq}
We consider a three-dimensional suspension of deformable particles (e.g. fluid-filled capsules) as shown in Fig. (\ref{fig:geom}),
where both the suspending fluid and the fluid enclosed by the particles
are assumed to be Newtonian and incompressible. The viscosity of the suspending
fluid is taken to be $\mu$, while the viscosity of fluid
enclosed by capsule $m$ is taken to be $\lambda_m \mu$, such that
$\lambda_m$ is the viscosity ratio of the interior and the exterior fluid
for this particular capsule. The Reynolds number for the problem is
assumed to be sufficiently small that the fluid motion is governed by the Stokes equation. Under these assumptions, one
may write the velocity at any point in the domain with an
integral expression involving only the boundary of the particles \citep{pozrikidis92}.
We first introduce the formulation that is most commonly used,
\begin{equation}\label{eq:BIa}
\begin{array}{ccc}
u_j(\mathbf{x}_0) =   \displaystyle\frac{2}{1+\lambda_m}u_{j}^{\infty}(\mathbf{x}_0) &   & \hspace{-10mm} - \;\;
\displaystyle \frac{1}{4\pi\mu(1+\lambda_m)} \displaystyle\sum_{n=1}^{N_p} \int_{S^n} \Delta f_i(\mathbf{x})\, G_{ji}(\mathbf{x}_0,\mathbf{x})\,dS(\mathbf{x}) \\
\\
&  & \hspace{-20mm} + \;\; \displaystyle\frac{1}{4\pi(1+\lambda_m)} \displaystyle\sum_{n=1}^{N_p} \displaystyle(1-\lambda_n)\int_{S^n} u_i(\mathbf{x})\, T_{ijk}(\mathbf{x},\mathbf{x}_0)\,n_k(\mathbf{x})\,dS(\mathbf{x}),
\end{array}
\end{equation}
where $\mathbf{u}(\mathbf{x}_0)$ is the fluid velocity at a point $\mathbf{x}_0$
lying on the boundary of particle $m$ (i.e. $\mathbf{x}_0 \in S^m$,
$S^m$ denotes the surface of particle $m$), $\mathbf{u}^{\infty}(\mathbf{x}_0)$ is the undisturbed fluid velocity
at the point $\mathbf{x}_0$, $\Delta \mathbf{f}(\mathbf{x})$ is the hydrodynamic traction jump
across the interface \citep{pozrikidis92}, and the sums are over all the $N_p$ particles
in the system. The Green's function and its
associated stress tensor are denoted by $\mathbf{G}$ and $\mathbf{T}$ respectively
in the above equation, and integrals involving them as the
kernel are typically referred to as the single layer integral
 and the double layer integral respectively \citep{kim_karrila,pozrikidis92}.
From here onwards, a principal value of the double layer
integral over a part of the boundary is assumed whenever
the target point $\mathbf{x}_0$ lies on that boundary. For example,
in the above equation, the double layer integral over $S^n$ is assumed
to denote the principal value when $n=m$. 
A crucial aspect of the above formulation is that the Green's function
$\mathbf{G}$ and associated stress tensor $\mathbf{T}$ are taken to satisfy
the boundary conditions imposed at the system boundaries, so the integrals above
only involve the internal (interfacial) boundaries; if the Green's function 
for any other geometry is employed (e.g. periodic), additional integrals
over the domain boundaries arise in Eq. (\ref{eq:BIa}).

The above form of the boundary integral equation (\ref{eq:BIa}), using the
free space Green's function (the Oseen-Burgers tensor) or the Green's function for a
triply periodic domain given by Hasimoto \citep{hasimoto59} is widely used in the literature
and is the basis for numerous numerical implementations, including the references
cited in the introduction. However, for reasons that will be discussed shortly,
this form is not amenable to numerical solution by an accelerated method
in an arbitrary domain when using the Green's function for that domain.
In the present effort, therefore, we employ an alternative formulation
in which the fluid velocity is expressed solely
in terms of the single layer integral with density $\mathbf{q}(\mathbf{x})$
as follows:
\begin{equation}\label{eq:bi_ua}
u_j(\mathbf{x}_0) =  u_j^{\infty}(\mathbf{x}_0) + \sum_{n=1}^{N_p}  \int_{S^n} q_i(\mathbf{x}) \, G_{ji}(\mathbf{x}_0,\mathbf{x}) \, dS(\mathbf{x}).
\end{equation}
The single layer density $\mathbf{q}(\mathbf{x}_0)$ satisfies (for $\mathbf{x}_0 \in S^m$)
\begin{equation}\label{eq:qa}
q_j(\mathbf{x}_0) \, + \,    \frac{\kappa_m}{4\pi} \, n_k(\mathbf{x}_0) \sum_{n=1}^{N_p} \int_{S^n} q_i(\mathbf{x}) \, T_{jik}(\mathbf{x}_0,\mathbf{x}) \, dS(\mathbf{x})
= -\frac{1}{4\pi\mu} \left(\frac{\Delta f_j(\mathbf{x}_0)}{\lambda_m+1} + \kappa_m f_j^{\infty}(\mathbf{x}_0) \right),
\end{equation}
where $\kappa_m$ is defined as
\begin{equation}
\kappa_m = \frac{\lambda_m-1}{\lambda_m+1},
\end{equation}
while $\mathbf{f}^{\infty}$ is the traction at a given point (computed with
the suspending fluid viscosity $\mu$) due to the stress generated in the fluid corresponding
to the undisturbed flow $\mathbf{u}^{\infty}$ (see \ref{sec:finf} for examples).
In \ref{sec:bi_derive}, this formulation is derived from the Lorentz reciprocal theorem for the case
of a single particle. This derivation follows closely the approach outlined in Chap. (5) of \citet{pozrikidis92}
and is provided here for completeness. 

We now clarify the motivation for employing  Eqs. (\ref{eq:bi_ua}) and (\ref{eq:qa})
rather than the more commonly employed formulation in Eq. (\ref{eq:BIa}).
We begin by noting that the first argument $\mathbf{x}_1$ of
$\mathbf{G}(\mathbf{x}_1,\mathbf{x}_2)$ and $\mathbf{T}(\mathbf{x}_1,\mathbf{x}_2)$
denotes the field (target) point of the functions, while the second
argument $\mathbf{x}_2$ denotes the location of the pole (source) of the singularity that drives the flow.
A close look at Eq. (\ref{eq:BIa}) reveals that the operand of
$\mathbf{G}(\mathbf{x}_0,\mathbf{x})$, $\Delta f_i(\mathbf{x})$,
is a function of the position of the pole of the singularity ($\mathbf{x}$)
and that the field point of the $\mathbf{G}$ tensor is same
as the target point of the overall boundary integral equation
($\mathbf{x}_0$). In other words, the operand of 
$\mathbf{G}$ is independent of its target point, and
consequently the same collection of  point forces can
be used to compute the velocity at any target point. This
requirement is essential to any accelerated method, as in
such methods a part of the calculation gives the velocity
(or other relevant quantities) simultaneously at \textit{all}
target points (e.g., boundary element nodes) due to \textit{all} the singularities
present in the system. This is
possible only if the operands of the singularities
are independent of the target points of the
singularities, and instead are functions of the location
of the pole of the respective singularities.

With this aspect clarified, it is seen that this important condition is not satisfied for the
double layer kernel $\mathbf{T}(\mathbf{x},\mathbf{x}_0)$ in Eq. (\ref{eq:BIa})
as its multiplicands $\mathbf{u}(\mathbf{x})$ and $\mathbf{n}(\mathbf{x})$  are functions
of its target point $\mathbf{x}$. Also note that no general relationship exists that would allow
one to switch the location of the pole and the field points in 
$\mathbf{T}$.  (This is possible for $\mathbf{G}$, since, by self-adjointness, $G_{ij}(\mathbf{x},\mathbf{x}_0) = G_{ji}(\mathbf{x}_0,\mathbf{x})$ \citep{pozrikidis92}).
Hence the above formulation (\ref{eq:BIa}) is not suitable for our purposes
here, though it can still be used for problems in which the
viscosity ratio is unity, as the double layer integral vanishes
in such a case \citep{pratik10}. In contrast, in the formulation
employed in this work (Eqs. \ref{eq:bi_ua} and \ref{eq:qa}),
the multiplicand $\mathbf{q}(\mathbf{x})$ of both $\mathbf{G}(\mathbf{x}_0,\mathbf{x})$
 and $\mathbf{T}(\mathbf{x}_0,\mathbf{x})$  is a function of the location of the 
 source point $\mathbf{x}$. Hence, it is amenable to numerical solution by an accelerated method.

We now describe the fast computation of the velocity
and pressure fields due to a collection of known point
forces, which is closely related to the
problem of computing the Green's function and its associated stress
tensor in the geometry of interest. Later in Secs. (\ref{sec:sl_comp})
and (\ref{sec:dbl_comp}), we employ this technique to
compute the single layer and double layer integrals.


\subsection{GGEM Stokes flow solver for a collection of point forces}\label{sec:greens_fn}
Consider the velocity field $\mathbf{u}(\mathbf{x})$ and the pressure field $p(\mathbf{x})$ due to a collection of
$N_s$ point forces, such that the strength and location of the $\nu^{th}$ point force
are given by $\mathbf{g}^\nu$ and $\mathbf{x}^\nu$ respectively. The velocity and pressure
fields above are obtained from the solution of the Stokes and the continuity equation
as shown below:

\begin{subequations}\label{eq:stokes_g}
\begin{equation}
-\bm{\nabla} p(\mathbf{x})  + \mu \nabla^2 \mathbf{u}(\mathbf{x})
+ \sum_{\nu=1}^{N_s}\mathbf{g}^\nu \delta(\mathbf{x}-\mathbf{x}^\nu) = 0,
\end{equation}
\begin{equation}
\bm{\nabla} \cdot \mathbf{u}(\mathbf{x}) = 0,
\end{equation}
\end{subequations}
and subject to given boundary conditions on the system boundary $S^{E}$. 
By definition, the above velocity and pressure fields along with the associated stress tensor
$\mbox{\boldmath{$\sigma$}}$ can be written in terms of a Green's function $\mathbf{G}$, its pressure
vector $\mathbf{P}$ and stress tensor $\mathbf{T}$ as
\begin{subequations}\label{eq:u_g}
\begin{equation}
u_i(\mathbf{x}) = \frac{1}{8\pi\mu} \sum_{\nu=1}^{N_s} G_{ij}(\mathbf{x},\mathbf{x}^\nu) g^\nu_j,
\end{equation}
\begin{equation}
p(\mathbf{x}) = \frac{1}{8\pi} \sum_{\nu=1}^{N_s} P_j(\mathbf{x},\mathbf{x}^\nu) g_j^\nu,
\end{equation}
\begin{equation}
\sigma_{ik}(\mathbf{x}) = \frac{1}{8\pi} \sum_{\nu=1}^{N_s} T_{ijk}(\mathbf{x},\mathbf{x}^\nu) g_j^\nu.
\end{equation}
\end{subequations}
The stress tensor $T_{ijk}$ in the above equation is obtained from $G_{ij}$ and $P_j$
from the Newtonian constitutive equation
\begin{equation}
T_{ijk}(\mathbf{x},\mathbf{x}^\nu) = - P_j(\mathbf{x},\mathbf{x}^\nu) \delta_{ik} +
\mu \left( \frac{\partial G_{ij}(\mathbf{x},\mathbf{x}^\nu)}{\partial x_k} +
\frac{\partial G_{kj}(\mathbf{x},\mathbf{x}^\nu)}{\partial x_i} \right).
\end{equation}
A close look at the boundary integral equations (\ref{eq:bi_ua}) and (\ref{eq:qa})
shows that to evaluate the integrals we do not explicitly need the Green's function $\mathbf{G}$ and its stress
tensor $\mathbf{T}$ but only their products with the density $\mathbf{q}$.
Put simply, our end goal is to quickly find in time $O(N_s)$ or $O(N_s\log N_s)$ 
the velocity $\mathbf{u}$ and the stress tensor $\mbox{\boldmath{$\sigma$}}$ due to a
given set of point forces -- explicit construction of $\mathbf{G}$ and $\mathbf{T}$ are not necessary.

One of the attractive features of the method presented here lies in the fact that
it is applicable to an \textit{arbitrary} geometry. For simplicity
and considering the interest of the present work, we provide a detailed
discussion only for a slit geometry (see Fig. \ref{fig:val_geom}); generalization of the formalism
for an arbitrary geometry is straightforward \citep{ortiz08}. For the present slit domain,
there is a no slip boundary condition at the two rigid walls at $y=0$
and $y=H$, while periodic boundary conditions are assumed
in the other two directions $x$ and $z$, with spatial periods 
$L_x$ and $L_z$, respectively. 

\begin{figure}[!t]
\centering
\subfigure[$\rho_g(x)$]{\includegraphics[width=0.45\textwidth]{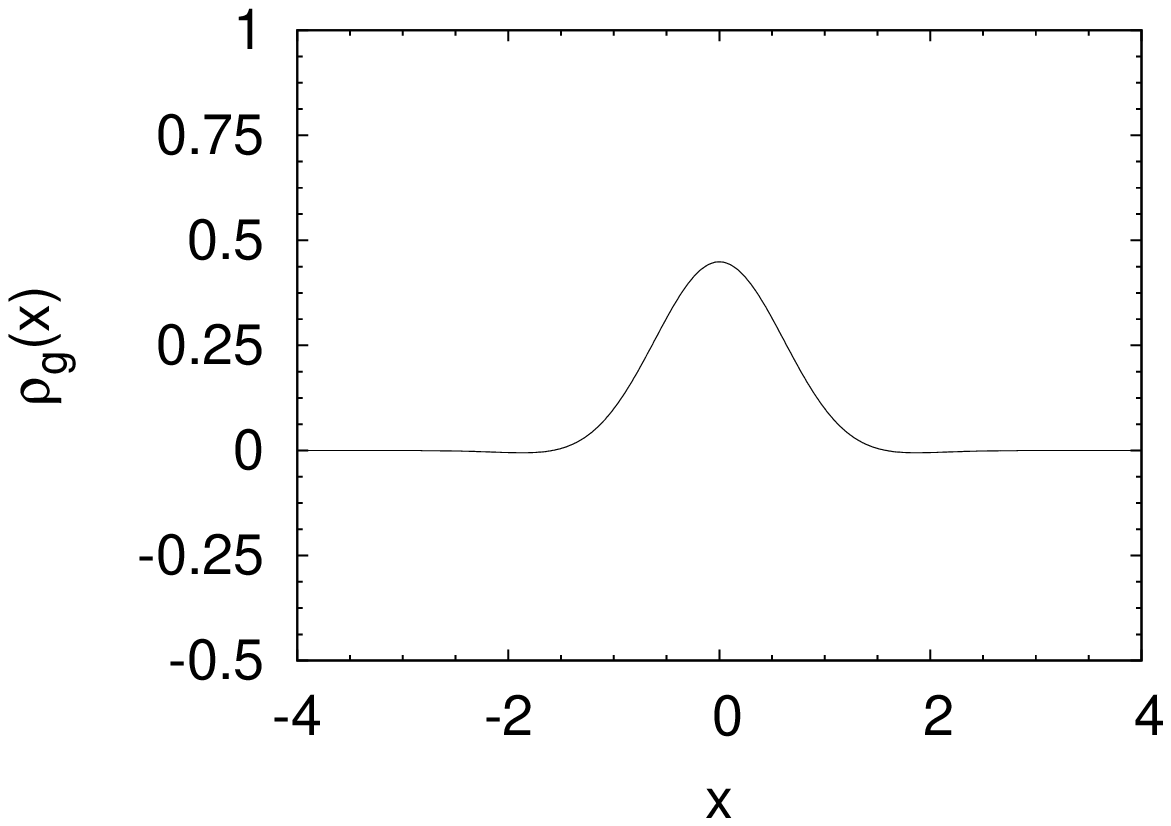}}
\subfigure[$\rho_l(x)$]{\includegraphics[width=0.45\textwidth]{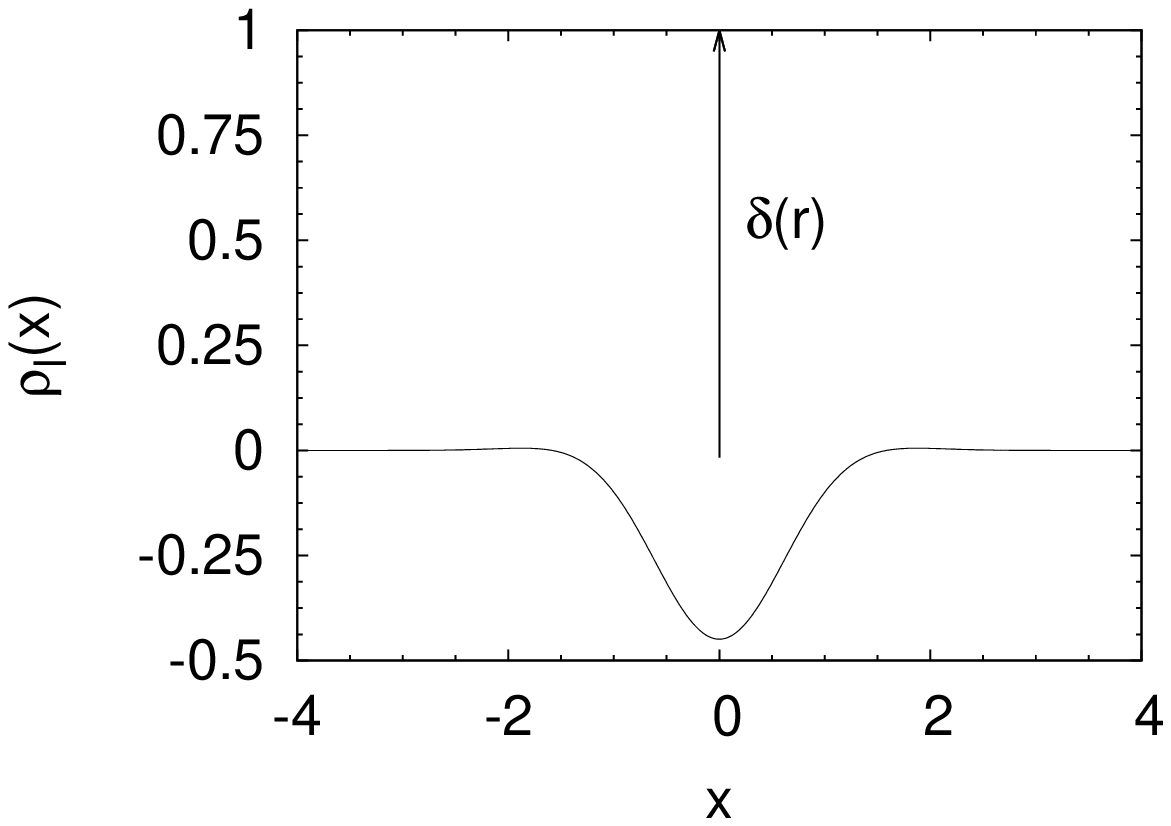}}
\caption{Variation of the global $\rho_g(x)$ and the local $\rho_l(x)$ force density 
along the $x$-axis, given the center of the force density is at the origin.
Note that both of these densities are functions only of the distance $r$ from the 
origin (see Eq. \ref{eq:rho_l_g}). Also note that $\rho_g(\mathbf{r}) + \rho_l(\mathbf{r}) = \delta (\mathbf{r})$.
In numerical calculations, we set $\rho_g(\mathbf{r}) = 0$ for $r > 4/\alpha$.
	For plotting  $\rho_g(x)$ and $\rho_l(x)$ here, we set $\alpha=1$.
}\label{fig:glob_loc_den}
\end{figure}

To achieve the computational complexity of $O(N_s\log N_s)$ alluded to above, we employ
the general geometry Ewald like method (GGEM) \citep{ortiz07} for computing
the velocity and stress fields due to a given collection of
point forces. We briefly describe GGEM next. In the GGEM methodology,
the Dirac-delta density in Eq. (\ref{eq:stokes_g}) is expressed as the sum of a 
smoothly varying quasi-Gaussian \textit{global} density $\rho_g(\mathbf{\hat{r}})$
characterized by a ``splitting parameter'' $\alpha$ and a second \textit{local} density
$\rho_l(\mathbf{\hat{r}})$ (see Fig. \ref{fig:glob_loc_den}). Here $\mathbf{\hat{r}}$ is
a position vector relative to the pole of the singularity,
$\mathbf{\hat{r}} = \mathbf{x} - \mathbf{x}^\nu$. The above global and  local densities
are respectively given by the following expressions:
\begin{subequations}\label{eq:rho_l_g}
\begin{equation}
\rho_g(\mathbf{\hat{r}}) = \frac{\alpha^3}{\pi^{3/2}}e^{-\alpha^2 \hat{r}^2} \left(\frac{5}{2}-\alpha^2 \hat{r}^2 \right),
\end{equation}
\begin{equation}
\rho_l(\mathbf{\hat{r}}) = \delta(\mathbf{\hat{r}})-\rho_g(\mathbf{\hat{r}}),
\end{equation}
\end{subequations}
where $\alpha^{-1}$ represents a length scale over which the delta-function
density has been smeared using the quasi-Gaussian form above, and consequently it
also represents the length scale beyond which both the global and the local
densities are effectively zero. It is important to emphasize 
that the total density remains a $\delta$-function, i.e.
$\rho_g(\mathbf{\hat{r}}) + \rho_l(\mathbf{\hat{r}}) = \delta(\mathbf{\hat{r}})$.
The motivation for this particular splitting of the $\delta$-function density into  $\rho_g(\mathbf{\hat{r}})$
and $\rho_l(\mathbf{\hat{r}})$ will be obvious below.

We next consider the solution of the Stokes and continuity equation with the above two force densities
as forcing functions. The solution driven by the local density, $\mathbf{u}^l(\mathbf{x})$, $p^l(\mathbf{x})$,
and $\mbox{\boldmath{$\sigma$}}^l(\mathbf{x})$ (velocity, pressure, and stress
  respectively) will be referred to as the local solution, and satisfies
  the local problem
\begin{subequations}\label{eq:stokes_l}
\begin{equation}
-\bm{\nabla} p^l(\mathbf{x})  + \mu \nabla^2 \mathbf{u}^l(\mathbf{x})
+ \sum_{\nu=1}^{N_s}\mathbf{g}^\nu \rho_l(\mathbf{x}-\mathbf{x}^\nu) = 0,
\end{equation}
\begin{equation}
\bm{\nabla} \cdot \mathbf{u}^l(\mathbf{x}) = 0.
\end{equation}
\end{subequations} 
This equation will be solved in an unbounded domain, i.e. the solution decays
to zero at infinity. The solution $\mathbf{u}^g(\mathbf{x})$,
$p^g(\mathbf{x})$, and $\mbox{\boldmath{$\sigma$}}^g(\mathbf{x})$ driven by the global density
will be referred to as the global solution, and satisfies the global
problem
\begin{subequations}\label{eq:stokes_gg}
\begin{equation}
-\bm{\nabla} p^g(\mathbf{x})  + \mu \nabla^2 \mathbf{u}^g(\mathbf{x})
+ \sum_{\nu=1}^{N_s}\mathbf{g}^\nu \rho_g(\mathbf{x}-\mathbf{x}^\nu) = 0,
\end{equation}
\begin{equation}
\bm{\nabla} \cdot \mathbf{u}^g(\mathbf{x}) = 0.
\end{equation}
\end{subequations}
The boundary conditions for the global problem are set so that the total velocity field $\mathbf{u}(\mathbf{x})=\mathbf{u}^l(\mathbf{x})+\mathbf{u}^g(\mathbf{x})$ satisfies the specified boundary conditions for the overall problem. Once the local and the global solutions are known, the solution to the overall problem is
obtained as
\begin{subequations}\label{eq:tot_l_g}
\begin{equation}
\mathbf{u}(\mathbf{x})=\mathbf{u}^l(\mathbf{x})+\mathbf{u}^g(\mathbf{x}),
\end{equation}
\begin{equation}
p(\mathbf{x})=p^l(\mathbf{x})+p^g(\mathbf{x}),
\end{equation}
\begin{equation}
\boldsymbol{\sigma}(\mathbf{x})=\boldsymbol{\sigma}^l(\mathbf{x})+\boldsymbol{\sigma}^g(\mathbf{x}).
\end{equation}
\end{subequations}
We next discuss the solution procedures for the local and the global
problems.

\subsubsection{Local solution}\label{sec:local_sol}
Consider first the local problem. The solution to this problem, 
$\mathbf{u}^l(\mathbf{x})$, $p^l(\mathbf{x})$,  and 
$\mbox{\boldmath{$\sigma$}}^l(\mathbf{x})$ is expressed by a set of equations
similar to that in Eqs (\ref{eq:u_g}), which, for the simplicity of nomenclature, is
called the local Green's function $\mathbf{G}^l$ and  its
associated quantities. In short, we append the superscript $l$
to the previously defined quantities to denote the solution associated
with the local density as the forcing function, and these are given by the
following: 
\begin{subequations}\label{eq:local}
\begin{equation}
G^l_{ij}(\mathbf{x},\mathbf{x}^\nu) = \left( \frac{\delta_{ij}}{\hat{r}} + \frac{\hat{x}_i\hat{x}_j}{\hat{r}^3} \right)\textnormal{erfc}(\alpha \hat{r})
-  \frac{2\alpha}{\sqrt{\pi}} \left(\delta_{ij} -  \frac{\hat{x}_i\hat{x}_j}{\hat{r}^2} \right)e^{-\alpha^2 \hat{r}^2},
\end{equation}

\begin{equation}
\\P^l_j (\mathbf{x},\mathbf{x}^\nu) = \frac{2\hat{x}_j}{\hat{r}^3}\,\textnormal{erfc}(\alpha \hat{r}) + \frac{4\alpha \hat{x}_j}{\sqrt{\pi}}
\left( \frac{1}{\hat{r}^2} - \alpha^2 \right) e^{-\alpha^2 \hat{r}^2},
\end{equation}
\begin{equation}
\begin{array}{c}
T^l_{ijk} (\mathbf{x},\mathbf{x}^\nu)  =  -\displaystyle \frac{6 \hat{x}_i \hat{x}_j \hat{x}_k}{\hat{r}^5} \textnormal{erfc}(\alpha \hat{r})
-\frac{12\alpha}{\sqrt{\pi}}  \frac{\hat{x}_i\hat{x}_j\hat{x}_k}{\hat{r}^4} e^{-\alpha^2 \hat{r}^2} \\  \\
+ \; \displaystyle\frac{4\alpha^3}{\sqrt{\pi}}
\displaystyle\left(\delta_{jk}\hat{x}_i + \delta_{ik}\hat{x}_j + \delta_{ij}\hat{x}_k - \frac{2\hat{x}_i\hat{x}_j\hat{x}_k}{\hat{r}^2}\right) e^{-\alpha^2 \hat{r}^2},
\end{array}
\end{equation}
\end{subequations}
where $\mathbf{\hat{x}} = \mathbf{x} - \mathbf{x}^\nu$, while $\hat{r} = |\mathbf{\hat{x}}|$.
The velocity and stress fields are then obtained as:
\begin{subequations}\label{eq:u_loc}
\begin{equation}
u^l_i(\mathbf{x}) = \frac{1}{8\pi\mu} \sum_{\nu=1}^{N_s} G^l_{ij}(\mathbf{x},\mathbf{x}^\nu) g^\nu_j,
\end{equation}
\begin{equation}
\sigma^l_{ik}(\mathbf{x}) = \frac{1}{8\pi} \sum_{\nu=1}^{N_s} T^l_{ijk}(\mathbf{x},\mathbf{x}^\nu) g_j^\nu.
\end{equation}
\end{subequations}

The solution in Eq. (\ref{eq:local}) has been obtained with \textit{free-space} boundary
conditions, i.e. all of them decay to zero at infinity. In other words, 
the local solution is independent of the geometry of
interest. The violation of the boundary condition
requirements of the domain by employing free space boundary conditions above
will be corrected by appropriately choosing the boundary conditions
for the flow problem associated with the global force densities as the forcing
function. 

An important observation at this point is that the local solutions in (\ref{eq:local}) are
short ranged, decaying approximately as $e^{-\alpha^2\hat{r}^2}$. Consequently, the contribution from the local
solution can be neglected beyond a length scale $\sim\alpha^{-1}$ from the origin of the corresponding local density.
In this work, this cutoff length was taken as $r_{cut}=4/\alpha$ throughout.
The near neighbor list required for the efficient computation of the local solution is 
generated by the $O(N_s)$ cell-linked list algorithm \citep{allen89}.

It is important to point out that the $\mathbf{G}^l$ in Eq. (\ref{eq:local}a)
has the same functional form as the real space term in the periodic Stokeslet (Green's function)
provided by Hasimoto \citep{hasimoto59}. In other words, Hasimoto's solution 
for the periodic Stokeslet  can also be obtained by first splitting the 
$\delta$-function density into the local and global densities as in
Eq. (\ref{eq:rho_l_g}); the local problem is then solved as described 
above, while the global problem is solved with a Fourier Galerkin method with
the appropriate assumptions described in \citet{hasimoto59}. Since PME
accelerated methods (e.g. \citep{kumar11}) for Stokes flow employ the
periodic Stokeslet given by Hasimoto \citep{freund10}, this observation 
illustrates a connection between PME like methods and GGEM. A very important
distinction, though, is that the performance of PME like methods is tied to the
use of discrete Fourier transforms and thus periodic domains, which is not the case
with GGEM (discussed below) and hence the latter's much broader applicability. 

\subsubsection{Global solution}\label{sec:global_sol}
We now describe the solution to the global problem, i.e.
the flow problem associated with the collection of global force
densities. We first discuss the boundary
conditions for the global problem. As was mentioned earlier,
the overall solution for a given collection of point
forces is the sum of the corresponding quantities from the
local and the global solutions, see Eq. (\ref{eq:tot_l_g}).
It is obvious that the same should be true for boundary
conditions. Consequently, to satisfy any type of boundary condition
(e.g. Dirichlet) at an arbitrary location, we  set the boundary
condition for the global part so that its sum with the known contribution
from the local part (above) adds up to the required value. Again, we
employ the example of the slit geometry, noting that this scheme is equally applicable to other
geometries. To satisfy the no-slip condition at the two rigid walls of the slit,
we require the following at $y=0$ and $y=H$:
\begin{equation}
\mathbf{u}^g = - \mathbf{u}^l.
\end{equation}
Note that in the present formulation the \textit{static} no-slip condition is always imposed at the rigid walls for
computing the velocity field due to the Green's function (or point forces).
This is true even in problems where the walls may not be at rest,
a common example being simple shear flow. The effect of the undisturbed flow enters
the boundary integral equation via $\mathbf{u}^\infty$ and
$\mathbf{f}^\infty$ in Eqs. (\ref{eq:bi_ua}) and (\ref{eq:qa}).
To satisfy the periodic boundary conditions in $x$ and
$z$ directions, we impose equivalent periodic boundary
conditions in the global calculation. As far as the local
solution is concerned, we require that it decays 
to a negligible value over a length scale equal
to half of the spatial period in $x$  and $z$ directions or smaller:
i.e. $r_{cut} < L_x/2$ and $r_{cut} < L_z/2$. Given the above
choice $r_{cut}=4/\alpha$, we require that $\alpha L_{x}>8,\alpha L_{z}>8$. This fact,
coupled with the minimum image convention \citep{deserno98} employed in the
computation of the local solution ensures its periodicity.

Before proceeding further, we note that, in general, two subtleties arise in considering
the behavior of the global solution near boundaries. The first is the issue of boundary 
shape. In the present work we take the boundary to be smooth on the scale of the suspended
particle size. If that is not the case, it will be necessary to resolve the length scales
of the boundary roughness. For such a boundary it might be convenient to revert to a
conventional accelerated method and explicitly discretize the boundary. Alternately,
in the present context, the global solution could be obtained using a locally-refined
mesh near the domain boundary to capture its features (see, e.g. \citet{Hulsen}), 
without destroying the scaling of the method with the number of suspended particles. 

Another issue arises in principle if a particle very closely approaches a (smooth) boundary.
This does not arise in the context of the present application as deformable particles
migrate away from solid surfaces in shear flow due to the hydrodynamic dipole interaction
between the particle and the wall \cite{Smart:1991vp}. Nevertheless, it can happen in
principle and leads to the situation where the boundary condition for the global problem that must be satisfied
becomes nearly singular. This is because, at a no-slip boundary, we require $\mathbf{u}^g = - \mathbf{u}^l$
and a point force a distance $\epsilon$ away from the boundary leads to a local velocity  $\mathbf{u}^l$
of $O(1/\epsilon)$ on the boundary.  This situation can be addressed by adding the image system for a plane wall \cite{Blake:1971vj} (and splitting it into local and global parts), in which case the effects of the singularities cancel on the wall. See \citet{swan:2011fh} for a related discussion in the Stokesian dynamics context.

Having discussed the boundary conditions for the global problem,
we turn to the solution procedure of the Stokes and
the continuity equation with the given collection
of global force densities as the forcing function; see Eq. (\ref{eq:stokes_gg}).
For an arbitrary geometry one may employ any desired discretization scheme
for the solution of the global problem. If a finite
difference or a finite element scheme is used, then the solution
can be obtained at a cost of $O(N)$ when the resulting sparse 
matrix equations are solved iteratively with proper preconditioners;
the multigrid preconditioner for Stokes flow is an attractive choice \citep{esw05,silvester94}.
Section \ref{sec:algo_com} contains further discussion of the scaling of computation time with problem size. 
For the slit problem of interest here, past work \citep{ortiz08,pratik10}
employed discrete Fourier series approximation in the periodic $x$ and $z$ directions,
while a second order finite difference discretization was employed in the wall
normal $y$ direction. In the present work, we develop a fully spectral solution
procedure by employing the discrete Chebyshev polynomial approximation \citep{canuto06} in the 
wall normal direction, while the discrete Fourier series approximation is used 
in the periodic $x$ and $z$ directions. For example, the $x$-component $u^{g}$ of the 
global velocity $\mathbf{u}^g = (u^g,v^g,w^g)$ in Eq. (\ref{eq:stokes_gg}) is expressed as 
\begin{equation}\label{eq:u_expn}
u^g(\mathbf{x}) = \displaystyle \sum_{l=-N_x/2}^{N_x/2-1} \; \sum_{m=-N_z/2}^{N_z/2-1} \; \sum_{n=0}^{N_y-1} \;   \hat{u}^g_{lmn} \; T_n(\bar{y}) \;  e^{i 2 \pi l x/L_x} \; e^{i 2\pi m z/L_z},
\end{equation}
where $T_n(\bar{y})=\cos(n \cos^{-1}\bar{y})$ is the Chebyshev polynomial of the $n^{th}$ degree \citep{canuto06},  $\bar{y}$ represents the mapping from $[0,H]$ to $[-1,1]$: $\bar{y} = 2y/H -1$, while $N_x$, $N_y$, and $N_z$ respectively
denote the number of terms (modes) in the corresponding series approximation. Similar expressions
are written for other components of the velocity and the pressure. {An important implication of this
representation, particularly with regard to the pressure, is that the pressure drop associated with this 
point force solution is
always zero over the spatial period of the domain, while the mean flow is (in general) non-zero.
This ensures that the pressure drop obtained from the boundary integral
method always equals the pressure drop  specified in the imposed bulk flow
(i.e. in the absence of the particles).} Returning to the expression in Eq. (\ref{eq:u_expn}),
we note that the use of the Fourier series approximation in $x$ and $z$ directions 
ensures that the periodic boundary conditions in these directions are inherently satisfied.
The Chebyshev polynomials, on the other hand,
do not automatically satisfy the boundary conditions in the wall normal direction;
the satisfaction of these boundary conditions was accomplished by employing
the tau method \citep{canuto06,peyret02}.
In the tau method, the equations for the highest two modes in the series approximation are replaced by equations
representing the two boundary conditions; see, e.g., \citet{canuto06} or \citet{peyret02} for details.
An attractive feature of the discrete Chebyshev polynomial approximation is that FFTs can
be used for rapidly transferring information from the physical to spectral space and vice-versa \citep{canuto07}.
A major drawback, though, with solving differential equations with Chebyshev polynomial 
approximation is that the differentiation matrix is full in both the spectral and the physical
space \citep{peyret02} (in contrast, the Fourier differentiation matrix is diagonal in the spectral space).
Due to the full nature of the Chebyshev differentiation matrix, a straightforward 
implementation for solving the Stokes flow problem will lead to an $O(N_y^3)$ method.
For the incompressible Stokes flow problem here, though, an alternate approach exists
in which the solution to the Stokes equation is obtained from the solution of a
series of Helmholtz equations \citep{canuto07}.
In this case, with a little manipulation, a quasi-tridiagonal system of equations
results, for which a direct $O(N_y)$ algorithm exists \citep{peyret02}. This approach 
for solving the incompressible Stokes equation, or more generally the incompressible
Navier-Stokes equations, is popularly known in the literature as the Kleiser-Schumann
influence matrix method \citep{canuto07}. A detailed discussion of this
approach including the equations being solved and their respective boundary
conditions is presented in \ref{sec:inf_mat}. Here we only sketch out the main
computational aspects of this approach. 
To begin, each of variables appearing in the Stokes and the continuity
equation are expanded in a truncated Fourier series in $x$ and $z$ directions;
see, e.g., Eq. (\ref{eq:u_expn}). These expressions are then substituted in the 
Stokes and the continuity equations. Subsequently, by the application of the Galerkin method,
a set of coupled ordinary differential equations (ODE) in $y$ is obtained for each of the
Fourier modes of all the unknown variables (velocity components and pressure).
These coupled ODEs are solved with the Chebyshev-tau
influence matrix method, which involve Chebyshev transformations,
quasi-tridiagonal matrix equation solves, and inverse 
Chebyshev transformations in that order. The solution thus obtained yields the
Fourier coefficients of the velocity components and the pressure. An inverse Fourier transform 
then leads to the solution for the velocity and the pressure in the physical space.
The computation of the stress tensor $\boldsymbol{\sigma}^g$ requires the derivatives
of the velocities; these differentiations are performed in the transform space \citep{canuto06,peyret02}.
All of the above Fourier and the Chebyshev transforms along with their inverse transforms
are performed using the FFT algorithm. Thus,  the 
asymptotic computational cost of the solution procedure for the global problem 
scales as $N\log N$, where $N=N_x N_y N_z$. Assuming $N \sim N_s$ (see Sec. \ref{sec:algo_com}),
we obtain the asymptotic computational cost of the global solution as $O(N_s \log N_s)$. A further discussion on the
computational complexity of the algorithm is provided in Sec. (\ref{sec:algo_com}).

We next introduce some of the important parameters associated with this solution 
procedure. Associated with each of the $N_x$ and $N_z$ Fourier modes, there 
are $N_x$ and $N_z$ equispaced trapezoidal quadrature points; the corresponding
spacings are denoted by $\Delta x_m = L_x/N_x$ and $\Delta z_m = L_z/N_z$.
Similarly, associated with the $N_y$ Chebyshev polynomials, there are $N_y$ 
Chebyshev Gauss-Lobatto quadrature points, the $j^{th}$ of which is
given by $y_j = H/2(1+\cos(\pi (j-1) /(N_y-1)))$;
the mean mesh spacing in this case is denoted by $\Delta y_m = H/(N_y-1)$.
Unless otherwise mentioned, the mean mesh spacings in all three 
directions are kept equal in simulations, i.e. $\Delta x_m = \Delta y_m = \Delta z_m$.
For computing any of the above transforms, the value of the corresponding physical variable 
is required only at the corresponding quadrature points. Likewise, as is customary, the final
solution for the velocity, pressure and stress are also computed only at these quadrature 
points. This last step is essential in maintaining the optimal computational complexity
of $O(N\log N)$ alluded to above. The velocity and stress at any point not on the mesh 
is obtained via interpolation; here we employ $4^{th}$ order 
Lagrange interpolation for which the error decays as $h^5$, where $h$ is the 
characteristic mesh spacing. The error, therefore, is expected to decay exponentially fast
with the number of modes for any point on the mesh, while it is expected to decay
as $h^5$ for any point not on the mesh. It is appropriate to pointÊout here that 
exponential convergence of the solution is possible even at a non-mesh point while maintaining 
the computational complexity of $O(N\log N)$ -- this can be achieved by employing the basic principles 
of non-uniform FFT calculations; see, e.g.,  \citep{greengard04,tornberg10} for details.

\begin{figure}[!t]
\centering
\subfigure[Observation point is a mesh point]{\includegraphics[width=0.45\textwidth]{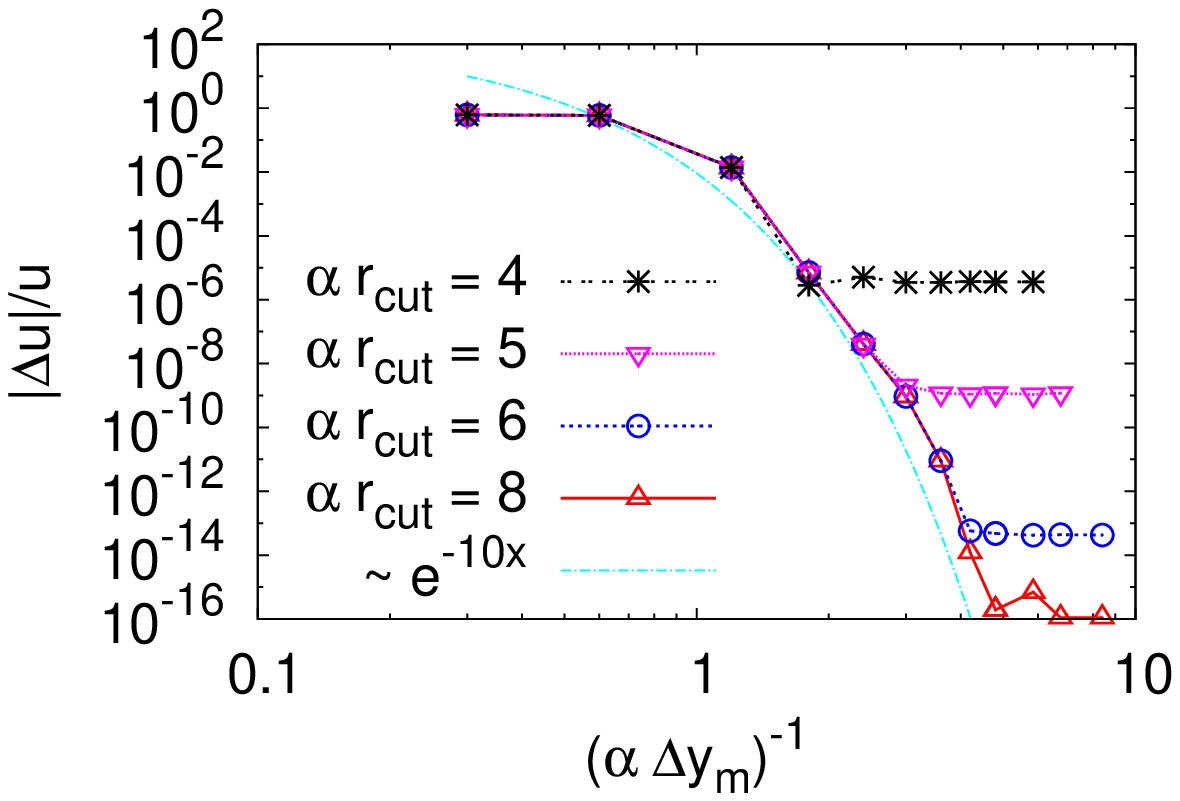}}
\subfigure[Observation point is not a  mesh point]{\includegraphics[width=0.45\textwidth]{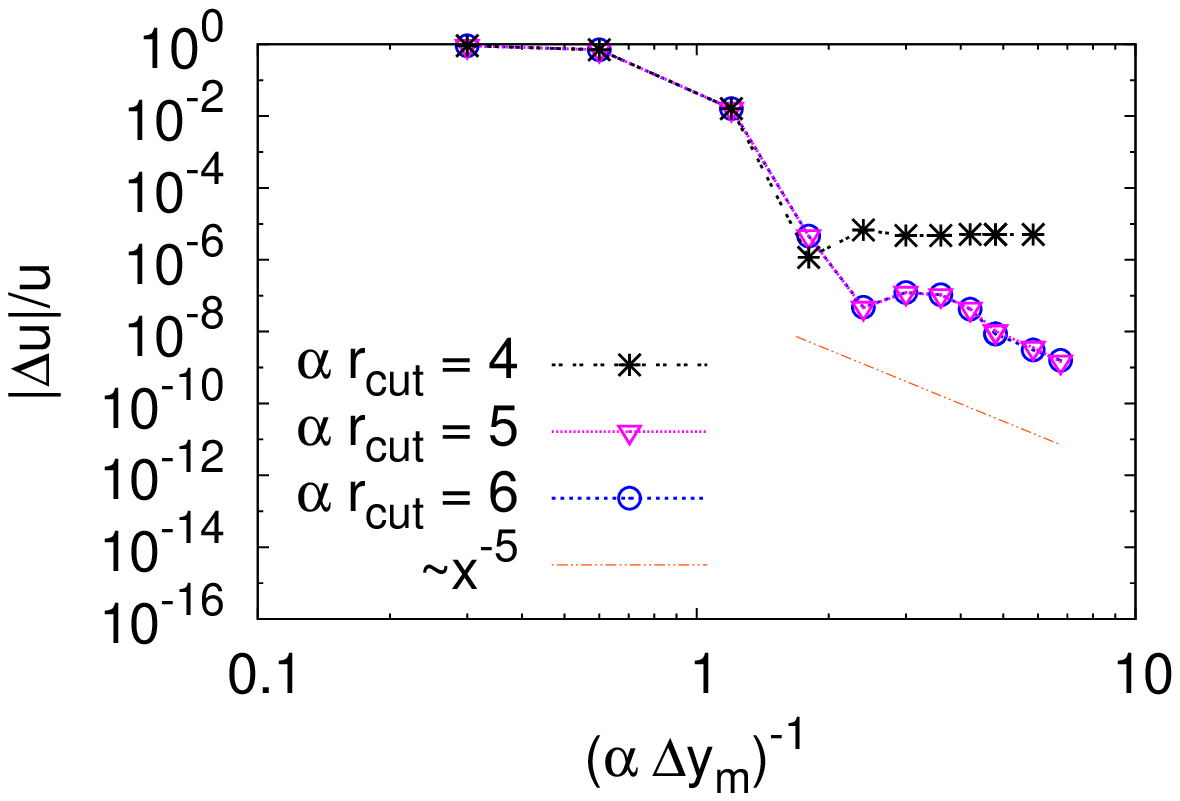}}
\caption{Relative error in the velocity ($\Delta u /u $) due to a point force  at a given observation
point in a slit geometry with $L_x=H=L_z=L$. The abscissa in the plots is $(\alpha \Delta y_m)^{-1}$,
while different curves are for different values of $r_{cut}$ as labeled in the key.
The strength of the point force is given by $(F,F,F)$, while its coordinates are $(0.25L,0.25L,0.25L)$.
(a) The observation point is $(0.5L,0.5L,0.5L)$, which is always maintained on 
a mesh point, and (b) the observation point is $(0.4312L,0.3734L,0.5234L)$ which is not 
a mesh point. In (a) we also plot the function $y \sim e^{-10x}$ (chosen to  closely match other curves 
on the plot) to demonstrate the exponential convergence of the solution,
while in (b) we plot the function $y \sim x^{-5}$ to demonstrate
the algebraic convergence due to dominant interpolation errors at higher values of $(\alpha \Delta y_m)^{-1}$.}\label{fig:ggem_conv}
\end{figure}

\subsubsection{Convergence of the GGEM solution}\label{sec:ggem_test}
We demonstrate next the convergence behavior of the GGEM Stokes flow solution presented above. 
It will be helpful to begin this section with a discussion of various sources of error in the solution procedure. It should be obvious that the overall error in the solution results from errors in both the local and the global solution procedures. The error in the local solution arises due to its truncation beyond a distance of $r_{cut}$ from the source of the singularity -- this error scales as $e^{-\alpha^2 r_{cut}^2}$ (Sec. \ref{sec:local_sol}, \citep{tornberg10}). Typically, we set $r_{cut}=4/\alpha$, which is expected to result in an error of $O(10^{-7})$. Smaller error in the local solution can be obtained by increasing the value of $r_{cut}$. The error in the global solution has three different sources. The first source of error is the truncation of the Fourier-Chebyshev series expansion at some finite number of modes; see Eq. (\ref{eq:u_expn}). The error due to this truncation is expected to decay exponentially fast on the mesh points with the 
parameter $(\alpha \Delta y_m)^{-1}$ -- this convergence rate results due to the spectral
nature of the global solution procedure employed here. For other solution procedures,
such as a finite difference scheme \citep{ortiz08,pratik10}, an 
algebraic convergence with the parameter $(\alpha \Delta y_m)^{-1}$ will be obtained.
Next, for any point not on the mesh, the global solution has to be obtained by interpolation from nearby 
mesh points, which introduces an additional error scaling as $(\alpha \Delta y_m)^5$
in the current work. Lastly, there is also an error associated with the assignment
of the global force density on the mesh points -- this is required for computing its Fourier and Chebyshev transforms. 
	The global density decays approximately as $e^{-(\alpha r)^2}$ with distance $r$ from the 
	origin of the density (Eq. \ref{eq:rho_l_g}a). Therefore, just like the local solution,
	we truncate the global density beyond a distance of $r_{cut} = 4/\alpha$ from the 
	origin of the singularity. It is expected that the error due to this truncation 
	will scale as $e^{-(\alpha r_{cut})^2}$. Again, a smaller truncation error can be obtained
	by setting a larger value of $r_{cut}$ -- in such cases the cost associated with the global force density assignment on the mesh points can be large and the fast Gaussian gridding algorithm  \citep{greengard04,tornberg10,tornberg11} is recommended.
	
	Having discussed the various sources of errors in the solution procedure, we turn to verifying the expected
	convergence behavior with a test problem.  In this test problem, we compute the velocity field 
due to a point force in a slit of side $L$, i.e. $L_x=L_z=H=L$.
The point force is located at coordinates $(0.25L,0.25L,0.25L)$, while its
strength is given by $(F,F,F)$. The velocity due to this point force
will be computed at two target points. The first target point is 
located at the center of the box $(0.5L,0.5L,0.5L)$, which is easy to maintain
on a mesh point, while the second target point is a randomly chosen point with 
coordinates $(0.4312L,0.3734L,0.5234L)$, which is unlikely to be a mesh point.
In this test study, we will keep the value of $\alpha$ fixed at $\alpha =80/3L$, while the value 
of $r_{cut}=C/\alpha$ ($C$ is a constant) and the mean mesh spacing $\alpha \Delta y_m$ will be varied.
Figure  (\ref{fig:ggem_conv}a) shows the relative error in the $x$ component of the velocity $\Delta u/u$ 
as a function of $(\alpha \Delta y_m)^{-1}$ for several different values of $r_{cut}$
for the first target point, while the same is shown for the second target point
in Fig. (\ref{fig:ggem_conv}b). The data points in these plots 
were obtained by varying $N_y$ between $17$ and $181$, while
the solution computed with $N_y=225$ and $r_{cut}=0.5L$ (i.e., $C=40/3$) is taken as the
reference for computing the relative error. Focusing first on Fig. (\ref{fig:ggem_conv}a),
we observe an exponential convergence in the velocity with increasing $(\alpha \Delta y_m)^{-1}$,
though it eventually levels off at a value depending on the choice of $r_{cut}$.
For the typical value of $r_{cut}=4/\alpha$, an error of $O(10^{-6})$ is obtained. 
We next focus on the velocity convergence at the second target point
(Fig. \ref{fig:ggem_conv}b), which is a non-mesh point.
For the choice $r_{cut}=4/\alpha$, we again observe an exponential convergence 
initially with increasing $(\alpha \Delta y_m)^{-1}$ that eventually levels off
at approximately the same value as for the first target point.
For higher values of $r_{cut}$ ($C=5$ and $C=6$), we observe an exponential convergence 
initially, though a convergence rate scaling as $(\alpha \Delta y_m)^5$ is observed at higher
values of $(\alpha \Delta y_m)^{-1}$. The latter convergence rate results from interpolation errors becoming dominant
at higher values of $(\alpha \Delta y_m)^{-1}$. We also note that an exponential convergence
is observed only when the length scale of the global
force density is well resolved by the numerical mesh. Since the length scale 
	of the global force density is represented
by $\alpha^{-1}$, the requirement for an exponential convergence is quantitatively expressed by
the condition $(\alpha \Delta y_m)^{-1} > 1$  (i.e. $\alpha \Delta y_m < 1$).
This requirement on the mesh spacing is more 
easily appreciated if one interprets $(\alpha \Delta y_m)^{-1}$ as the number of mesh points
per unit smearing length represented by $\alpha^{-1}$; therefore the larger is
$(\alpha \Delta y_m)^{-1}$, the higher is the resolution of the numerical scheme.
Based on extensive numerical tests presented in this paper, 
$\alpha \Delta y_m=0.5$ is a recommended value as convergence was usually
observed at this resolution.

\begin{figure}[!t]
\centering
\subfigure[Triangulation]{\label{fig:triangulation} \includegraphics[width=0.2\textheight]{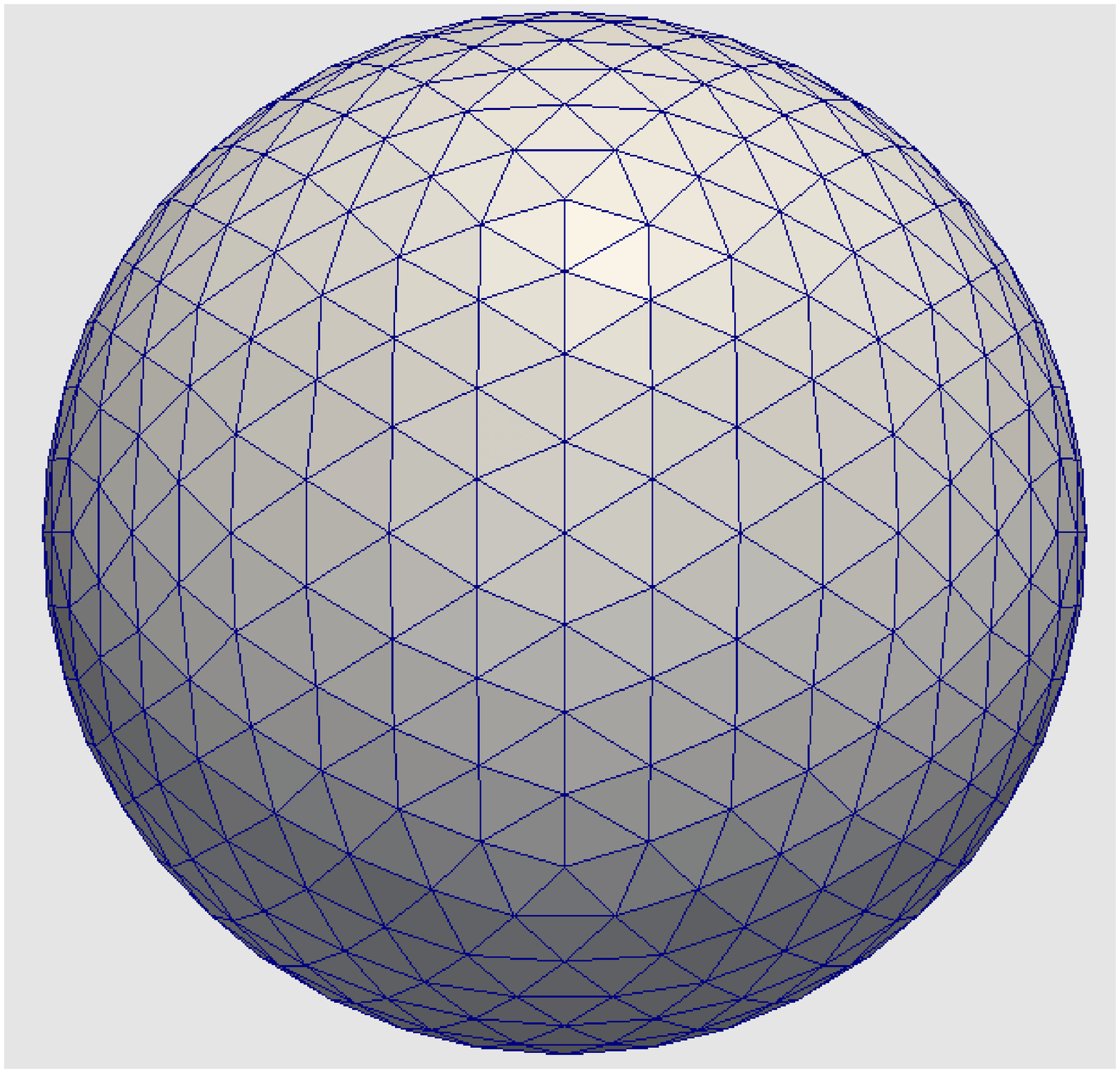}}
\hspace{10mm}
\subfigure[Parent Triangle]{\label{fig:nat_crd} \includegraphics[width=0.2\textheight]{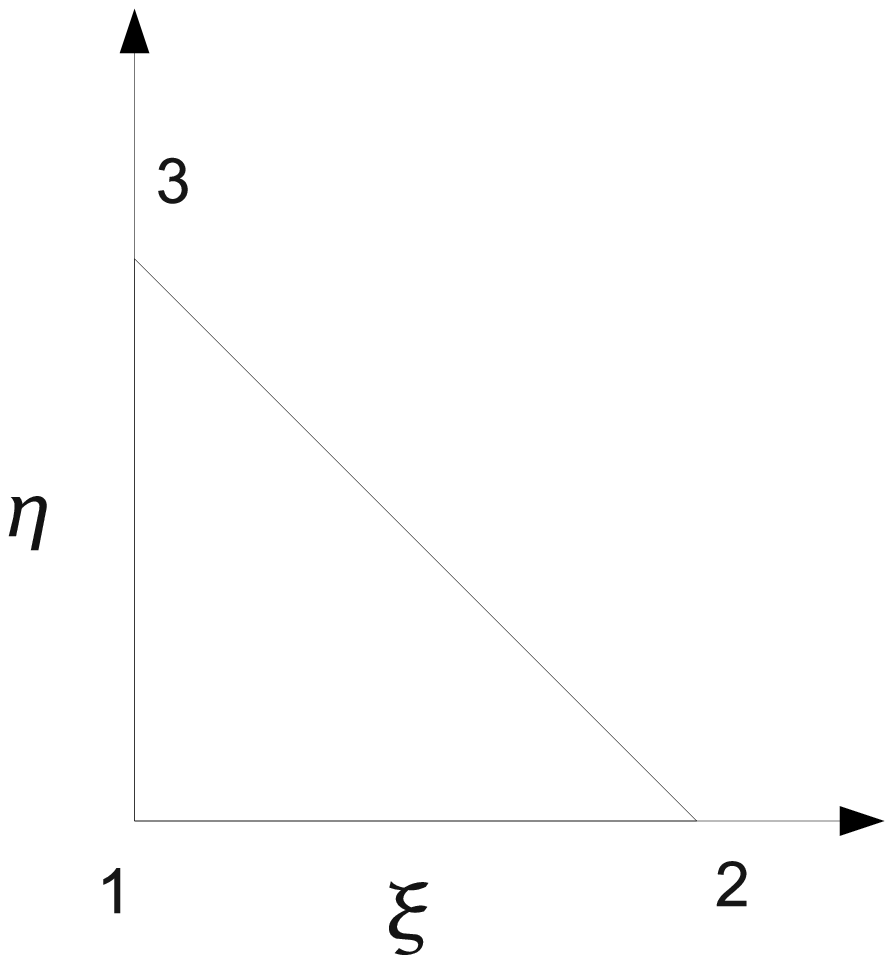}}
\caption{(a) Discretization of a sphere into triangular elements.
The number of triangular elements and the number of vertices
for the discretization shown in the figure are
$N_{\Delta}=1280$ and $N_b=642$ respectively, (b)
A schematic of the parent triangle. Edges 1-2 and
1-3 are of unit length.}
\end{figure}


\subsection{Surface discretization}\label{sec:surf}
Having described the procedure for the fast computation of the velocity and the stress
fields associated with a given collection of point forces, we now turn to the 
numerical solution of the boundary integral equation introduced in Sec. (\ref{sec:BI_eq}).
In this section, we describe the discretization of the particle's surface into
elements along with the basis functions employed over each element. Following this, 
we describe the numerical implementation of the single and the
double layer integrals present in the boundary integral equation. It should be emphasized
that accelerated approach described here is not limited to the specific surface discretization
used here; this discretization was chosen because it has been used in past works on the dynamics
of fluid-filled elastic capsules and drops in flow \cite{bagchi09,kumar11b,pratik10,hinch96,zinchenko2000}.

In the present work, the surface of a capsule is discretized into triangular elements. Triangulation
of a sphere is achieved by mapping the vertices of an
icosahedron, which has 12 vertices and 20 triangular faces, to the
surface of the inscribed sphere \citep{pozrikidis98}. This procedure will, therefore,
give 20 elements on the surface of the sphere. Further
refinement is obtained by subdividing each triangular
face of the icosahedron recursively into 4 equal
triangular elements, with all the vertices (and consequently
the elements) again being mapped to the surface of the inscribed sphere
as described above. The number of elements ($N_\Delta$)
and the number of vertices ($N_b$) obtained by this procedure can be
expressed as $N_\Delta = 20\cdot4^k$ and $N_b = N_\Delta/2+2$,
where $k$ is the level of refinement ($k=0$ corresponds to the
original icosahedron). Note that the 12 original vertices
of the icosahedron have a coordination number 5, while
the remaining vertices have a coordination number of 6.
As an example, a sphere subdivided into $N_\Delta=1280$ elements with
$N_b=642$ vertices is shown in Fig. (\ref{fig:triangulation}).

\subsubsection{Basis functions over elements}
Linear basis functions are used over each element. All
computations over a triangular element is performed
by mapping to or from the parent triangle \citep{hughes87}. The parent triangle
employed in this work is shown in Fig. (\ref{fig:nat_crd}),
where $\xi$ and $\eta$ denote the natural coordinates. The basis
functions associated with the nodes 1, 2, and 3 are respectively given
in natural coordinates by
\begin{subequations}
\begin{equation}
\phi_1(\xi,\eta) =  1 - \xi - \eta,
\end{equation}
\begin{equation}
\phi_2(\xi,\eta) =  \xi,
\end{equation}
\begin{equation}
\phi_3(\xi,\eta) =  \eta.
\end{equation}
\end{subequations}

As an example, the position vector $\mathbf{x}$ as a function of natural
coordinates $\mathbf{x}(\xi,\eta)$ is obtained as
\begin{equation}\label{eq:basis_int}
\mathbf{x}(\xi,\eta) = \phi_1(\xi,\eta) \, \mathbf{x}_1 + \phi_2(\xi,\eta) \,  \mathbf{x}_2
+ \phi_3(\xi,\eta) \,  \mathbf{x}_3,
\end{equation}
where $\mathbf{x}_1$, $\mathbf{x}_2$, and $\mathbf{x}_3$ are the real space
positions of vertices 1,2, and 3 respectively. The same procedure
is employed to obtain the value of any physical variable (e.g. velocity)
at coordinates ($\xi,\eta$) over the domain of the parent triangle.


\subsection{Single layer integral}\label{sec:sl_comp}
Let the single layer integral over the surface $S$ be denoted by
\begin{equation}\label{eq:sl_t}
w_j(\mathbf{z}) = \int_{S} q_i(\mathbf{x})\, G_{ji}(\mathbf{z},\mathbf{x})\,dS(\mathbf{x}),
\end{equation}
where $ q_i(\mathbf{x})$ is the single layer density, while
$\mathbf{w}(\mathbf{z})$ is assumed to represent the velocity at point $\mathbf{z}$.
In order to employ GGEM as discussed in Sec. (\ref{sec:greens_fn})
to compute the above integral, we write this equation
in the form
\begin{equation}\label{eq:sl_t1}
\displaystyle w_j(\mathbf{z}) = \int_{S} \int_{V} q_i(\mathbf{x}) \delta(\mathbf{y}-\mathbf{x}) \,
G_{ji}(\mathbf{z},\mathbf{y})\,dS(\mathbf{x}) \,dV(\mathbf{y}),
\end{equation}
where $V$ represents the volume of the domain and $\delta$ is the
three dimensional Dirac delta function. It is easy to see that
both the expressions for $\mathbf{w}(\mathbf{z})$ in Eqs. (\ref{eq:sl_t})
and (\ref{eq:sl_t1}) are identical. Next, we write the Dirac-delta
function as a sum of the local and the global density introduced
in Sec. (\ref{sec:greens_fn}); see Eq. (\ref{eq:rho_l_g}). Consequently,
we have 
\begin{equation}\label{eq:sl_t2}
\displaystyle w_j(\mathbf{z}) = \int_{S} \int_{V} q_i(\mathbf{x}) \left(\rho_l(\mathbf{y}-\mathbf{x})
+\rho_g(\mathbf{y}-\mathbf{x}) \right)   \,
G_{ji}(\mathbf{z},\mathbf{y})\,dS(\mathbf{x}) \,dV(\mathbf{y}),
\end{equation}
Next, we separate the integrals associated with the local and global densities,
and write the contribution due to the local density as
\begin{equation}\label{eq:sl_la}
w_j^l(\mathbf{z}) = \int_{S}  q_i(\mathbf{x})  \, \left( \int_{V} \rho_l(\mathbf{y}-\mathbf{x})
G_{ji}(\mathbf{z},\mathbf{y}) dV(\mathbf{y}) \right) \,dS(\mathbf{x}).
\end{equation}
It is easy to see that the above integral can be written as
\begin{equation}\label{eq:sl_l}
w_j^l(\mathbf{z}) = \int_{S} q_i(\mathbf{x})\, G^l_{ji}(\mathbf{z},\mathbf{x})\,dS(\mathbf{x}),
\end{equation}
where $\mathbf{G}^l$ has been defined in Eq. (\ref{eq:local}). This follows
from the fact that the local Green's function $\mathbf{G}^l(\mathbf{z},\mathbf{x})$
can also be constructed by the superposition of Green's function
$\mathbf{G}(\mathbf{z},\mathbf{y})$ weighted by the
density $\rho^l(\mathbf{y}-\mathbf{x})$, i.e.
\begin{equation}\label{eq:gl_g}
G^l_{ji}(\mathbf{z},\mathbf{x}) = \int_{V} \rho_l(\mathbf{y}-\mathbf{x})
G_{ji}(\mathbf{z},\mathbf{y}) dV(\mathbf{y}).
\end{equation}
It is important to emphasize that the domain  was assumed
to be unbounded in arriving at Eq. (\ref{eq:gl_g}). This
is \textit{always} the case for the local problem as discussed in Sec. (\ref{sec:greens_fn});
any error in the boundary condition introduced due to this assumption will be
accounted for in the global calculation. 
Next, consider the contribution from the global density in Eq. (\ref{eq:sl_t2}), which we write as
\begin{equation}\label{eq:sl_ga}
w_j^g(\mathbf{z}) = \int_{S} \int_{V}  q_i(\mathbf{x})  \rho_g(\mathbf{y}-\mathbf{x})
G_{ji}(\mathbf{z},\mathbf{y}) \, dV(\mathbf{y}) \,dS(\mathbf{x}).
\end{equation}
It can shown that $\mathbf{w}^g(\mathbf{z})$ satisfies
\begin{subequations}\label{eq:stokes_gga}
\begin{equation}
-\bm{\nabla} p^{w^g}(\mathbf{z})  + \mu \nabla^2 \mathbf{w}^g(\mathbf{z})+ \mu \, \boldsymbol{\Pi}^g(\mathbf{z}) = 0,
\end{equation}
\begin{equation}
\bm{\nabla} \cdot \mathbf{w}^g(\mathbf{z}) = 0,
\end{equation}
\end{subequations}
where the density $\boldsymbol{\Pi}^g(\mathbf{z})$ is given by
\begin{equation}\label{eq:rho_int}
\boldsymbol{\Pi}^g(\mathbf{z}) = 8\pi \int_{S} \mathbf{q}(\mathbf{x})  \rho_g(\mathbf{z}-\mathbf{x}) \,dS(\mathbf{x}).
\end{equation}
The boundary condition for the global solution comes from the known local
solution ($\mathbf{w}^l(\mathbf{z})$ for $\mathbf{z}$ at the
domain boundary) and the given overall boundary conditions, such that the
sum of the local and the global solution satisfies the overall boundary
condition; see Sec. (\ref{sec:greens_fn}). Having expressed
the single layer integral in a form suitable for its computation
with the GGEM technique, we next describe its numerical implementation.
This includes the numerical solution of the local problem represented by Eq. (\ref{eq:sl_l})
and the global problem represented by Eq. (\ref{eq:stokes_gga}).

\subsubsection{Local contribution}\label{sec:sl_l}
We first consider the contribution from the local Green's function to
the velocity at a given point $\mathbf{z}$, which typically is one of the
nodes of the elements. We discretize the surface integral in Eq. (\ref{eq:sl_l})
as
\begin{equation}
w_j^l(\mathbf{z}) = \sum_{k=1}^{N_\Delta N_p} \int_{S_k} q_i(\mathbf{x})\, G_{ji}^l(\mathbf{z},\mathbf{x})\,dS(\mathbf{x}),
\end{equation}
where the summation is over all the triangular elements $N_\Delta N_p$ present in the
system, while $S_k$ denotes an integral over the element $k$.
For convenience, all the integrals are performed
over the parent triangle. To accomplish this, we write the above
equation as
\begin{equation}\label{eq:int_l}
w_j^l(\mathbf{z}) = \displaystyle \sum_{k=1}^{N_\Delta N_p} \int_0^1 \int_0^{\xi(\eta)} q_i\left(\mathbf{x(\xi,\eta)}\right)\; G_{ji}^l\left(\mathbf{z},\mathbf{x}(\xi,\eta)\right)\; \omega \; d\xi d\eta,
\end{equation}
where the differential area element $dS$ has been replaced by its equivalent
expression
\begin{equation}
dS \,= \, \omega \, d\xi \, d\eta = |\mathbf{x}_\xi \times \mathbf{x}_\eta | \, d\xi \, d\eta.
\end{equation}
As noted earlier, the value of any quantity can be obtained at coordinates
$(\xi,\eta)$ by the usual interpolation from the corresponding values at the
nodes of the triangle, e.g. see Eq. (\ref{eq:basis_int}).
The double integral in Eq. (\ref{eq:int_l}) is
evaluated using the product of two one-dimensional Gauss-Legendre quadrature
rule (one for $\xi$ and the other for $\eta$). This proved competitive in
terms of computational cost for a given accuracy with Gaussian quadrature rules
available for a triangular element \citep{hughes87}, perhaps due to the
fact that the integrands are not polynomials. In most cases a 4x4 product
rule is found to be sufficient for accurate integration
over a triangular element. In addition, if the vertex at which the velocity
is being computed is a member of the triangular element over which
the integration is being performed, then the integral in Eq. (\ref{eq:int_l})
over the parent triangle is further transformed to polar coordinates
$(r,\theta)$ \citep{pozrikidis98}. This transformation makes the integrand
non-singular and hence ensures sufficient accuracy with the same low order product
integration rule discussed above. Lastly, we note that for computing the
contribution to velocity at any given point $\mathbf{z}$
due to the local Green's function, only triangular
elements within a distance of $ r_{cut} \sim \alpha^{-1}$ from the
point  $\mathbf{z}$ need to be considered (typically, $r_{cut}=4/\alpha$). This is justified due to the exponentially
decaying contribution from the integral over an element at separations
larger than $ O(\alpha^{-1})$ from the point of interest.
As mentioned in Sec. (\ref{sec:local_sol}), the near neighbor list required for the local 
calculation is generated in $O(N)$ time via the cell-linked list method \citep{allen89}.

\subsubsection{Global Contribution}\label{sec:sl_g}
Our goal here is to find $\boldsymbol{\Pi}^g(\mathbf{z})$ in Eq. (\ref{eq:stokes_gga}),
for which we need to compute the integral in Eq. (\ref{eq:rho_int}).
We being by discretizing the integral in Eq. (\ref{eq:rho_int})
and write it as 
\begin{equation}\label{eq:rho_g_int}
\boldsymbol{\Pi}^g(\mathbf{z}) = \sum_{k=1}^{N_\Delta N_p} \int_{S_k} q_i(\mathbf{x})\, \rho_g(\mathbf{z}-\mathbf{x})\,dS(\mathbf{x}).
\end{equation}
The above integral can be evaluated with any desired quadrature
rule, though, due to the smoothly varying nature of the integrand
a simple trapezoidal rule proves sufficient. Note that the trapezoidal
integration rule essentially reassigns the contribution
from the surface of the triangular element to its three vertices in equal proportions.
Consider first the integral over an element $S_k$ in Eq. (\ref{eq:rho_g_int}),
which as per the trapezoidal rule is expressed as a sum of
contributions from its three vertices as
\begin{equation}\label{eq:rhog_sum}
\left[\Pi_i^g(\mathbf{z})\right]_{S_k} =  \sum_{p=1}^{3} \left(\frac{A_k q_i(\mathbf{x}^{k_p})}{3}\right)  \; \rho_g(\mathbf{z}-\mathbf{x}^{k_p}),
\end{equation}
where $A_k$ is the area of the triangular element $S_k$, $p$ denotes the vertex
number of the given element $k$, $\mathbf{x}^{k_p}$ denotes
the coordinate of the $p^{th}$ vertex of the triangular element $k$, and
$\left[\boldsymbol{\Pi}^g(\mathbf{z})\right]_{S_k}$ denotes the density
at $\mathbf{z}$ due to the integration over the element $S_k$ only.
The term in the parenthesis in Eq. (\ref{eq:rhog_sum}) can be
considered as the strength of the global density at the node $\mathbf{x}^{k_p}$
due to the element $k$; by summing it over all the elements $k$ to which
a given node belongs (let us say that this node is \textit{globally} represented
by $\mathbf{x}^b$), one obtains the total strength of the global density at this node,
say $\mathbf{q}^{\mathbf{x}^b}$. The overall density at a point
$\mathbf{z}$ is then obtained by adding contributions from all the nodes present in the system,
\begin{equation}
\Pi_i^g(\mathbf{z}) = \sum_{p=1}^{N_b} q_i^{\mathbf{x}^b}  \; \rho_g(\mathbf{z}-\mathbf{x}^{b}),
\end{equation}
where $N_b$ is the total number of nodes in the system.
We also note that due to the exponentially decaying nature of
$\rho_g(\mathbf{z}-\mathbf{x}^{b})$ as a function of distance from $\mathbf{x}^{b}$,
we consider only those nodes for computing the density at a point $\mathbf{z}$ which are within
a distance $r_{cut} \sim \alpha^{-1}$ from it. Once $\Pi^g(\mathbf{z})$ is evaluated, we solve the set of equations in (\ref{eq:stokes_gga}) using the procedure described in detail in Sec. (\ref{sec:greens_fn}).
This gives us $\mathbf{w}^g(\mathbf{z})$ at the mesh points. The velocity $\mathbf{w}^g(\mathbf{z})$
at any point not on the mesh is obtained using $4^{th}$ order Lagrange interpolation.
Once $\mathbf{w}^g(\mathbf{z})$ is known, the overall single layer integral $\mathbf{w}(\mathbf{z})$ is 
obtained as:
\begin{equation}
\mathbf{w}(\mathbf{z}) = \mathbf{w}^l(\mathbf{z}) + \mathbf{w}^g(\mathbf{z}).
\end{equation}

\subsection{Double layer integral}\label{sec:dbl_comp}
We now describe the evaluation of the double layer integral.
We denote the double layer integral over a surface $S$
(similar to that in Eq. \ref{eq:qa}) by
\begin{equation}\label{eq:wdbl}
v_j(\mathbf{z}) = n_k(\mathbf{z}) \int_S q_i(\mathbf{x}) \, T_{jik}(\mathbf{z},\mathbf{x}) \, dS(\mathbf{x}).
\end{equation}
If we define a stress tensor $\mbox{\boldmath{$\sigma$}($\mathbf{z}$)}$  as (note that one needs to multiply
it by $\mu$ to get units of stress)
\begin{equation}\label{eq:st_dbl}
\sigma_{jk}(\mathbf{z}) = \int_S q_i(\mathbf{x}) \, T_{jik}(\mathbf{z},\mathbf{x}) \, dS(\mathbf{x}),
\end{equation}
then we have the following relationship between $\mathbf{v}(\mathbf{z})$
and $\mbox{\boldmath{$\sigma$}($\mathbf{z}$)}$:
\begin{equation}\label{eq:vsigma}
v_j(\mathbf{z}) = n_k(\mathbf{z})\sigma_{jk}(\mathbf{z}).
\end{equation}
The motivation for introducing the stress field $\mbox{\boldmath{$\sigma$}($\mathbf{z}$)}$
in Eq. (\ref{eq:st_dbl}) should be clear now, as it is the stress field associated with the
following velocity field
\begin{equation}\label{eq:u_dbl}
w_j(\mathbf{z}) = \int_S q_i(\mathbf{x}) \, G_{ji}(\mathbf{z},\mathbf{x}) \, dS(\mathbf{x}).
\end{equation}
As described in the previous section (\ref{sec:sl_comp}), we write the above
velocity field as the sum of a local $\mathbf{w}^l(\mathbf{z})$ and  global $\mathbf{w}^g(\mathbf{z})$ velocity
fields, and denote the corresponding stress fields by $\mbox{\boldmath{$\sigma^l$}($\mathbf{z}$)}$
and $\mbox{\boldmath{$\sigma^g$}($\mathbf{z}$)}$
respectively. Using (\ref{eq:vsigma}), we obtain the corresponding local and global
contributions to the double layer integral as
\begin{subequations}
\begin{equation}\label{eq:vsigma_l}
v^l_j(\mathbf{z}) = n_k(\mathbf{z})\sigma^l_{jk}(\mathbf{z}),
\end{equation}
\begin{equation}\label{eq:vsigma_g}
v^g_j(\mathbf{z}) = n_k(\mathbf{z})\sigma^g_{jk}(\mathbf{z}).
\end{equation}
\end{subequations}
We describe the computation of the global contribution
to the double layer integral first, as it is a straightforward extension of the procedure
presented in Sec. (\ref{sec:sl_comp}). This will be followed by a discussion
of the procedure for computing the local contribution to the double layer integral.

\subsubsection{Global contribution}
Here we describe the procedure to compute the global contribution
to the double layer integral. Consider the global velocity
field $\mathbf{w}^g(\mathbf{z})$ and the pressure
field $\mathbf{p}^{w^g}(\mathbf{z})$ associated with the
global force density $\boldsymbol{\Pi}^g(\mathbf{z})$; see Eq. (\ref{eq:stokes_gga}).
The procedure to compute the velocity and pressure field for this \textit{global}
distribution of density has been discussed in detail in Sec. (\ref{sec:sl_comp}).
Once these are known, one can obtain the stress field $\boldsymbol{\sigma}^g(\mathbf{z})$ from
the usual Newtonian constitutive equation as
\begin{equation}
\sigma_{jk}^g(\mathbf{z}) = - \frac{p^{w^g}(\mathbf{z})}{\mu}\,\delta_{jk} +
 \left( \frac{\partial w^g_j(\mathbf{z})}{\partial x_k} +
\frac{\partial w^g_{k}(\mathbf{z})}{\partial x_j} \right).
\end{equation}
As was mentioned in Sec. (\ref{sec:greens_fn}), the differentiations required
in the above expression are performed in the transform space \citep{canuto06,peyret02};
consequently, the stress field is known with spectral accuracy at the mesh points.
Once the stress tensor is obtained at the mesh points, the stress at the nodes 
of the elements are obtained using $4^{th}$ order Lagrange interpolation. Lastly,
$\mathbf{v}^g(\mathbf{z})$ is obtained from Eq. (\ref{eq:vsigma_g}).

\subsubsection{Local contribution}\label{sec:dbl_l}
Consider the velocity field due to the local density,
which is written as
\begin{equation}\label{eq:vel_l}
w^l(\mathbf{z}) = \int_S q_i(\mathbf{x}) \, G^l_{ji}(\mathbf{z},\mathbf{x}) \, dS(\mathbf{x}).
\end{equation}
It is trivial to show that the stress field associated with the above
velocity field is given by the following
\begin{equation}\label{eq:st_dbl_l}
\sigma_{jk}^l(\mathbf{z}) = \int_S q_i(\mathbf{x}) \, T^l_{jik}(\mathbf{z},\mathbf{x}) \, dS(\mathbf{x}),
\end{equation}
and consequently the local contribution to the double layer integral
is given by
\begin{equation}\label{eq:dbl_la}
v^l_j(\mathbf{z}) = n_k(\mathbf{z})\int_S q_i(\mathbf{x}) \, T^l_{jik}(\mathbf{z},\mathbf{x}) \, dS(\mathbf{x}).
\end{equation}
The above integral is discretized in the same fashion as in Sec. (\ref{sec:sl_comp}),
and is written as a sum of integrals over the triangular elements. 
Again, as in Sec. (\ref{sec:sl_l}),  all the integrals are performed
over the parent triangle as follows
\begin{equation}\label{eq:int_dbl_l}
v_j^l(\mathbf{z}) =  n_k(\mathbf{z})  \displaystyle \sum_{k=1}^{N_\Delta N_p} \int_0^1 \int_0^{\xi(\eta)} q_i\left(\mathbf{x(\xi,\eta)}\right)\; T_{jik}^l\left(\mathbf{z},\mathbf{x}(\xi,\eta)\right)\; \omega \; d\xi d\eta.
\end{equation}
The integral in the above equation is performed with
the product of two one dimensional Gauss-Legendre quadrature rule;
see Sec (\ref{sec:sl_l}) for details. In addition, when the vertex at $\mathbf{z}$
is a member of the element over which integration is performed,
the integrand is singular and requires special treatment. At this
point it is worth noting that the double layer integrand has the
same $1/\hat{r}$ singularity as the single layer integral \citep{higdon95},
as $\hat{\mathbf{x}} \cdot \mathbf{n}(\mathbf{x}_0) \approx \hat{r}^2$
for small $\hat{r}$. So in principle one may employ the polar
coordinate transformation to regularize the singular double
layer integral. In practice, because we employ flat elements,
the condition  $\hat{\mathbf{x}} \cdot \mathbf{n}(\mathbf{x}_0) \approx \hat{r}^2$
is not valid. Note that the normal vector at any given vertex is
taken as the area averaged normal vector of triangles to which
it belongs, from which it follows that the numerical
double layer has a $1/\hat{r}^2$ singularity. This stronger singularity 
can be avoided by employing surface elements that yields a continuously varying
normal vector such as splines \citep{lac07}.

In the literature, desingularization of the double layer integral is usually achieved by singularity subtraction \citep{pozrikidis92}. This desingularization scheme is also applicable for the current formulation (\ref{eq:wdbl}), though
this procedure will require 9 times the computational effort of performing the double layer integral
itself; we omit the details here. We therefore look elsewhere
for more computationally efficient schemes for performing the
singular part of the double layer integral. The first point to note
is that the singularity in the stress tensor $\mathbf{T}(\mathbf{z},\mathbf{x})$ is contained entirely in its local part $\mathbf{T}^l(\mathbf{z},\mathbf{x})$. Moreover, desingularization is necessary only when the target point $\mathbf{z}$
is one of the vertices of the element over which integration is being
performed. If this is the case, we proceed by replacing the normal vector $\mathbf{n}(\mathbf{z})$
outside the integral in Eq. (\ref{eq:int_dbl_l}) by a vector $\tilde{\mathbf{n}}(\mathbf{x})$
inside the integral, which leads to the following expression for the integral over the current element $k$: 
\begin{equation}\label{eq:dbl_l1}
\left[v_j^l(\mathbf{z})\right]_{S_k} =\int_{S_k} q_i(\mathbf{x}) \, T^l_{jik}(\mathbf{z},\mathbf{x})  \tilde{n}_k(\mathbf{x}) \, dS(\mathbf{x}).
\end{equation}
The vector $\tilde{\mathbf{n}}(\mathbf{x})$ at any point $\mathbf{x}(\xi,\eta)$ on the element is defined 
as per the following equation:
\begin{equation}\label{eq:ntilde}
\tilde{\mathbf{n}}(\mathbf{x}) = \mathbf{n}_\Delta \, \phi_1(\xi,\eta) + \mathbf{n}(\mathbf{z}) \, \phi_2(\xi,\eta)  +
\mathbf{n}(\mathbf{z}) \, \phi_3(\xi,\eta),
\end{equation}
where $\mathbf{n}_\Delta$ refers to the normal vector of the current triangular element. In writing
the above equation, we have assumed that the target point $\mathbf{z}$ is mapped to 
the vertex labeled 1 of the parent triangle; see Fig. \ref{fig:nat_crd}. It is immediately clear
from Eq. (\ref{eq:ntilde}) that $\tilde{\mathbf{n}}(\mathbf{x})$ will take the value $\mathbf{n}_\Delta$
when the source point coincides with the target point (i.e., when $\mathbf{x}=\mathbf{z}$),
while it will tend to $\mathbf{n}(\mathbf{z})$ as the source point $\mathbf{x}$ moves away from the 
target point $\mathbf{z}$ over the current element. Now, if we substitute the expression of
$\mathbf{T}^l$ from Eq. (\ref{eq:local}) into Eq. (\ref{eq:dbl_l1}), one immediately sees that
the singular parts of the integrand tends to zero as the source point approaches 
the target point since $\hat{\mathbf{x}}\cdot \mathbf{n}_{\Delta}=0$. This is
due to the fact that elements are flat and $\hat{\mathbf{x}}$ lies
in the plane of the element, while $\mathbf{n}_\Delta$ is
normal to the element. In order to estimate the error introduced due to this approximation,
one can show by simple Taylor series expansion that $||\tilde{\mathbf{n}}(\mathbf{x})-\mathbf{n}(\mathbf{z})|| \sim || \nabla \mathbf{n}(\mathbf{z})||\, h_s$, where $h_s$ is the characteristic surface mesh spacing scaling as $N_{\Delta}^{-1/2}$.
In addition, since this approximation is applied only when the target point is a member of the 
triangle over which the integration is being performed, the previous error gets multiplied by
the area of the triangle which is $O(N_{\Delta}^{-1})$. Therefore, we estimate the error introduced
in the solution  due to this approximation as $O(N_{\Delta}^{-3/2})$.
This completes the evaluation of the local contribution to the double layer integral. The total
double layer integral is then obtained as the sum of the local
and the global parts as:
\begin{equation}
v_j(\mathbf{z}) = v_j^l(\mathbf{z}) + v_j^g(\mathbf{z}).
\end{equation}

\begin{figure}[!t]
\centering
\subfigure[A patch on the membrane]{\includegraphics[width=0.27\textwidth]{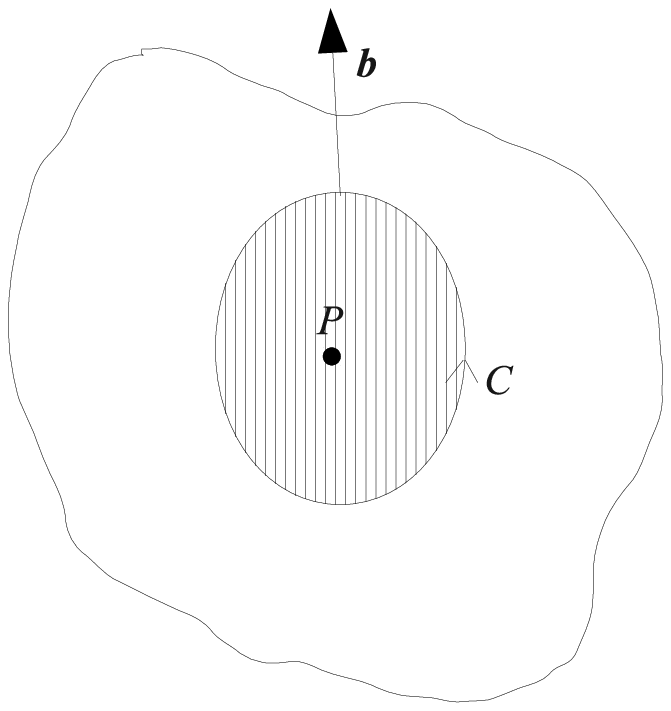}}
\hspace{10mm}
\subfigure[A patch on the discretized surface]{\includegraphics[width=0.25\textwidth]{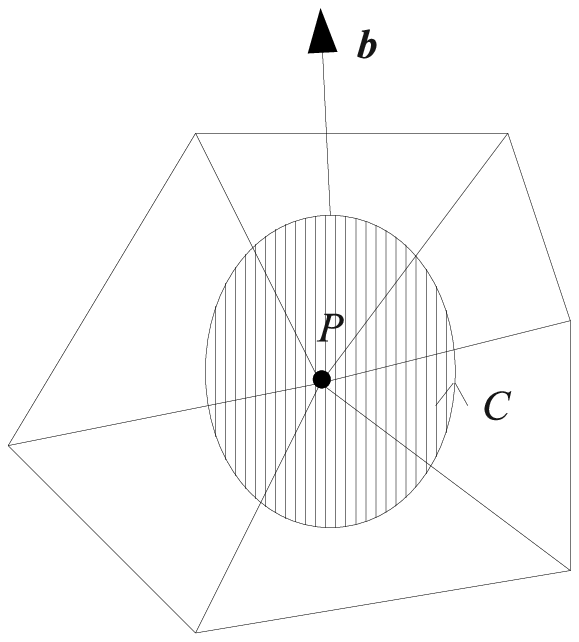}}
\caption{Defining the contour $C$ enclosing an area (hatched) containing the point of interest $P$. $\mathbf{b}$
is the in-plane  normal to the contour $C$.}\label{fig:cntr}
\end{figure}

\section{Membrane Mechanics: Hydrodynamic traction jump}\label{sec:memb}
The solution of the boundary integral equation (\ref{eq:qa})
requires the knowledge of the hydrodynamic traction jump
across the interface $\Delta \mathbf{f}$. This jump
in traction is obtained from the membrane equilibrium
condition as discussed next. Consider first a patch of element
on the membrane's surface as shown in Fig. (\ref{fig:cntr}a).
The forces acting on this patch are the hydrodynamic stresses
on the inner and the outer surface and the membrane tension at
the boundary denoted by contour $C$. Now, let the membrane tension tensor be
given by $\mbox{\boldmath{$\tau$}}$, then the force balance on
the membrane patch is given by
\begin{equation}\label{eq:cntr}
\int_{S_C} \Delta \mathbf{f} \; dS + \int_C \mathbf{b} \, \cdot \, \mbox{\boldmath{$\tau$}} \; dl = 0,
\end{equation}
where $S_C$ denotes the area enclosed by the contour $C$.
Using the divergence theorem, one can convert to
contour integral to a surface integral, which in the
limit of infinitesimal area yields
\begin{equation}\label{eq:div_tau}
\Delta \mathbf{f} = - \mathbf{\nabla}_s \cdot \mbox{\boldmath{$\tau$}},
\end{equation}
where $\mathbf{\nabla}_s$ is the surface divergence operator \citep{biesel10}.
The above equation (\ref{eq:div_tau}) has been directly employed in several
boundary integral implementations to obtain the traction jump $\Delta \mathbf{f}$; see e.g. \citet{lac04}.
For flat elements as in this work, we note that $\mbox{\boldmath{$\tau$}}$
is constant over each element such that its surface divergence is
identically zero, though there is a jump in its value across elements
and consequently the contour integral in the Eq. (\ref{eq:cntr})
is generally expected to be non-zero. The Contour integral can therefore
be used to obtain the traction jump as \citep{pozrikidis98}
\begin{equation}\label{eq:cntr1}
\Delta \mathbf{f}  = - \frac{1}{A_C} \int_C \mathbf{b} \, \cdot \, \mbox{\boldmath{$\tau$}} \; dl,
\end{equation}
where the $A_C$ is the area enclosed by the contour $C$;
Fig. (\ref{fig:cntr}b) shows an example of the contour for the discretized surface.
To proceed with our implementation, we first interpret the contour integral
on the right of (\ref{eq:cntr1}) as the reaction force on a given node
obtained under the condition that the entire elastic energy stored in the particle membrane has been reassigned
to the vertices of the discretized triangular elements. We then use the
principal of virtual work as presented by \citet{CharrierSW89}
to compute the reaction force at the vertices. Once the total reaction
force $\mathbf{F}_p$ at a given vertex $P$ is known, we obtain the traction
discontinuity at that vertex as
\begin{equation}\label{eq:cntr2}
\Delta \mathbf{f}_p  = - \frac{\mathbf{F}_p}{A_p}
\end{equation}
where $A_p$ is area assigned to the vertex, which is essentially
the area enclosed by a hypothetical contour around the vertex $P$;
see Fig. (\ref{fig:cntr}b). We call the contour hypothetical as we
never explicitly define it here. For the area assignment to the vertex,
we use a very simple rule where each vertex of the triangular
element is assigned a third of the triangular element's area.
Therefore the total area $A_p$ is obtained as (1/3) of the total
area of the triangular elements of
which the given vertex $P$ is a member. We believe the method
outlined here is substantially simpler to implement than employing
the contour integral explicitly. In the remainder of this section,
we outline the procedure employed for computing the reaction force at the
vertices of the triangular elements.

We begin by introducing the formalism for describing
the kinematics of the membrane deformation. 
This formalism is mostly clearly presented for deformations
in a plane, which for the moment is taken to be the $xy$ plane.
Let $(x, y)$ and $(X, Y)$ denote respectively the undeformed and
deformed coordinates of a material point, with respect to a fixed
set of Cartesian axes.  If $u$ and $v$ denote the displacements of
the material point in $x$ and $y$ directions respectively, then
\begin{equation}
\begin{array}{l}
\displaystyle X = x + u,\\
\displaystyle Y = y + v.
\end{array}
\label{eq_displacement}
\end{equation}
The relationship between an infinitesimal line segment
before and after the deformation can be expressed as
\begin{equation}
\left [ \begin{array}{ccc}
dX\\ dY
\end{array}\right]
= \left [
\begin{array}{ccc}
1 + \partial u / \partial x & \partial u / \partial y \\
\partial v / \partial x & 1 + \partial v/\partial y
\end{array} \right]
\left [\begin{array}{ccc} dx\\ dy
\end{array}\right],
\label{eq_defmormation1}
\end{equation}
or compactly as
\begin{equation}
d{\bf X} = {\bf F}\cdot d{\bf x}.
\label{eq_defmormation2}
\end{equation}
where $\mathbf{F}$ is the deformation gradient tensor.
The square of the distance between the two neighboring points
after deformation is given by
\begin{equation}
\begin{array}{l}
\displaystyle dS^2 = d{\bf X}\cdot d{\bf X} = d{\bf x}\cdot {\bf G}\cdot d{\bf x},\\
\displaystyle {\bf G} = {\bf F}^{T}\cdot {\bf F},
\end{array}
\label{eq_displacement3}
\end{equation}
where ${\bf G}$ is a symmetric positive definite matrix.
We denote the eigenvalues of the $\bf{G}$ by $\lambda_1^2$ and
$\lambda_2^2$, such that $\lambda_1$ and $\lambda_2$ are
the principal stretch ratios. For a thin membrane that
displays no resistance to bending, the strain energy
density $W$ of the membrane is a function of
$\lambda_{1}$ and $\lambda_{2}$. Here we consider
the capsule to be an infinitely thin neo-Hookean membrane,
for which the strain energy density is defined as \citep{Barthes-BieselDD02}
\begin{equation}
W_{\mbox{\tiny NH}} = \frac{G}{2}\left[\lambda_1^2 + \lambda_2^2 + \frac{1}{\lambda_1^2 \lambda_2^2} -3 \right].
\label{eq_WNH}
\end{equation}
Here $G$ is the two-dimensional shear modulus for the
membrane, having units of force per unit length.
To compute the reaction force at the nodes,
we adopt the finite element approach of \citet{CharrierSW89}.
Only the briefest account will be given here; for details
the reader is referred to the original reference.
In the approach of \citet{CharrierSW89}, the
membrane is discretized into flat triangular
elements such that the strain is uniform over an element.
Moreover, it is assumed that an element remains flat even
after deformation. The forces at the nodes are then
determined from the knowledge of the displacement of the vertices
of the element with respect to the undeformed element
followed by the application of the principal of virtual
work, such that the computed forces and the known displacements are
consistent with the strain energy stored in the element.
For an arbitrarily oriented element, rigid body rotations
and translations can be defined to make the deformed
and undeformed state in the same plane. Note that the rigid
body rotations and translations have no effect on the strain
energy and consequently the forces. The forces are then
computed using the coplanar formalism discussed
above. Finally, these forces are transformed back to the
frame of reference in the deformed state by applying the
inverse transformation. The total reaction force at a
node is obtained as a sum of reaction forces at that
node due to contributions from all the triangular elements
of which it is a member. Once the reaction force at any
given node is known, the hydrodynamic traction discontinuity
at that node  is obtained from Eq. (\ref{eq:cntr2}) as detailed
earlier.

\section{Solution procedure and parameters}\label{sec:sol}
In this section we describe solution methods and parameters
for the equations resulting from the formulation just presented.

The heart of the computation is the determination of the fluid velocity
at the element nodes using equations (\ref{eq:bi_ua}) and (\ref{eq:qa}).
To compute the velocity from Eq. (\ref{eq:bi_ua}),
we first need to compute the single layer density
$\mathbf{q}(\mathbf{x})$. This is obtained from the solution of the integral equation (\ref{eq:qa}).
Upon discretization of the double layer integral in Eq. (\ref{eq:qa}),
which was discussed in Sec. (\ref{sec:dbl_comp}), we obtain
a linear coupled system of equations for $\mathbf{q}^b$, where
$\mathbf{q}^b$ is a vector of length $3N_b$ denoting the value
of $\mathbf{q}(\mathbf{x})$ at the element nodes; $N_b$ is
the number of element nodes in the system. We express
this linear system of equations as
\begin{equation}
(\mathbf{I} + \frac{\mbox{\boldmath{$\kappa$}}}{4\pi} \cdot \mathbf{D}^I) \cdot \mathbf{q}^b
= \mathbf{b},
\end{equation}
where $\mbox{\boldmath{$\kappa$}}$ is a diagonal matrix of size
$3N_b \times 3N_b$ denoting the value of $\kappa_m$ in Eq. (\ref{eq:qa})
at each element node, $\mathbf{D}^I$ denotes the discretized
double layer operator of size $3N_b\times 3N_b$, while $\mathbf{b}$ is a $3N_b$ vector
denoting the known right hand side of Eq. (\ref{eq:qa}) at the element nodes.
The above system of equations is solved
iteratively using the GMRES algorithm \citep{saad}. An important
benefit of this iterative procedure is that the matrix in the parenthesis
above is never explicitly computed; at each iteration step only the
product of the above matrix with a known vector generated by the algorithm
is to be computed. The procedure to compute this matrix
vector product is similar to computing  $\mathbf{v}(\mathbf{z})$ in Sec. (\ref{sec:dbl_comp})
(see Eq. \ref{eq:wdbl}) at the element nodes for a known $\mathbf{q}^b$. We take the initial guess for
$\mathbf{q}^b$ either from  the previous time step, or from the
previous stage if a multistage method is employed as is the case here.
This leads to a substantial savings in the number of iterations required
for convergence. Iterations were terminated when the $L_2$ norm of the
current residual vector $\mathbf{S}$ relative to the norm of the right hand side $\mathbf{b}$
was less than $10^{-4}$; the residual vector $\mathbf{S}$ is defined as
\begin{equation}
\mathbf{S} = (\mathbf{I} + \frac{\mbox{\boldmath{$\kappa$}}}{4\pi} \cdot \mathbf{D}^I) \cdot \mathbf{q}^b
- \mathbf{b}.
\end{equation}
In addition to the convergence in the residual vector $\mathbf{S}$, which is obtained
naturally as part of the iterative procedure, it is also important
to investigate the convergence in the solution; the error in solution denoted by vector $\mathbf{S}_q$ is defined
as
\begin{equation}
\mathbf{S}_q = \mathbf{q}^b - \mathbf{q}^b_{ex}
\end{equation}
where we have defined the ``{exact}'' solution as $\mathbf{q}^b_{ex}$, which can be obtained by solving the above system of
 equations to a very small tolerance, e.g. $10^{-10}$. The above
 tolerance of $10^{-4}$ for the residual vector relative to the norm
 of the right hand side leads to an error of the same order for the
 error vector relative to the exact solution for well conditioned systems.
 This implies an accuracy of $0.01 \%$ for most cases. Even for the worst
 cases in the present work, the relative error in the solution
 was always less than $0.1\%$. In most cases convergence was
 achieved in less than $5$ iterations; see Sec. (\ref{sec:mult_part})
 for some examples. For systems with high viscosity contrast and/or with
 large number of particles, the matrix may become ill-conditioned and a
 preconditioner may become necessary. In the present study, no preconditioner
 was employed, though multigrid preconditioners for Stokes flow may be useful \citep{esw05,silvester94}.
 
The iterative procedure described above gives the $\mathbf{q}^b$
vector at the element nodes.  These are subsequently substituted
in Eq. (\ref{eq:bi_ua}) to compute the corresponding velocity at the element nodes;
the numerical procedure described in Sec. (\ref{sec:sl_comp}) is employed
to compute this single layer integral. We denote the velocity thus computed
at the element nodes by a $3N_b$ vector $\mathbf{u}^b$, which
is used to evolve the position of element nodes $\mathbf{x}^b$ as per the equation
\begin{equation}
\frac{d\mathbf{x}^b}{dt} = \mathbf{u}^b.
\end{equation}
The time integration in the above equation is performed via the second order midpoint method, which
belongs to the family of explicit Runge-Kutta integrators \citep{lambert97}.  The
time step $\Delta t$ employed in this work was set adaptively
using the rule \citep{hinch96,rallison81}:
\begin{equation}
\dot{\gamma} \Delta t = 0.5 \, Ca \,h_s^m/a,
\end{equation}
where $h_s^m$ is the minimum node-to-node separation in the
system (which does not have to be for two points on the same particle), $a$
quantifies the length scale of the particle such as the radius for a
spherical particle, $\dot{\gamma}$ is the shear rate, while $Ca$ is the capillary number, expressing
the ratio of viscous and interfacial stresses (for a spherical capsule
with radius $a$ and shear modulus $G$, the capillary number is defined
as $Ca=\mu \dot{\gamma}a/G$). This rule gave a stable evolution with time; a time step
twice this value led to an instability in some simulations presented later,
while a time step half this size gave nearly indistinguishable result.
The volume of the particle was found to be well conserved with time. For
example, for a capsule with $\lambda=5$ in shear flow at $Ca=0.6$, the volume
changed by an average of $10^{-4}$ of its original value over a unit
strain. In the present work, a volume correction was performed
only when the volume of the particle deviated by more than
$10^{-4}$ of its original value; the procedure employed for this
correction is described in \citep{freund07}.

Finally, we consider parameters related to the numerical solution procedure.
Many of these have already been introduced earlier, and are repeated here for completeness.
The first parameter is $N_{\Delta}$, which refers to the number of triangular elements
employed to discretize the surface of each of the $N_p$ particles in the system;
the number of element nodes per particle is denoted by $N_b$.
Next, there are several parameters associated with the GGEM methodology described in
Sec. (\ref{sec:greens_fn}). First is the length scale $\alpha^{-1}$
associated with the quasi-Gaussian global density. The local solution 
as well as the assignment of the global force density to mesh points
is truncated beyond a distance of $r_{cut}=4/\alpha$ from the origin of 
the singularity. The solution of the global problem  (Sec. \ref{sec:greens_fn}) requires one to define a three
dimensional mesh with $N_x$, $N_y$, and $N_z$ mesh points
in $x$, $y$ and $z$ directions respectively. This gives
a mean mesh spacing in the three directions as $\Delta x_m =L_x/N_x$,
$\Delta y_m = H/(N_y-1)$ and $\Delta z_m = L_z/N_z$; all three mean mesh
spacings are kept equal unless otherwise mentioned. An important
parameter denoting the resolution of the GGEM methodology
is $\alpha \Delta y_m$, with the resolution 
and hence the accuracy of the method increasing with
decreasing $\alpha \Delta y_m$. As a rule of thumb, we
require $\alpha \Delta y_m < 1$; see Sec. (\ref{sec:ggem_test}) for details.
Choices of above parameters are specified below in the descriptions of the test problems.

\begin{figure}[!t]
\centering
\includegraphics[width=0.35\textwidth,angle=-90]{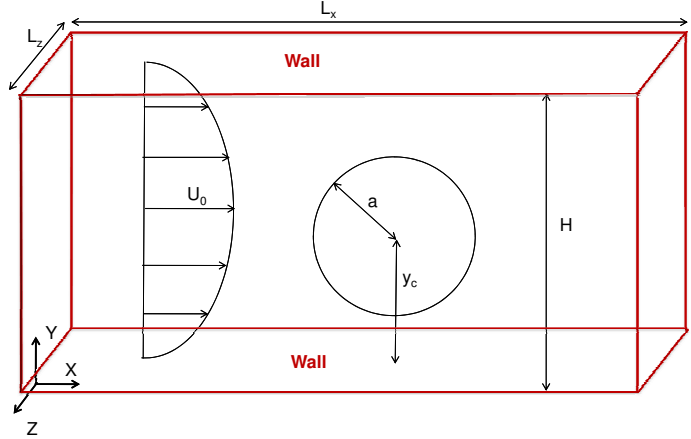}
\caption{Schematic of the slit geometry for the test problems in Secs. (\ref{sec:urigid}) and (\ref{sec:udrop}).
A single sphere, either a rigid particle or a drop, is placed in a slit geometry
with the channel height being $H$.  The radius of the sphere is $a$,
while its center is at $y_c$. A pressure driven flow is considered with $U_0$
being the centerline velocity. The confinement ratio is $2a/H$. Periodic boundary
  conditions are employed in $\mathbf{x}$ and $\mathbf{z}$ directions, with
  periodicity being $L_x$ and $L_z$ respectively.}\label{fig:val_geom}
\end{figure}

\begin{figure}[!t]
\centering
\subfigure[Velocity profile]{\includegraphics[width=0.45\textwidth]{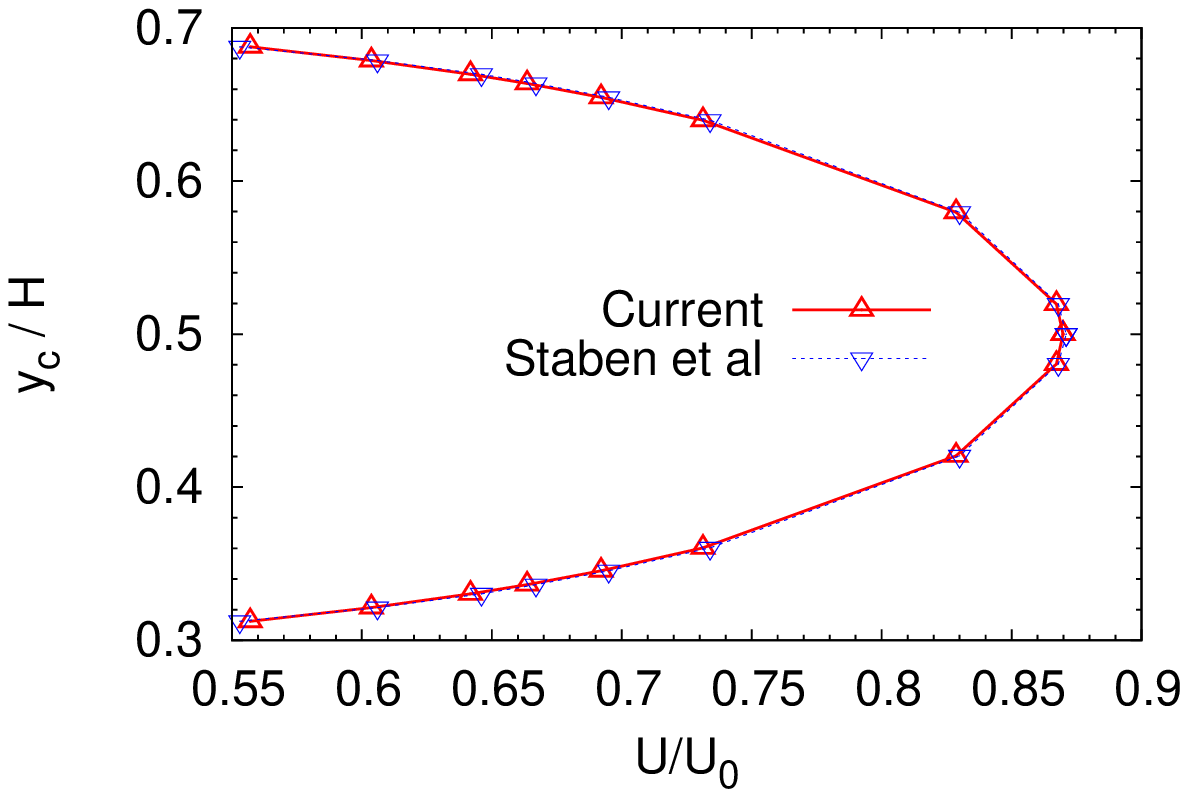}}
\subfigure[Convergence]{\includegraphics[width=0.45\textwidth]{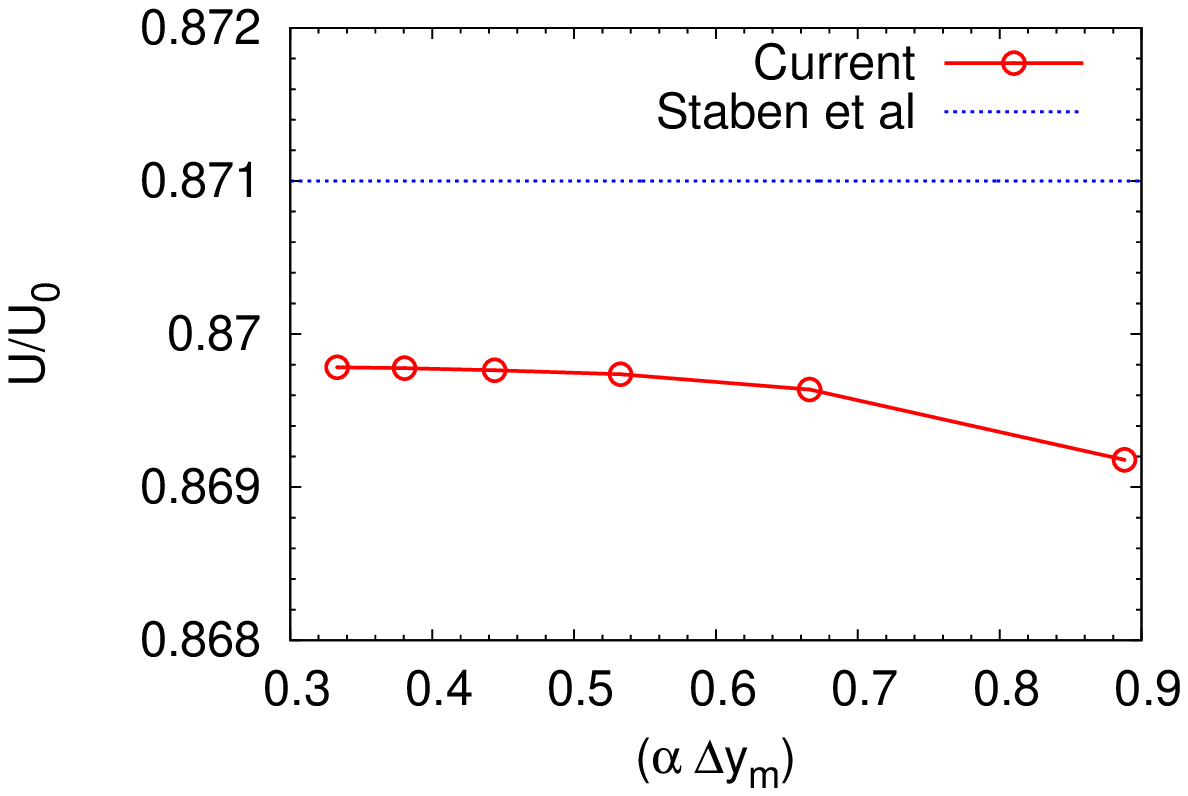}}
\subfigure[Effect of the spatial period $L$]{\includegraphics[width=0.45\textwidth]{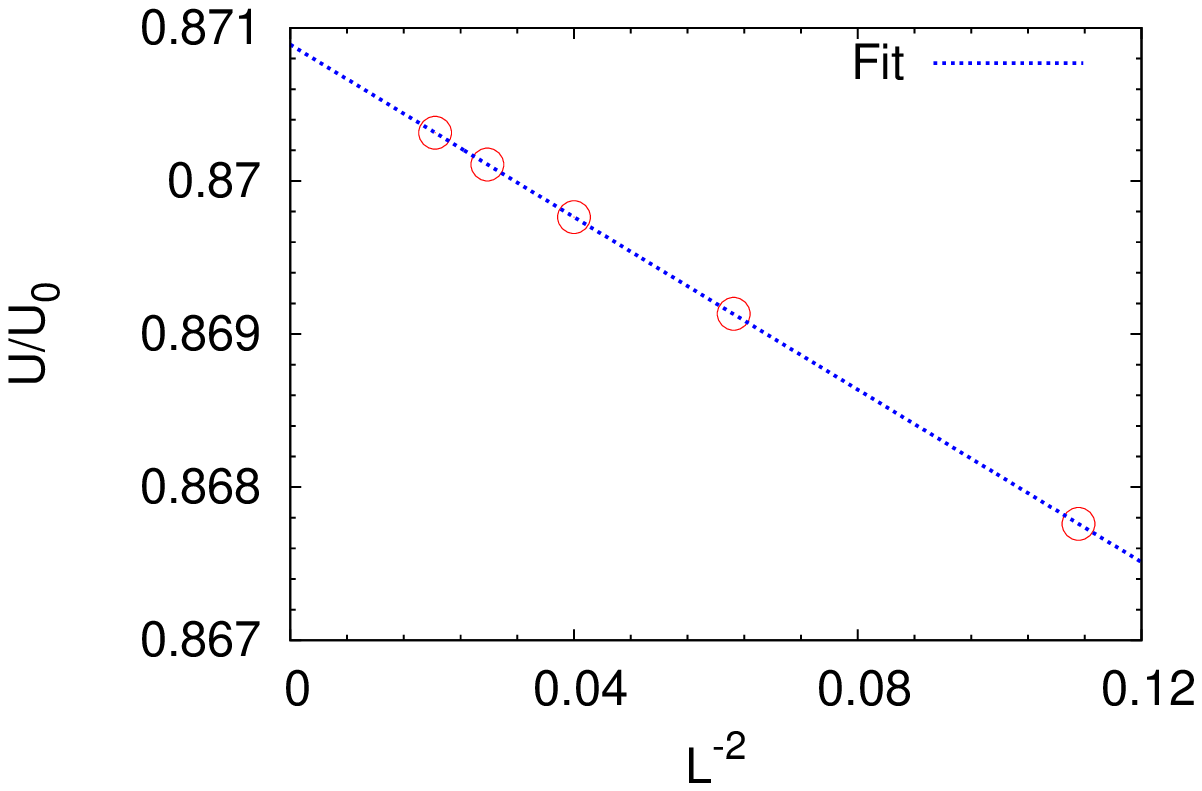}}
\caption{Rigid particle in a slit: (a) Comparison of the translational velocity of a rigid
sphere in a slit geometry with results of \citet{zinchenko03}.
The confinement ratio of the particle was $2a/H=0.6$.
The horizontal axis gives the velocity of the particle non-dimensionalized
by the centerline velocity of the undisturbed fluid, while the vertical axis
gives the location of the center of the sphere along the gradient
direction non-dimensionalized by the height of the channel. Walls
are present at $y=0$ and $y=H$. Simulation parameters were: $N_y=61$, $\alpha \Delta y_m = 0.44$,
$N_{\Delta}=5120$, $L_x=L_z=5H$, and $\Delta x_m = \Delta y_m = \Delta z_m$; see text for details. (b) Convergence of the velocity
for $y_c/H=0.5$ with $\alpha \Delta y_m$. For this calculation $r_{cut}=0.5a$ is kept fixed,
while $N_y$ is varied between 17 and 81. (c) Effect of the spatial period $L=L_x=L_z$ on the 
translational velocity of a particle with $y_c/H=0.5$. Also shown is a linear fit to the data in the plot.
All simulation parameters in (c) were the same as in (a) above
except for $L_x$ and $L_z$ which were varied. 
}\label{fig:rigid}
\end{figure}

\section{Numerical Results: Validation}\label{sec:val}
\subsection{Single layer validation: Rigid particle in a slit}\label{sec:urigid}
As a validation of the single layer integral implementation, we
consider a rigid sphere between two parallel walls and subject it to
a pressure driven flow with a centerline velocity $U_0$ as shown in
Fig. (\ref{fig:val_geom}). For a rigid particle, the velocity
at a point $\mathbf{x}_0$ on the surface satisfies the following integral equation
\citep{kim_karrila}

\begin{equation}\label{eq:sl_r}
u_j(\mathbf{x}_0) = u_j^{\infty}(\mathbf{x}_0) - \frac{1}{8\pi\mu}\int_{S} f_i(\mathbf{x})\, G_{ji}(\mathbf{x}_0,\mathbf{x})\,dS(\mathbf{x}),
\end{equation}
where $\mathbf{f}$ is the traction on the external surface of the sphere
due to the stresses in the fluid. The velocity at the surface of the particle
$\mathbf{u}(\mathbf{x}_0)$ in the above equation  can be written as
\begin{equation}
\mathbf{u}(\mathbf{x}_0) = \mathbf{U} + \mathbf{\Omega} \times (\mathbf{x}_0-\mathbf{x_c}),
\end{equation}
where $\mathbf{U}$ and $\mathbf{\Omega}$ represent the 
translational and rotational velocities of the particle, and
$\mathbf{x}_{c}$ denotes the center of the sphere.
Here we take the particle to be force and torque free and our goal
is to compute the velocity and angular velocity of the particle.
For this particular problem, the surface of the sphere was discretized
into $N_{\Delta}=5120$ triangular elements with $N_b=2562$ vertices. The unknowns
in the discretized system are $3N_b$ tractions at element vertices,
along with $\mathbf{U}$ and $\mathbf{\Omega}$. The force
and torque free condition along with the discretization of
Eq. (\ref{eq:sl_r}) gives $3N_b+6$ equations, which were solved
iteratively using the GMRES algorithm \citep{saad}.

Translational and rotational velocities of a rigid sphere
between two infinite parallel walls have been reported previously by \citet{zinchenko03}.
Here, we compare the translational velocity obtained in the present work
with their results for a fixed confinement ratio of $2a/H=0.6$
and for various positions of the sphere's center along the channel height $y_c$. This comparison
is shown in Fig. (\ref{fig:rigid}a), where the velocity of the particle has been non-dimensionalized
by the velocity of the undisturbed flow at the centerline $U_0$, while the
height of the sphere's center has been non-dimensionalized by the channel
height $H$. Very good agreement between the two results was observed, with
the discrepancy typically being less than $0.8\%$; the source of the 
slight discrepancy is discussed below. The GGEM parameters employed in the above
calculation were: $N_y=61$ and $r_{cut}=0.5a$, which gives $\alpha \Delta y_m = 0.44$; note that $r_{cut}=4/\alpha$.
The mean mesh spacing was equal in all three directions ($\Delta x_m = \Delta y_m = \Delta z_m$),
and the spatial period in both $x$ and $z$ directions were set to five times the wall spacing: $L_x=L_z=5H$.
Convergence of the particle velocity with respect to $\alpha \Delta y_m$ is demonstrated
next in Fig. (\ref{fig:rigid}b) for a particle placed at the centerline, i.e. $y_c/H=0.5$.
For this calculation $r_{cut}=0.5a$ was held constant, while $\alpha \Delta y_m$ was
varied by varying $N_y$ between $17$ and $81$ ($N_x$ and $N_z$ varied 
between 80 and 340). As could be seen in the figure, the velocity of particle
reaches its converged value for $\alpha \Delta y_m < 0.5$ and shows very little
variation with any further increase in mesh resolution.
Also shown in this plot is the result of \citet{zinchenko03} which reveals that
the velocity obtained in this work converges to a slightly lower value than the 
previous reference. The source of this discrepancy can be traced to 
the periodic boundary conditions employed in $x$ and $z$ directions in the present work;
\citet{zinchenko03} used an unbounded domain in these directions.
In the present case, we can easily estimate the result for an infinite box by observing its trend in
a series of simulations with varying spatial period $L$ ($L=L_x=L_z$).
This procedure is commonly used in triply periodic simulations to remove
the effects of the periodic boundary conditions; see, e.g. \citep{kumar11}.
In this particular example, we numerically find that the periodic image effects decay as $L^{-2}$
as shown in Fig. (\ref{fig:rigid}c), where we have plotted the translational velocity 
of the particle against $L^{-2}$. The y-intercept of the linear fit 
through the data points in the previous plot then gives an estimate of
the particle velocity in an infinite slit. This value comes out to be $0.871U_0$ (rounded
to three significant digits), which is exactly equal to the value reported by \citet{zinchenko03}. Recall that their boundary integral formulation is based on the slit Green's function of \citet{liron76}; the  agreement of our results with
those of \citet{zinchenko03} thus implicitly validates our Green's function implementation
for a slit with respect to the Green's function provided by \citet{liron76}. 
In addition to the periodic image effects, a slight discrepancy between our results
and those of \citet{zinchenko03} can also be expected in cases where the particle-wall
separation is very small. In such problems,
a large lubrication pressure can develop in the region around the small gap \citep{kim_karrila,kumar11a}. To obtain 
accurate solutions in this case, the surface discretization of the particle near the small gap must be
adaptively refined  as was done by \citet{zinchenko03}. No fundamental changes to the present formalism would be required to implement adaptive refinement.

\begin{figure}[!t]
\centering
\subfigure[Effect of spatial period $L$]{\includegraphics[width=0.45\textwidth]{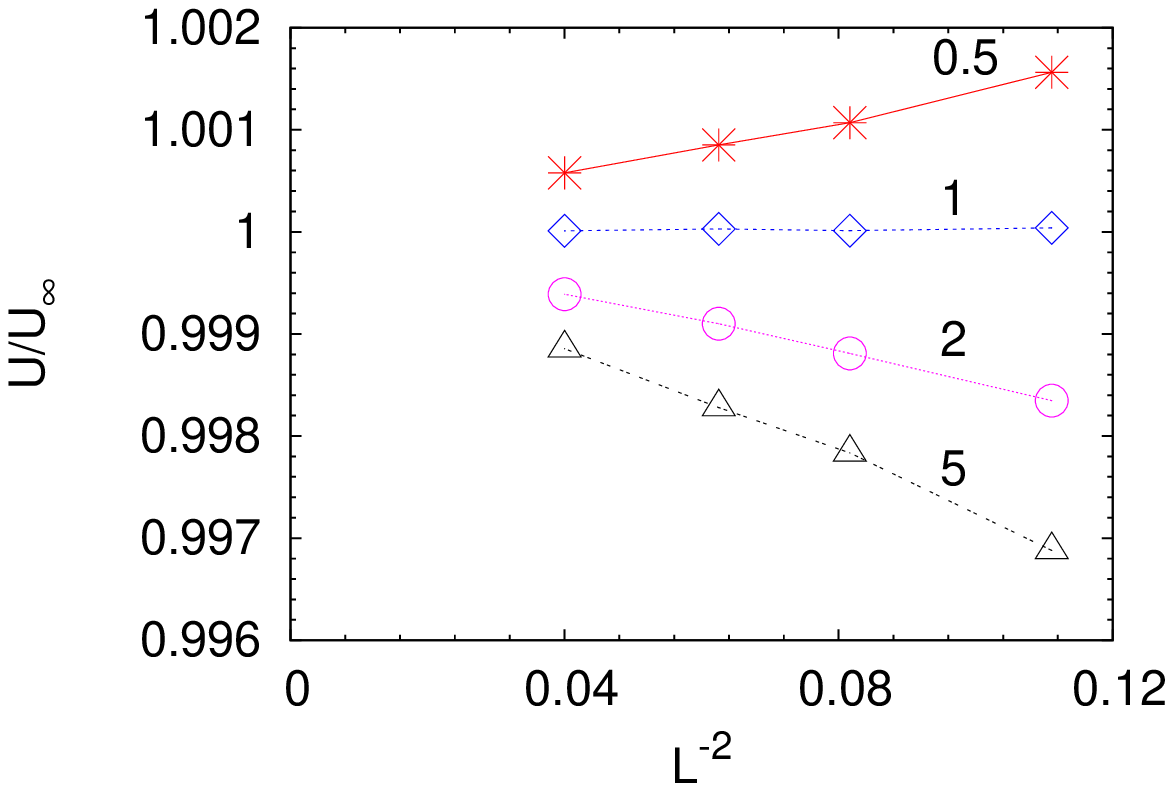}}
\subfigure[$U/U_0$]{\includegraphics[width=0.45\textwidth]{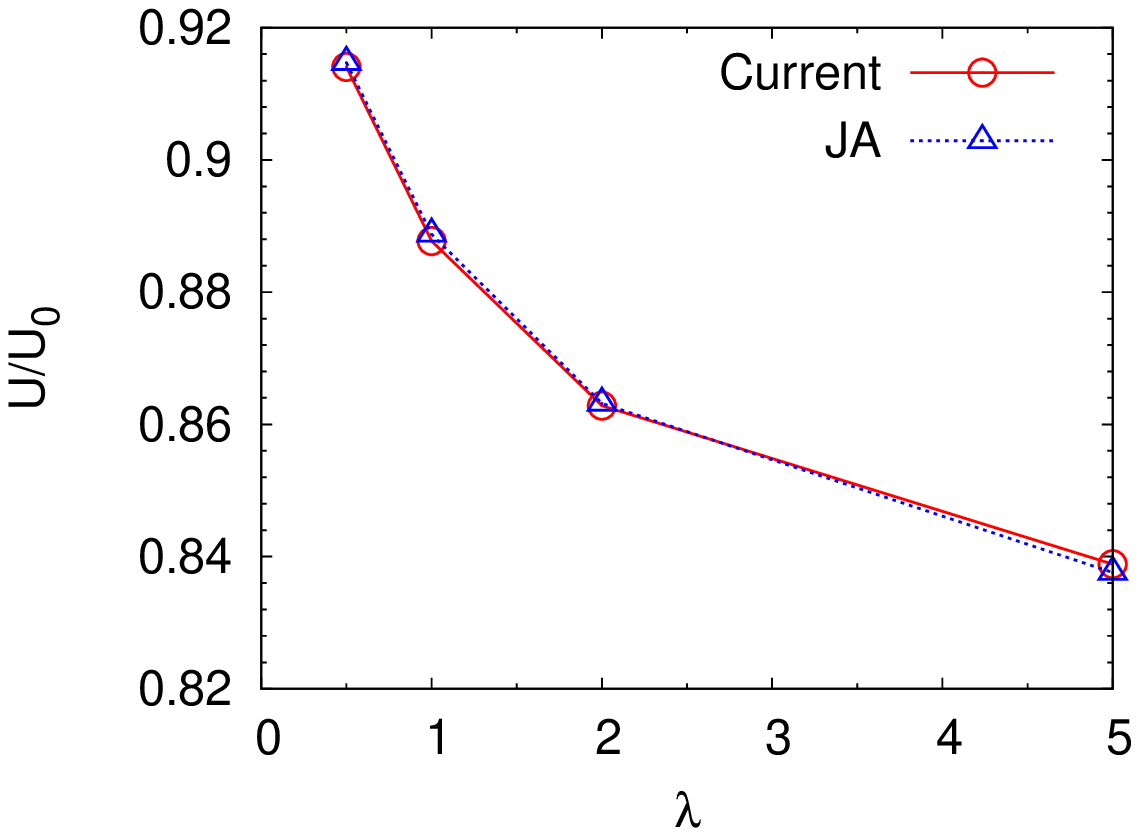}}
\caption{ Volume averaged translational velocity of a spherical drop in a slit.
The confinement ratio of the drop is $2a/H=0.6$, while its center is at $y_c/H=0.4$.
The simulation parameters are: $N_y=61$, $\alpha \Delta y_m = 0.44$,
$N_{\Delta}=5120$, and $\Delta x_m = \Delta y_m = \Delta z_m$.
(a) Effect of the spatial period $L=L_x=L_z$ on the drop's velocity for 
different viscosity ratios $\lambda$. Note the velocity has been
non-dimensionalized by $U_{\infty}$, which is the estimated velocity of the
drop as $L \rightarrow \infty$. This is obtained by fitting a straight line to the data which have not been
non-dimensionalized by $U_{\infty}$; the y-intercept of this fit gives $U_{\infty}$.
(b) Comparison of the volume averaged translational velocity of a 
drop as a function of viscosity ratio with the results of \citet{janssen10} (JA).
The velocity in the current work corresponds to $U_{\infty}$ in (a) above.}\label{fig:ui_drop}
\end{figure}


\subsection{Single and double layer validation: A Drop in a slit}\label{sec:udrop}
Having validated the single layer integral, we next move on to the
validation of the double layer integral. For this, we consider
the same geometry and bulk flow as in the above test case (Fig. \ref{fig:val_geom}),
but now consider a spherical drop instead of a rigid sphere. The
motion of the drop can be obtained by first solving Eq. (\ref{eq:qa})
for $\mathbf{q}(\mathbf{x})$, which upon substitution in Eq. (\ref{eq:bi_ua})
gives the velocity on the surface of the drop. We first point out that
for a spherical drop, the interfacial traction jump $\Delta \mathbf{f}$
is inconsequential. This is due to the fact that $\Delta \mathbf{f}$
is uniform in strength and acts radially everywhere, which when combined
with the incompressibility of the fluid implies zero velocity contribution
from this term. Once the velocity at the surface of the drop is known, we
compute the volume averaged velocity of the drop as
\begin{equation}
U_i = \frac{1}{V} \int_V u_i dV =  \frac{1}{V} \int_S (u_jn_j) \, x_i \, dS,
\end{equation}
where $V$ represents the volume of the drop, while $\mathbf{n}$ is the unit normal
vector at the surface. We computed the instantaneous volume averaged
velocity of a spherical drop placed at $y_c/H=0.4$ and with a confinement
ratio of $2a/H=0.6$ for different drop viscosity ratios $\lambda$.
The simulation parameters for this calculation were kept the same as in the 
previous section, namely $N_y=61$, $N_\Delta=5120$, $\alpha \Delta y_m = 0.44$, 
and $\Delta x_m = \Delta y_m = \Delta z_m$. Just like in the case of rigid
particle above, we again find the same $L^{-2}$ scaling of the periodic
image effects on the drop velocity. This is shown in Fig. (\ref{fig:ui_drop}a)
for drops of different viscosity ratios $\lambda$, where the velocity of the 
drop has been non-dimensionalized by the corresponding velocity estimated
for an infinite slit using the procedure outlined above in Sec. (\ref{sec:urigid}).
Interestingly, for a drop with $\lambda=1$, there is no observable periodicity
effect,  while the drop velocity increases with increasing $L$ for $\lambda>1$ and 
decreases with increasing $L$ for $\lambda<1$. In Fig. (\ref{fig:ui_drop}b),
we compare the results in the present work corrected for periodicity effects with those of \citet{janssen10}
 for several different viscosity ratios ($\lambda \in [0.5,1,2,5]$).
A very good agreement between our results and those reported by \citet{janssen10}
is evident at all viscosity ratios (error $< 0.15\%$), thereby validating
our implementation of the single and the double layer integral.


\begin{figure}[!t]
\centering
\psfrag{gdott}[c][Bl][1.0]{$\dot{\gamma}t$}
\subfigure[D: $Ca=0.30$]{\includegraphics[width=0.45\textwidth]{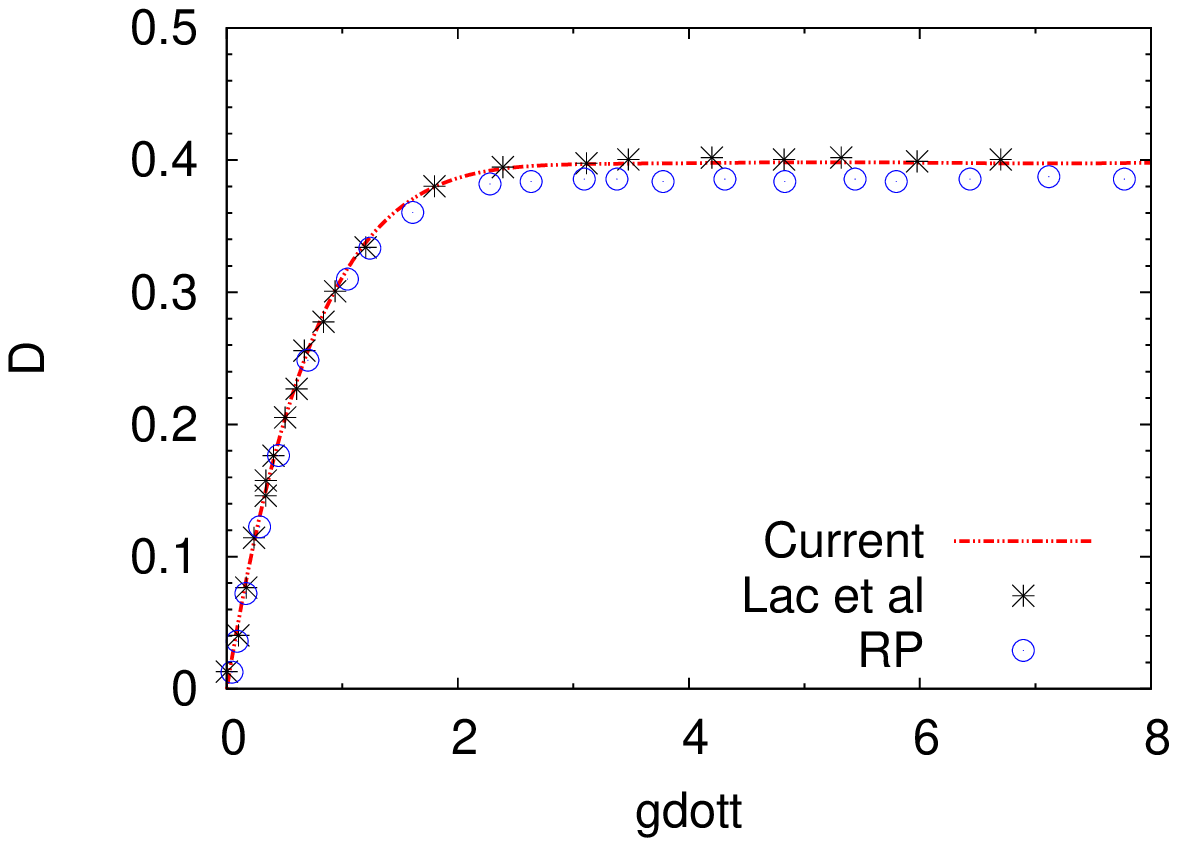}}
\subfigure[D: $Ca=0.60$]{\includegraphics[width=0.45\textwidth]{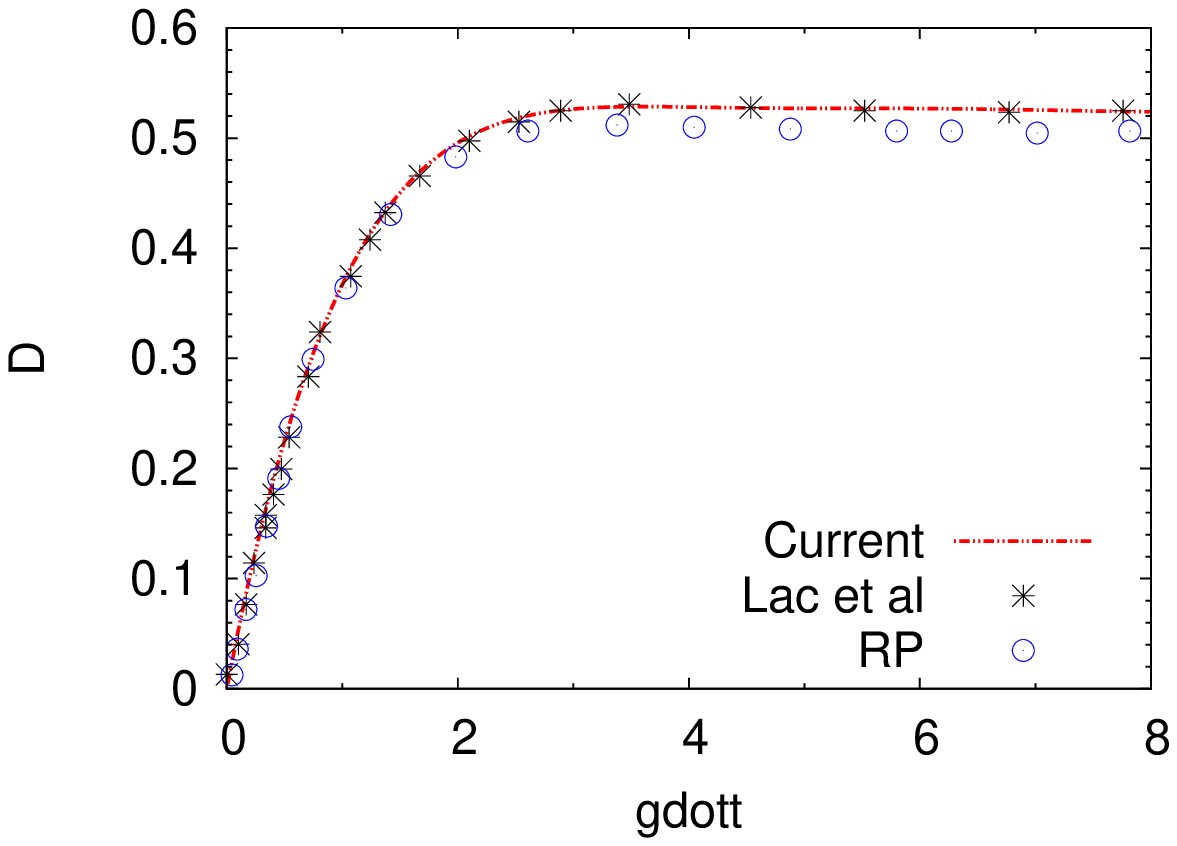}}
\subfigure[Convergence with $\alpha \Delta y_m$]{\includegraphics[width=0.45\textwidth]{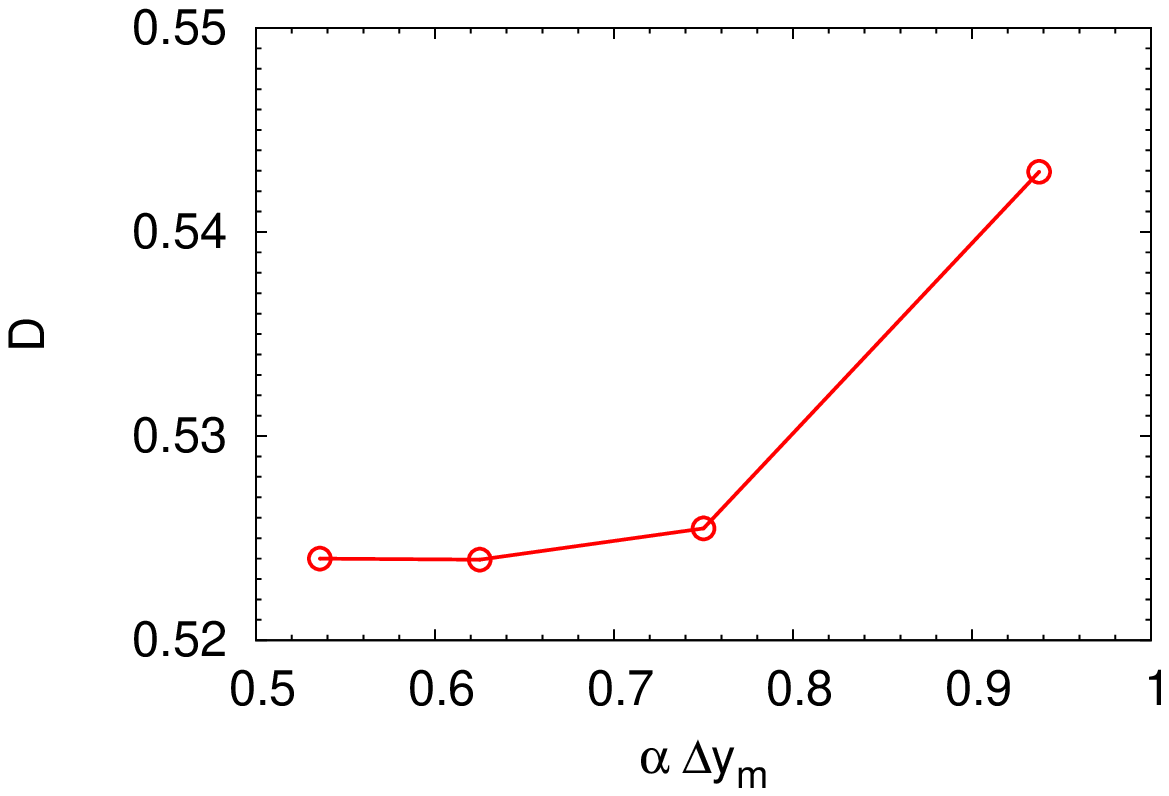}}
\subfigure[Convergence with $N_\Delta$]{\includegraphics[width=0.45\textwidth]{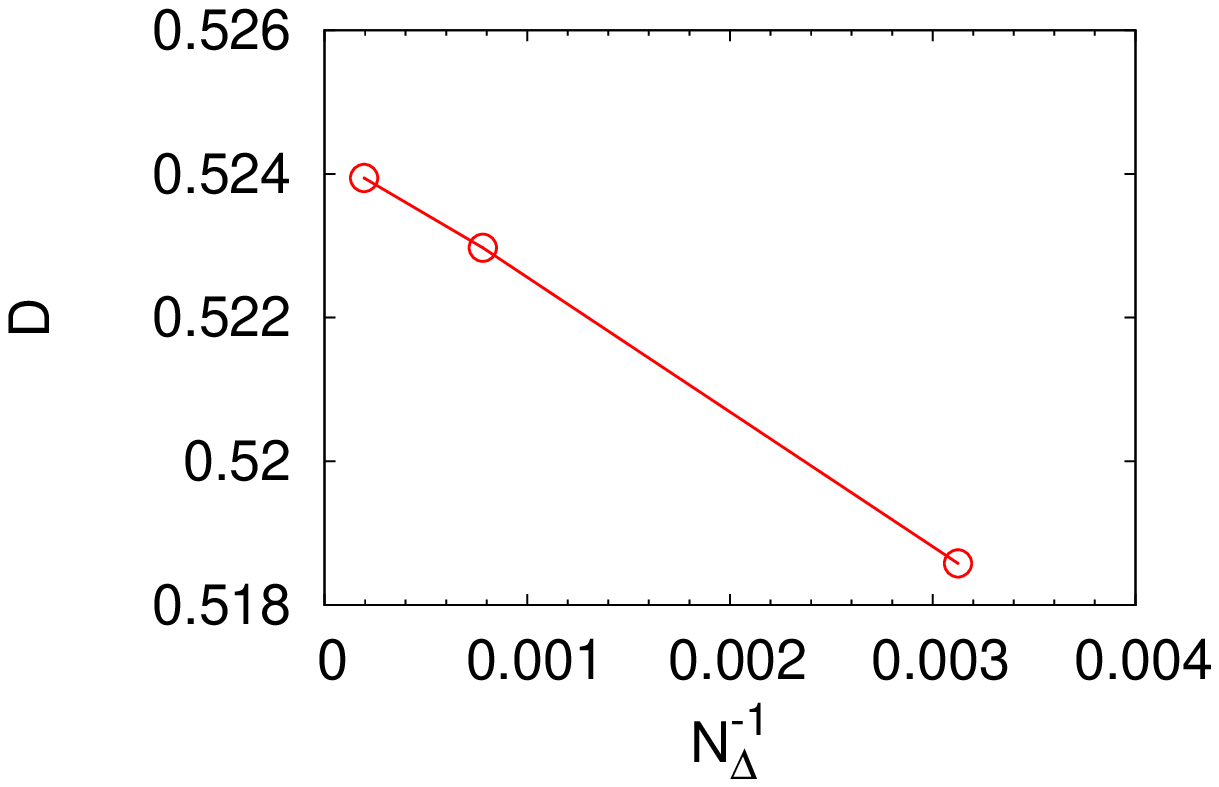}}
\caption{$\lambda=1$: Time evolution of the Taylor deformation parameter $D$ at (a) $Ca=0.3$
and (b) $Ca=0.6$. Lac et al in the plot refers to the results of \citet{lac04},
while RP refers to the results of \citet{pozrikidis98}. Simulation
parameters were: $L_x=L_y=H=15a$, $N_y=97$, $r_{cut}=a$, $\alpha \Delta y_m =0.625$, $N_\Delta=5120$, and
$\Delta x_m = \Delta y_m = \Delta z_m$. (c) Convergence of $D$ at $Ca=0.6$ with respect to 
	$\alpha \Delta y_m$. In this study $r_{cut}=a$ was held fixed, while $N_y$ was varied.
	(d) Convergence of $D$ at $Ca=0.6$ with respect to $N_{\Delta}$. Rest of the parameters 
	are the same as in (a) and (b).}\label{fig:Dl1}
\end{figure}

\subsection{Validation of the capsule membrane mechanics and the
overall implementation}\label{sec:dprm}
We next consider a capsule in a simple shear flow. 
To enable comparison with literature results in an unbounded 
domain, we consider a large simulation box with $L_x=H=L_z=15a$, where $a$ is the radius of the
initially spherical capsule  placed at the center of the box. Other simulation
parameters were: $N_y=97$, $r_{cut}=1.0$, $\alpha \Delta y_m =0.625$, $N_\Delta = 5120$, and
$\Delta x_m = \Delta y_m = \Delta z_m$; a convergence study with these
parameters will be presented later in this section.
The system is subjected to  simple shear flow and we follow the evolution of the
shape of capsule at various capillary numbers ($Ca=\mu\dot{\gamma}a/G$) and
viscosity ratios ($\lambda$). \citet{lac04} showed that a membrane
lacking bending resistance buckles at high or low
$Ca$; the origin of this buckling has been shown to be numerical \citep{sarkar08}.
We saw a similar behavior  and therefore restrict the results reported here to $0.3 \leq Ca \leq 0.6$.
In this regime, the capsule shape evolution appeared to be stable
with no apparent buckling. To characterize the shape of the deformed
capsule, we introduce the commonly employed Taylor deformation
parameter $D$ defined as
\begin{equation}
D = \frac{L-B}{L+B},
\end{equation}
where $L$ and $B$ are the maximum and the minimum distance in the
shear plane of a point on the surface of the capsule from its center.
We use this as the definition of $D$, though some authors,
e.g. \citet{pozrikidis98}, instead find a triaxial ellipsoid
with the same inertia tensor as the given capsule, and then take
$L$ and $B$ as the major and minor axis of that ellipsoid. 

The evolution of the deformation parameter $D$ for a capsule with unit viscosity
ratio ($\lambda=1$) is shown in Figs. (\ref{fig:Dl1}a) and (\ref{fig:Dl1}b)
at two different capillary numbers $Ca$. The time in the figures has
been non-dimensionalized by the shear rate. For comparison,
$D$ values reported by \citet{pozrikidis98}
and \citet{lac04} are also plotted. Note that \citet{pozrikidis98}
have used a zero-thickness shell model for their capsules, though
that gives only marginally lower deformation than a neo-Hookean
capsule at the same $Ca$ \citep{pozrikidis98}. Moreover,
they used the Young's modulus for computing their $Ca$,
such that our results should be compared with their results
at a $Ca$ which is (1/3) of the $Ca$ in this work.
The data in  Figs. (\ref{fig:Dl1}a) and (\ref{fig:Dl1}b) both show that the evolution of $D$ in this work
is in very good agreement with the corresponding results of \citet{lac04}.
Our results are also close to the values reported by \citet{pozrikidis98},
though the latter consistently display a slightly lower $D$. 
The broad agreement of $D$ between our values and
the literature values validates our implementation of the membrane mechanics along
with other aspects of our method such as time-stepping and the
already validated single layer integral.

Next, we demonstrate the convergence of the steady state $D$ at $Ca=0.6$ with respect to the GGEM parameter
$\alpha \Delta y_m$ in Fig. (\ref{fig:Dl1}c). For this calculation, $r_{cut}=a$ was held fixed, while $N_y$ was
varied between 65 and 113. As could be seen, a convergence in $D$ is observed
at $N_y=97$ corresponding to $\alpha \Delta y_m=0.625$ -- the parameter set for which all results
are presented in this section. The convergence of the steady state $D$ with respect to the 
number of triangular elements $N_{\Delta}$ is demonstrated in Fig. (\ref{fig:Dl1}d) for a $Ca=0.6$ capsule.
The three data points in this plot correspond to simulations with $N_\Delta=320$,
$1280$ and $5120$ elements. It is clear from this plot that the solution converges
linearly with $N_\Delta ^{-1}$, which is expected for linear elements.

\begin{figure}[!t]
\centering
\psfrag{gdott}[c][Bl][0.8]{$\dot{\gamma}t$}
\subfigure[$Ca=0.30$]{\includegraphics[width=0.45\textwidth]{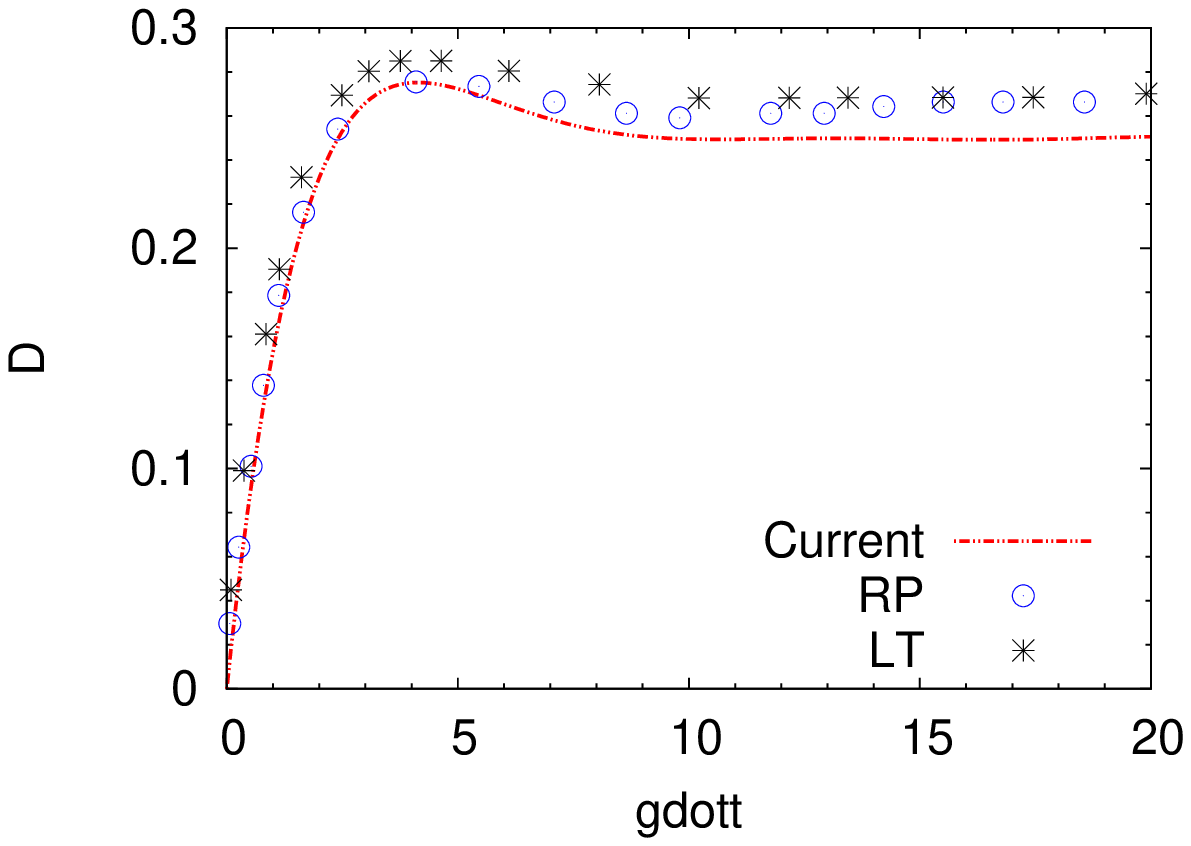}}
\subfigure[$Ca=0.60$]{\includegraphics[width=0.45\textwidth]{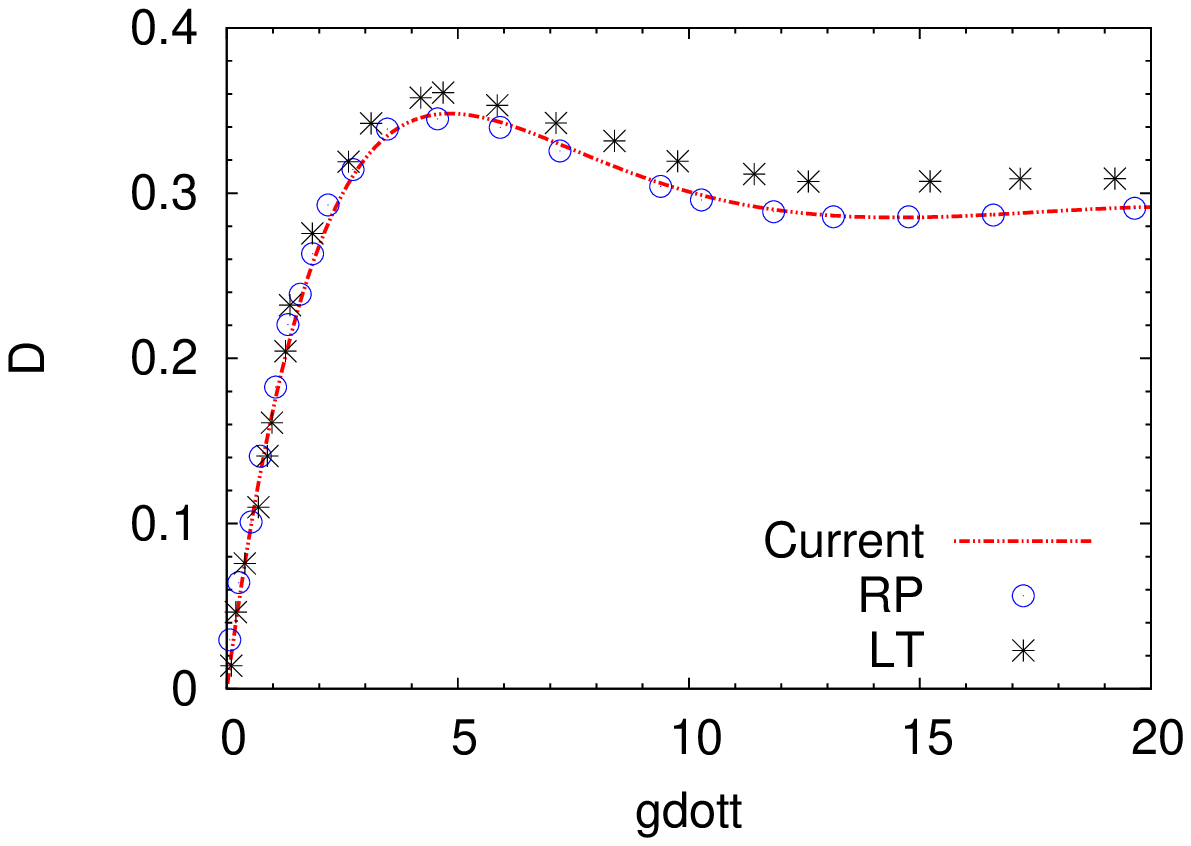}}
\subfigure[Convergence: $Ca=0.60$]{\includegraphics[width=0.45\textwidth]{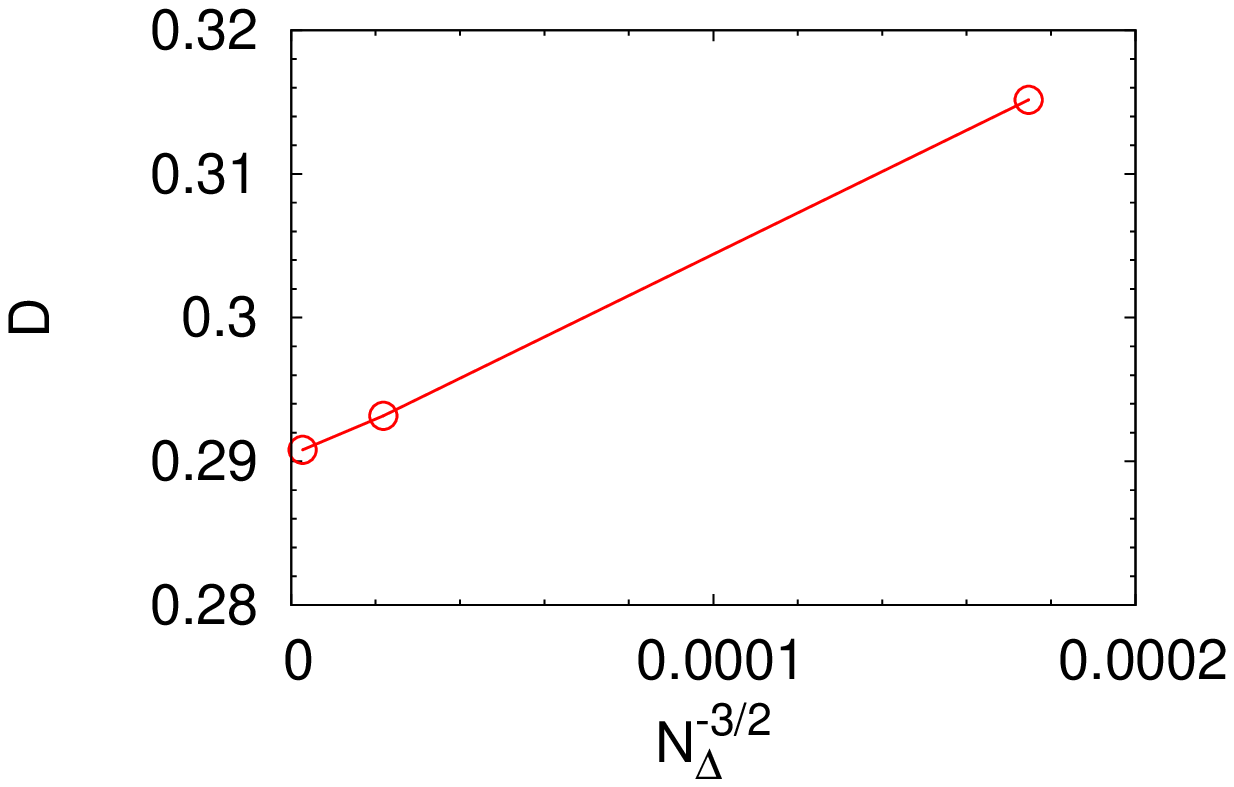}}
\caption{$\lambda=5$: Time evolution of the Taylor deformation parameter $D$ at (a) $Ca=0.3$
and (b) $Ca=0.6$. Results from the present study are compared with the work of \citet{pozrikidis98} (RP) and 
\citet{dvle10} (LT). Simulation parameters were the same as in Fig. (\ref{fig:Dl1}).
 (c) Convergence of D with $N_{\Delta}$ at $Ca=0.6$.}\label{fig:Dl5}
\end{figure}

\begin{figure}[!t]
\centering
\psfrag{gdott}[c][Bl][1.0]{$\dot{\gamma}t$}
\subfigure[$Ca=0.30$]{\includegraphics[width=0.45\textwidth]{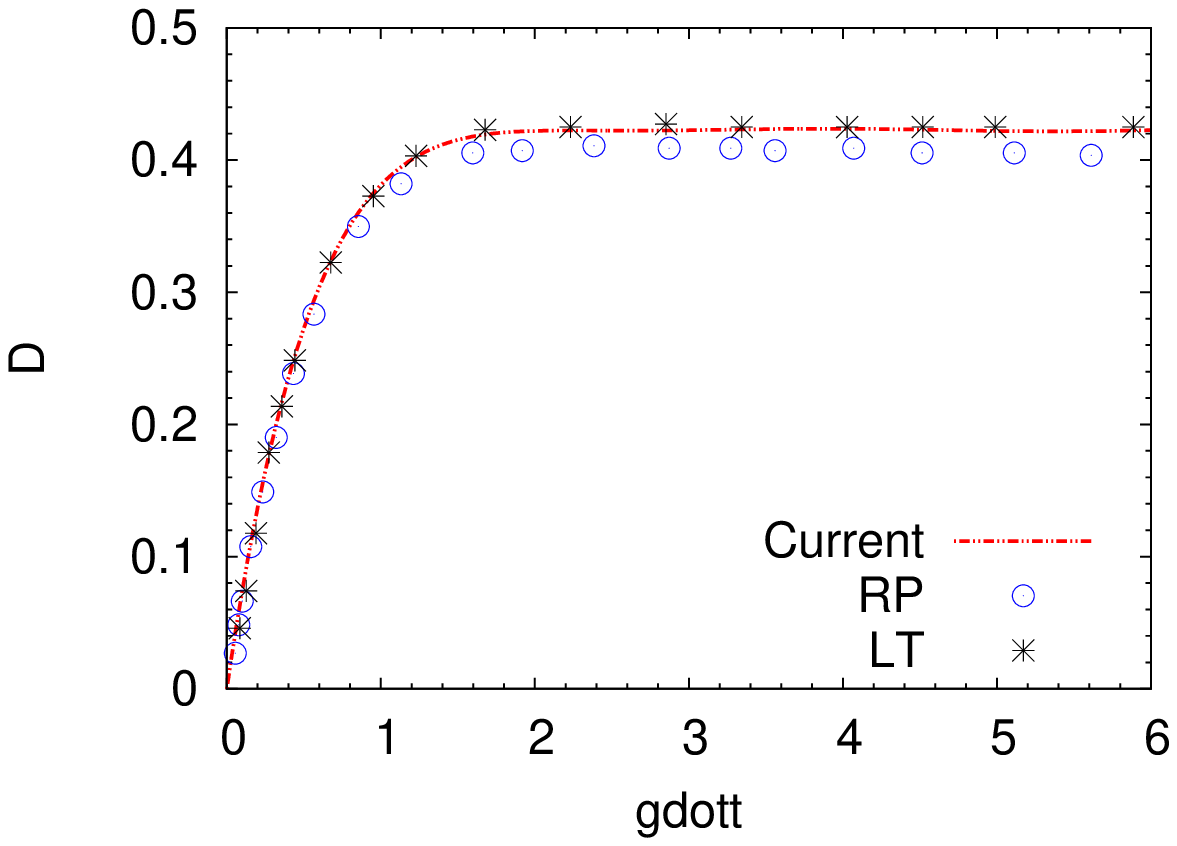}}
\subfigure[$Ca=0.60$]{\includegraphics[width=0.45\textwidth]{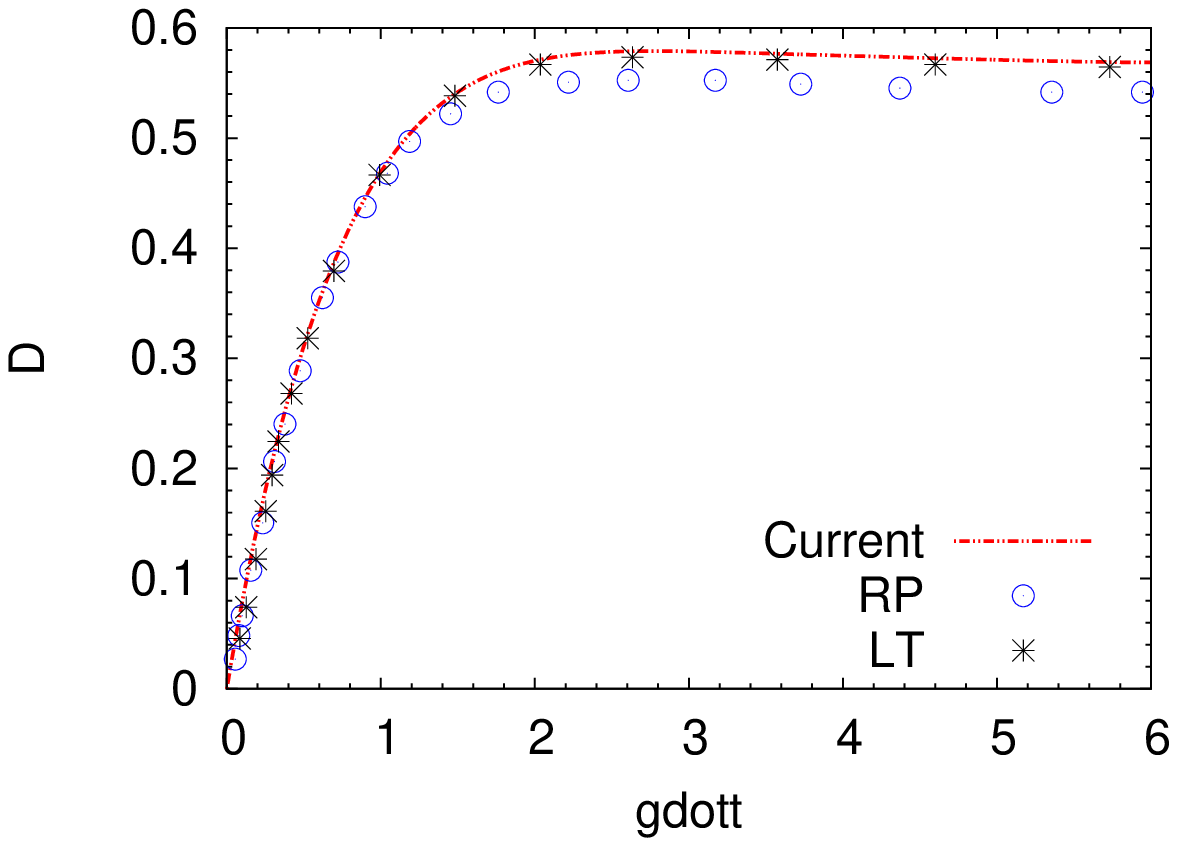}}
\subfigure[Convergence: $Ca=0.60$]{\includegraphics[width=0.45\textwidth]{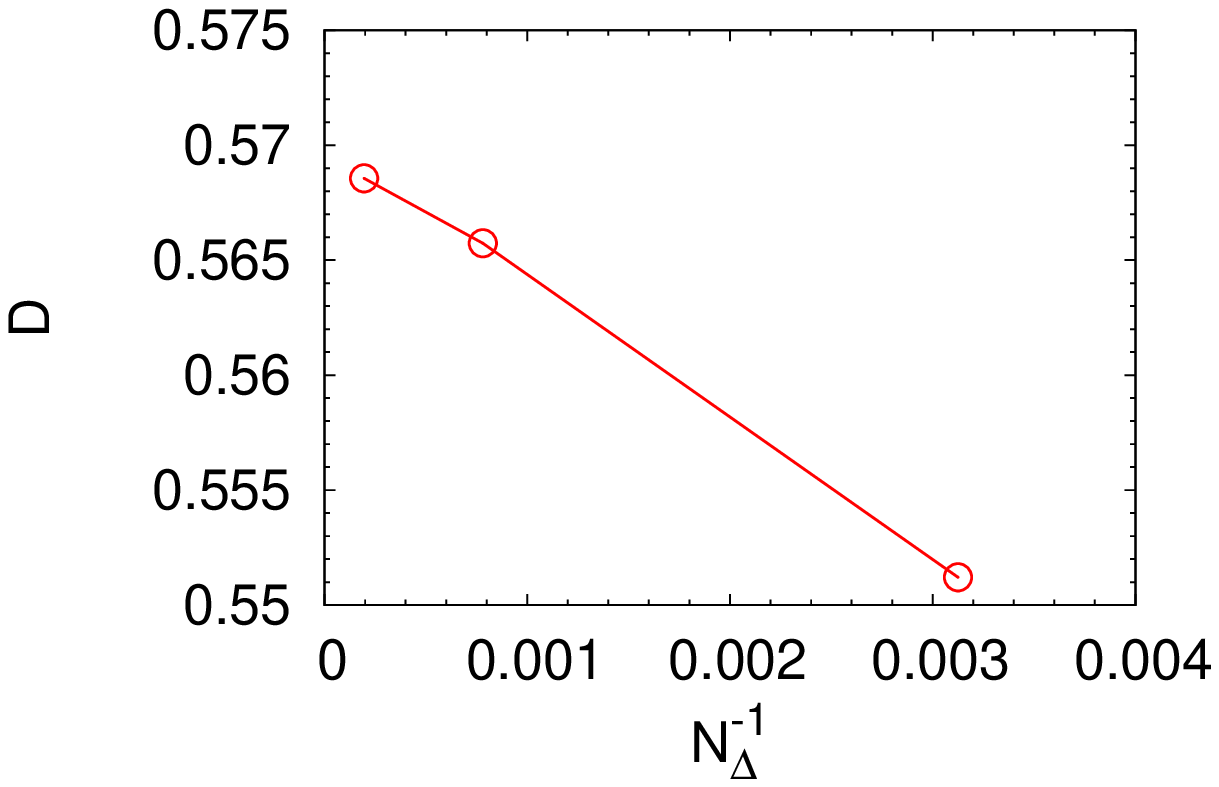}}
\caption{$\lambda=0.2$: Time evolution of the Taylor deformation parameter at $Ca=0.3$ and 
(b) $Ca=0.6$. Results from the current study are compared with the work of
\citet{pozrikidis98} (RP) and \citet{dvle10} (LT). Simulation parameters were the same as in Fig. (\ref{fig:Dl1}).
	(c) Convergence of $D$ at $Ca=0.6$ with $N_{\Delta}$.}\label{fig:Dlp2}
\end{figure}

We now briefly discuss the deformation parameter results for capsules with non-unit viscosity
ratios (i.e. $\lambda \neq 1$). These are reported in Fig. (\ref{fig:Dl5}) for $\lambda=5$
and in Fig. (\ref{fig:Dlp2}) for $\lambda=0.2$, and are compared with the results of \citet{pozrikidis98} and
\citet{dvle10}, with the latter reference employing the immersed boundary
technique for their simulations. At $\lambda=5$, the results 
for $D$ in the present work are slightly smaller than those reported by the previous
authors at $Ca=0.3$. At $Ca=0.6$, an excellent agreement with the results of \citet{pozrikidis98}
is observed, though our results for $D$ are slightly lower than those of \citet{dvle10}.
We next show the convergence of the steady state $D$ at $Ca=0.6$ with respect to $N_\Delta$ in Fig. (\ref{fig:Dl5}c).
These results indicate that the error decays as $N_{\Delta}^{-3/2}$, thereby implying
that the error incurred in the calculation of the singular double layer integral dominates 
the overall error; see  Sec. (\ref{sec:dbl_l}) for details.

We next discuss the results for the deformation parameter for capsules with $\lambda=0.2$ 
(Figs. \ref{fig:Dlp2}a and \ref{fig:Dlp2}b). In this case, we observe a very good agreement with the 
results of \citet{dvle10} at both $Ca=0.3$ and $Ca=0.6$. The values
reported by \citet{pozrikidis98} are also close, though they are marginally lower than
the results in the present study. Lastly, we show the convergence of the steady state $D$ 
at $Ca=0.6$ with respect to $N_\Delta$ in Fig. (\ref{fig:Dlp2}c). In this case the 
error is observed to decay at the expected rate of $N_{\Delta}^{-1}$ -- this 
probably implies that the error incurred in the calculation of the singular double 
layer integral is dominant only at high values of $\lambda$.


\begin{figure}[!t]
\centering
\subfigure[Trajectory]{\includegraphics[width=0.45\textwidth]{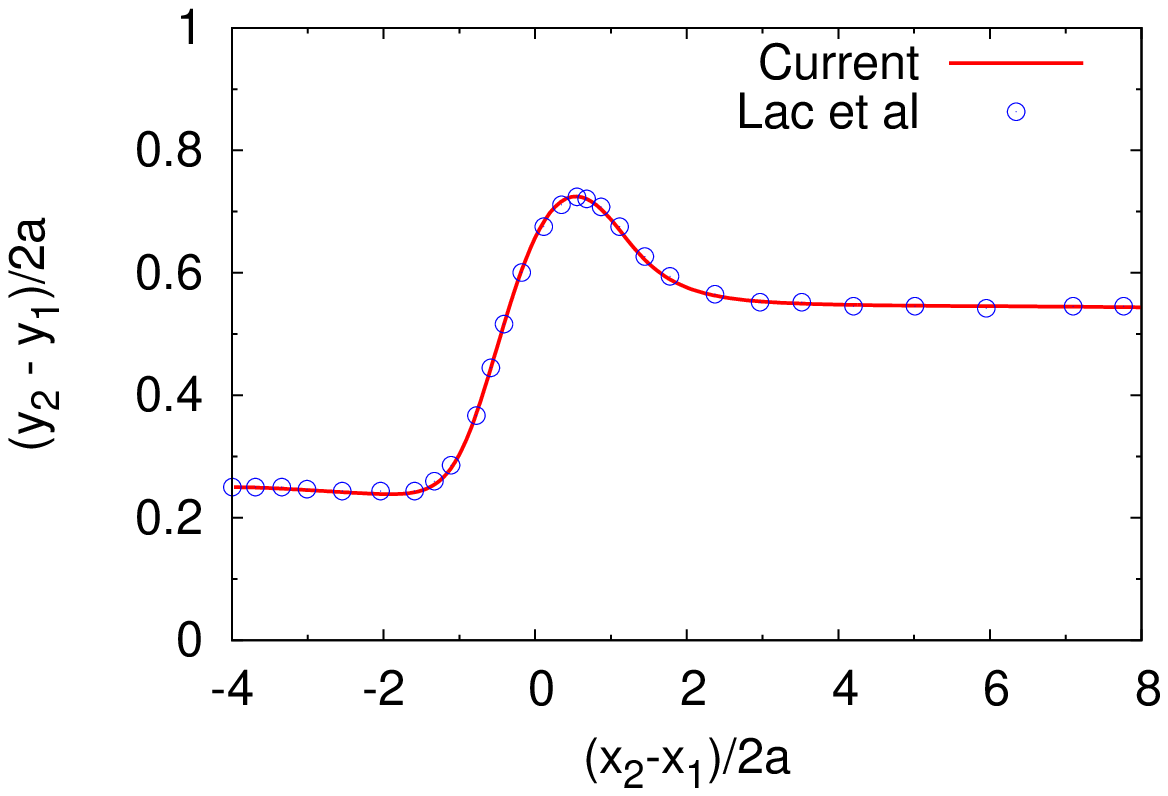}}
\subfigure[Convergence with $\alpha \Delta y_m$]{\includegraphics[width=0.45\textwidth]{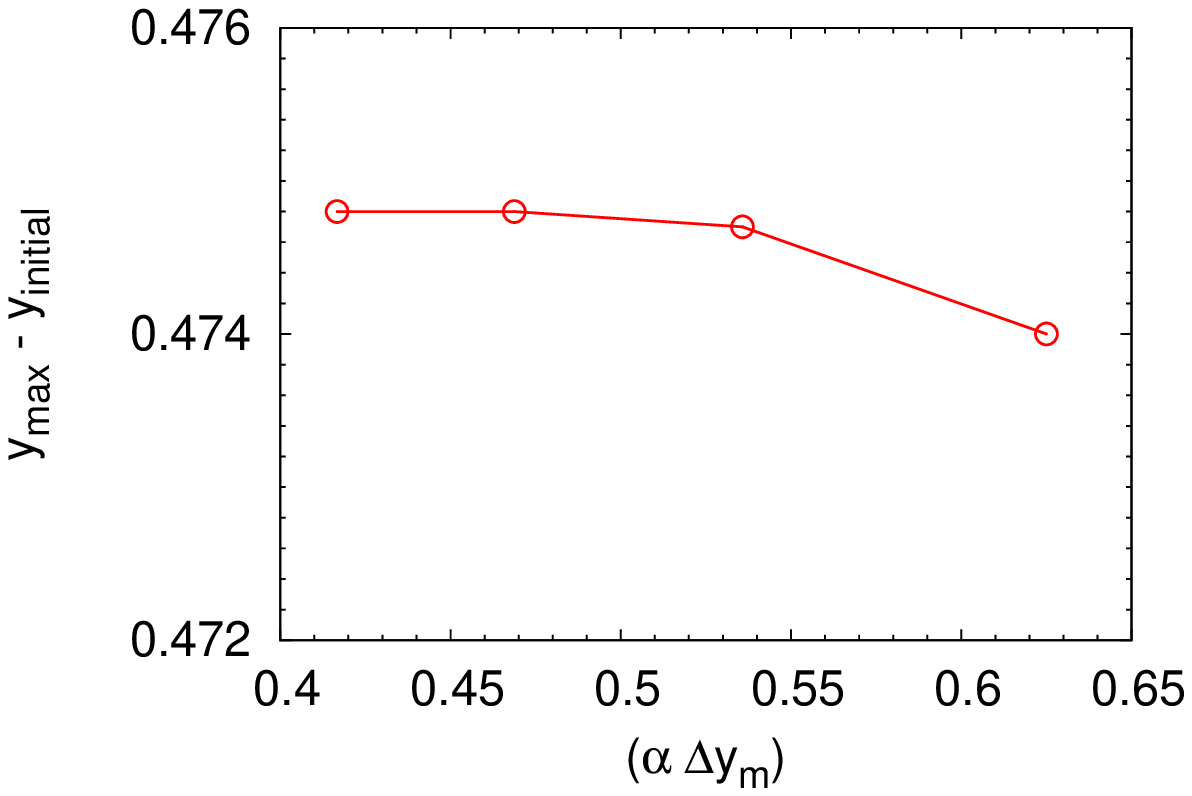}}
\subfigure[Convergence with $N_\Delta$]{\includegraphics[width=0.45\textwidth]{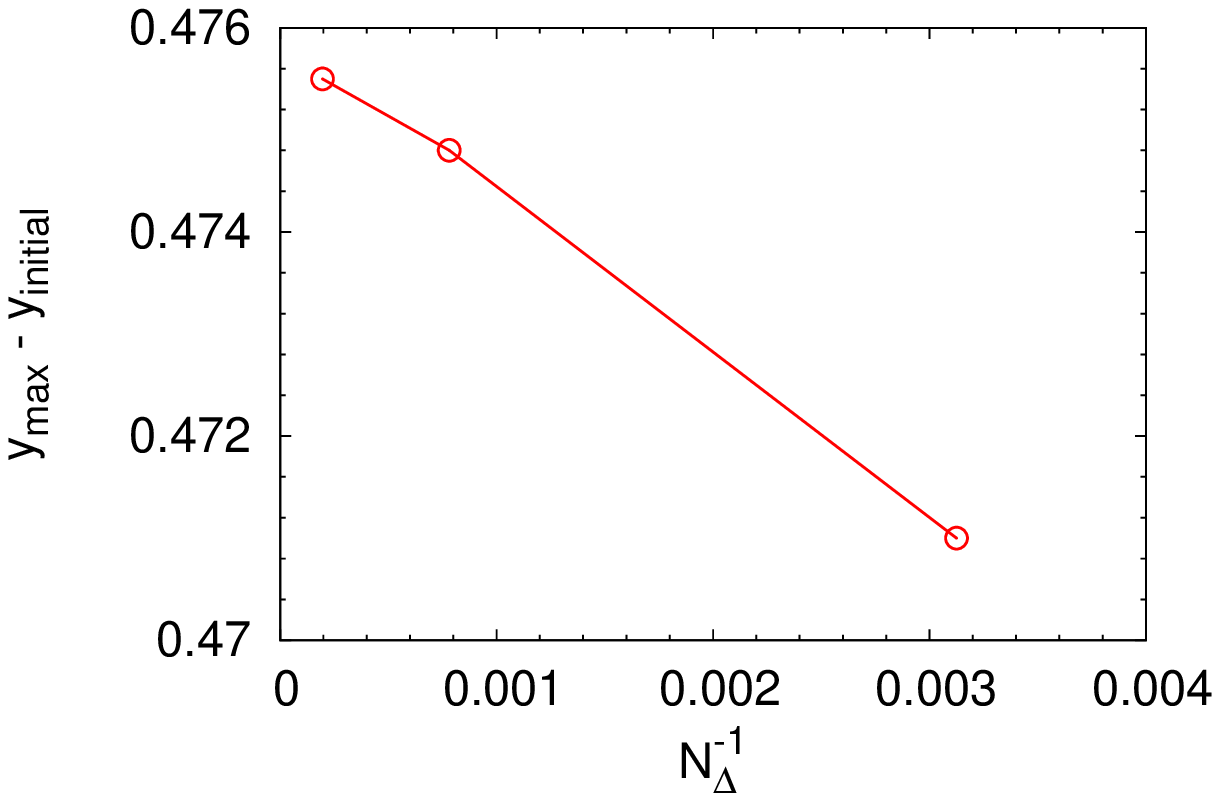}}
\caption{Pair collision: $\lambda=1$. (a) Shows the separation between the centers of mass
  of capsules in the gradient direction as a function of their separation in the flow
  direction. Also plotted is the corresponding result from \citet{lac07}. Simulation
  parameters were: $L_x=L_y=H=30a$, $N_y=129$, $r_{cut}=2a$, $\alpha \Delta y_m = 0.469$,
	$N_{\Delta} = 1280$,  and $\Delta x_m = \Delta y_m = \Delta z_m$. 
	(b) Convergence of the maximum displacement of either particle (absolute value)
  from its initial position along the gradient direction ($y$) as a function
  of $\alpha \Delta y_m$. Data points in this curve were obtained by holding 
  $r_{cut}=2a$ fixed and varying $N_y$. (c) Convergence of the maximum absolute 
	displacement of a particle in the gradient direction with $N_{\Delta}$.}\label{fig:pcollision}
\end{figure}
\subsection{Pair collision}\label{sec:pcoll}
As a final test problem, we consider the collision between a pair of capsules
with $\lambda=1$ in a simple shear flow and compare the results with the work
of \citet{lac07} in Fig. (\ref{fig:pcollision}a). We first describe the problem setup.
The size of the cubic box (slit) for this problem was set to $30a$ to approximate
an unbounded domain; simulations were performed with $N_y=129$, $r_{cut}=2a$,
$\alpha \Delta y_m = 0.469$, $N_{\Delta}=1280$, and $\Delta x_m = \Delta y_m = \Delta z_m$.
The two capsules were initially kept in the same
flow-gradient plane ($x-y$), such that the initial separation in the
flow direction was $x_2-x_1 = -8a$, while the initial offset in
the gradient direction was $y_2-y_1=0.5a$. As in \citet{lac07},
we preinflate the capsule by $5\%$, i.e. the radius of the
spherical capsule was increased by $5\%$ over its rest
value, and this new increased radius is denoted by $a$. For
a spherical shape, this inflation does not lead to any flow
due to the incompressibility condition as discussed above
in the case of a spherical drop. Nonetheless, this inflation
keeps the membrane in a state of tension at rest and, if sufficient,
will prevent buckling during the course of the collision. With these
preliminaries, we return to Fig. (\ref{fig:pcollision}a) where
we show the relative separation between the center of masses in the gradient direction
as a function of the corresponding separation in the flow direction.
A very good agreement with the results of \citet{lac07} is evident. Next, we show the convergence of the numerical
scheme with $\alpha \Delta y_m$ in Fig. (\ref{fig:pcollision}b),
where we plot the absolute value of the maximum displacement in the gradient direction
for either particle. For this calculation $r_{cut}=2a$ was held fixed, while $N_y$ was 
varied between 97 and 145. A convergence is seen at $\alpha \Delta y_m =0.469$ 
corresponding to $N_y=129$ -- the parameters for which results are reported in this section.
The convergence of the maximum displacement in the gradient direction with respect to 
$N_\Delta$ is demonstrated in Fig. (\ref{fig:pcollision}c), which confirms the expected error 
decay rate of $N_{\Delta}^{-1}$.

\section{Multiparticle Simulations: suspension apparent viscosity and computational complexity}\label{sec:mult_part}

\begin{figure}[!t]
\centering
\subfigure[Snapshot for $\lambda=1$ ]{\includegraphics[width=0.35\textwidth]{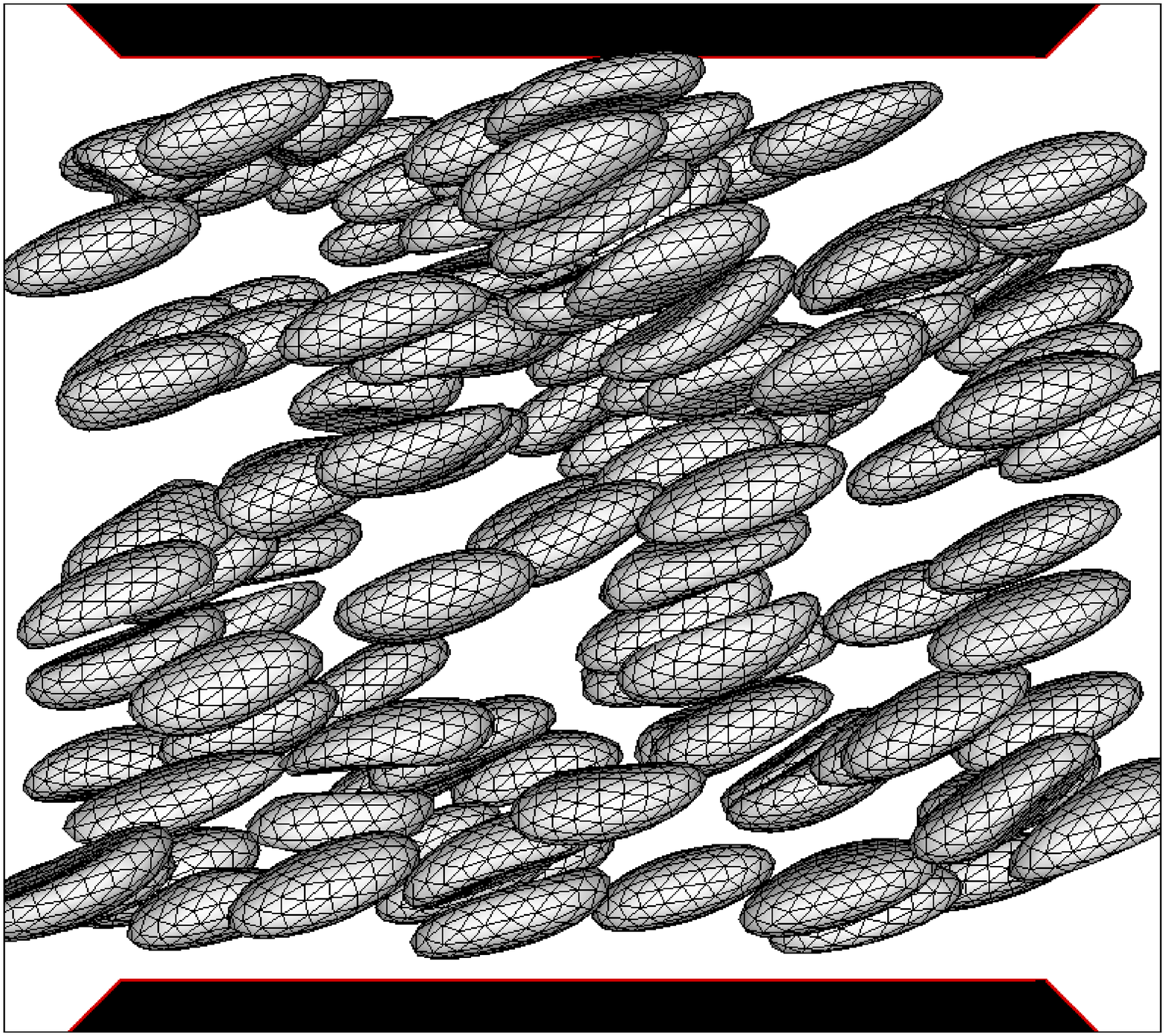}}
\subfigure[Snapshot for $\lambda=5$]{\includegraphics[width=0.35\textwidth]{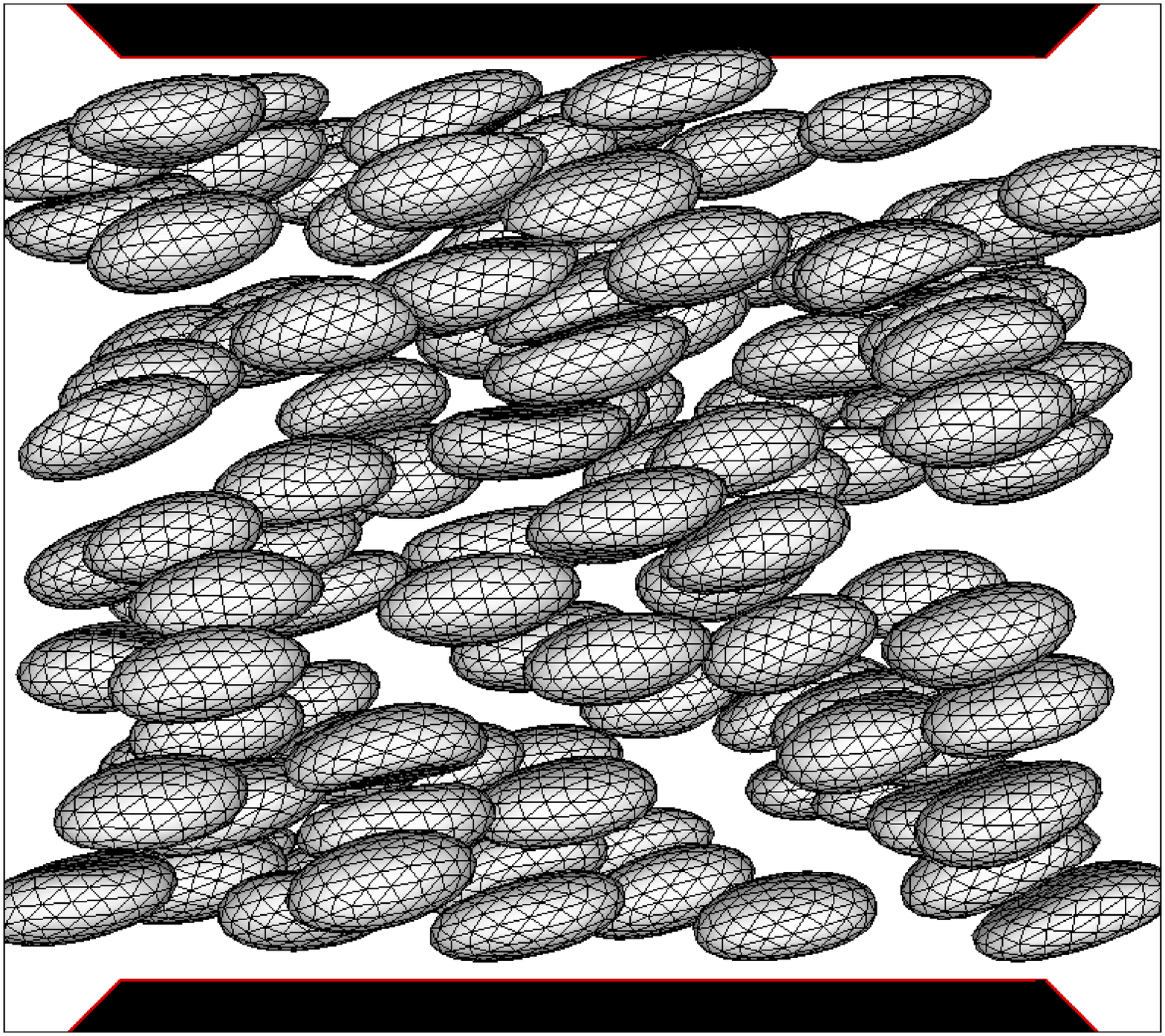}}
\subfigure[Convergence]{\includegraphics[width=0.4\textwidth]{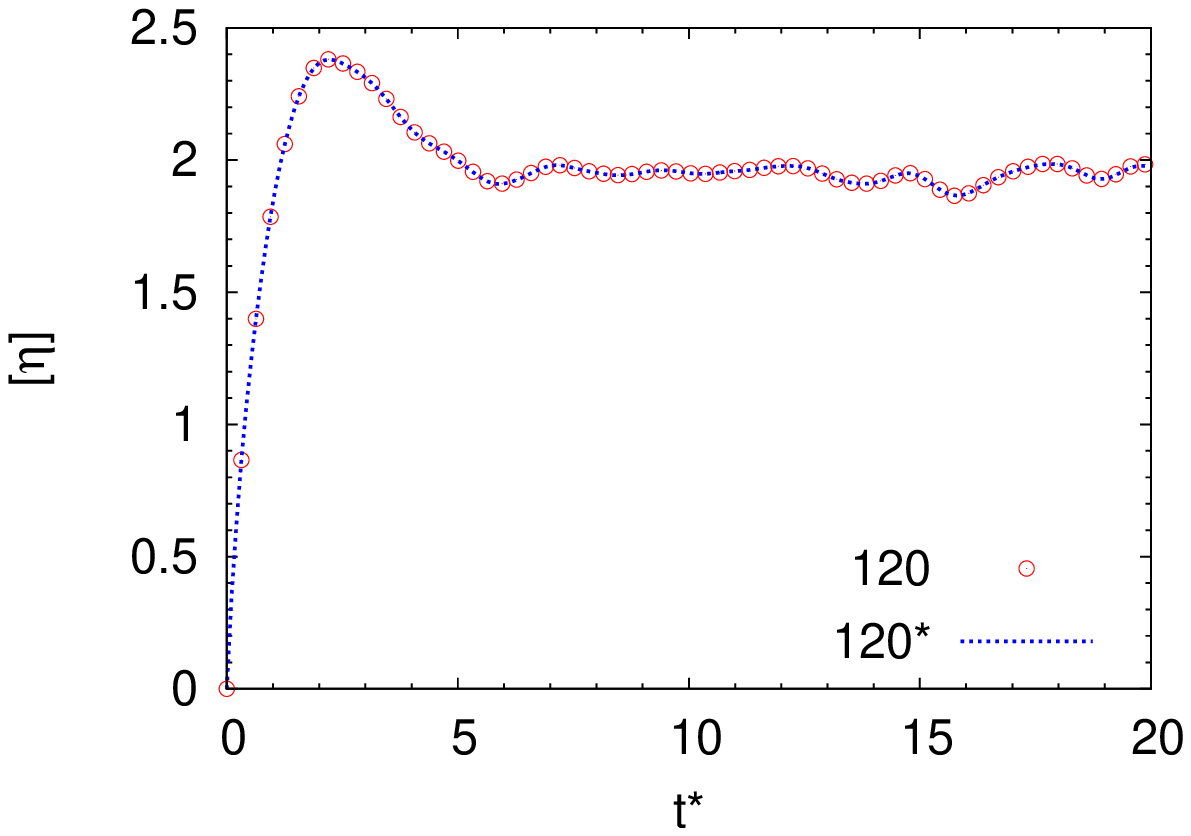}}
\subfigure[$\lambda$ dependence]{\includegraphics[width=0.4\textwidth]{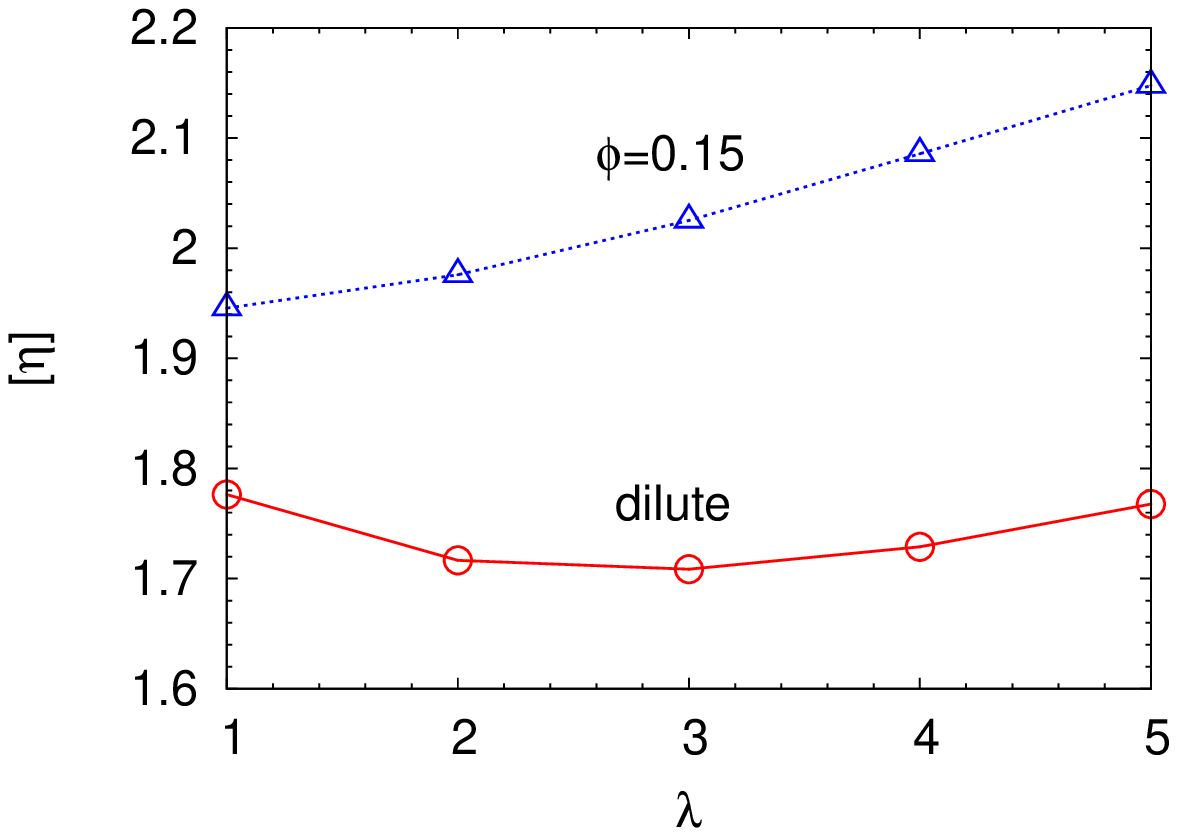}}
\caption{Suspensions of capsules at $Ca=0.5$ and volume fraction $\phi=0.15$:
(a) Snapshot of $\lambda=1$ capsule suspension with $N_p=120$, (b) Snapshot of $\lambda=5$ capsule suspension with $N_p=120$,
(c) Convergence of the apparent intrinsic viscosity for $\lambda=1$ capsules in $N_p=120$ particle system.
The simulation parameter $\alpha \Delta y_m$  was 0.5 for the run labeled 120, while it was 0.417 for the run
	labeled $120^*$, (d) Effect of 	$\lambda$ on the apparent intrinsic viscosity for $N_p=120$ particle system.
	The simulation parameters in all cases above were: $r_{cut}=a$, $\alpha \Delta y_m =0.5$,
	$N_\Delta=320$, and $\Delta x_m = \Delta y_m = \Delta z_m$ unless otherwise mentioned.}\label{fig:visc}
\end{figure}

In this section, we report results from several large scale simulations on multiparticle 
suspensions of capsules. In the first part of the section, we present results for the
suspension viscosity and discuss its dependence on the viscosity ratio. 
In the second part, we discuss the expected computational complexity of the 
algorithm and verify it with timing results from the multiparticle simulations.

\subsection{Suspension viscosity}\label{sec:susp_visc}
We consider here a suspension of Neo-Hookean capsules in a cubic slit
(Fig. \ref{fig:visc}a,b). These suspensions are subjected to a simple shear flow
at a capillary number of $Ca=0.5$; the volume fraction of the suspension is  $\phi=0.15$,
which is typical of the blood flow in the microcirculation \citep{fung96}. 
Capsules with five different viscosity ratios $\lambda$ are considered:
$\lambda=1$, 2, 3, 4, and 5. The surface of each of the capsules was discretized into $N_{\Delta}=320$
triangular elements, while four different system sizes were
considered with the number of particles being $N_p=15$, $N_p=30$, $N_p=60$, and $N_p=120$. 
In each of the problems, $r_{cut}=a$ was kept fixed, where $a$ is radius of the spherical capsule at rest.
Similar to Sec. (\ref{sec:pcoll}), the capsules were preinflated by $5\%$ 
to prevent membrane buckling; the radius after preinflation is denoted by $a$. The number
of mesh points $N_y$ for the global solution varied between $61$ and $121$ such that
$\alpha \Delta y_m = 0.5$ in all cases with $\Delta x_m = \Delta z_m = \Delta y_m$.
For testing the convergence of the results with respect to GGEM parameters, 
some simulations were also run with $20\%$ extra mesh points in each of the directions
corresponding to $\alpha \Delta y_m = 0.417$. We discuss next an important issue
in multiparticle simulations, which concerns the treatment of near singular integrals -- these 
integrals arise when the gaps between the particles become small. 

In a suspension of particles subjected to shear, it is not uncommon to find particle pairs
separated by a small gap. This is true, at least occasionally, even in suspensions with a moderate volume 
fraction such as $\phi=0.15$ studied here. When the gap between a particle pair 
becomes small, the interparticle contributions between such a pair from both the
single as well as the double layer integrals become nearly singular;
these nearly singular integrals require special treatment or the simulations may diverge \citep{zinchenko2002}.
In the literature, several techniques have been proposed to address this numerically difficult problem. 
A few among these are the near singularity subtraction technique \citep{hinch96,zinchenko2002}
and coordinate mapping techniques \citep{higdon95} for the evaluation of the near singular integrals.
These techniques will, however, require very high surface mesh resolution and very high 
order quadrature techniques for the accurate evaluation of the nearly 
singular integrals \citep{freund07,biros11} -- the cost associated with these requirements can be prohibitive. 
An alternative approach is the use of a short range repulsive force, which 
can prevent the formation of small gaps \citep{freund07}, though this is 
not very effective in three dimensional simulations \citep{freund10}.
The best approach appears to be an overlap correction in an auxiliary step \citep{freund10}.
This approach is also frequently used in suspensions of rigid particles \citep{foss2000,meng08,kumar10}.
In the present work, like in past efforts \citep{meng08,freund10}, we not only correct the
overlaps, but also maintain a minimum gap ($h_s^m$) between the surfaces of two particles.
This approach was also employed in our recent work \citep{kumar11b}. 
For simulations in the current section, we set the minimum gap parameter to a small value of $h_s^m=0.05a$.

The overlap correction procedure employed in this work involves moving a pair of overlapping particles 
apart along their line of centers like a rigid particle until the 
minimum gap requirement is satisfied. Translating  the capsules like
a rigid particle in this auxiliary step has the benefit that the
shapes of the particles remain unchanged, as is the orientation of
the particles with respect to the flow. In general, multiple steps
of the correction procedure is required, as the correction of overlap between one pair could result in other 
overlaps \citep{foss2000,meng08}. This overlap correction step involves 
minimal displacement of the particles, on the order of $h_s^m$. 
Given that the volume fraction studied in this work is $\phi=0.15$, this procedure was rarely required  --
on an average, less than $0.1$ particle pairs in the $N_p=120$ particle system exhibited minimum gap violations at any given
time. We finally remark that apart from correcting the minimum gap violations in the system,
no special treatment was accorded to the evaluation of the near singular integrals in the present effort.

Having discussed the procedure for controlling the minimum gap in the system, 
we now turn to the results from multiparticle simulations. All these simulations
were initiated by placing the particles randomly in the simulation box, and then they
were sheared for a total non-dimensional time of $t^* = \dot{\gamma}t = 20$.
We show some representative snapshots from $120$ particle simulations in Figs. (\ref{fig:visc}a)
and (\ref{fig:visc}b) for $\lambda=1$ and $\lambda=5$ capsules, respectively. It is immediately obvious from these 
snapshots that the more viscous capsules deform less and also have a smaller 
inclination angle with the flow direction; both of these observations are similar
to observations in isolated sheared capsules \citep{pozrikidis98}. 
The convergence of the simulation with respect to  GGEM parameters is demonstrated in 
Fig. (\ref{fig:visc}c), where we plot the suspension apparent intrinsic viscosity $[\eta]$
for $\lambda=1$ capsule and $N_p=120$ particle system for two different mesh resolutions
corresponding to  $\alpha \Delta y_m=0.5$ and $\alpha \Delta y_m = 0.417$.
The apparent intrinsic viscosity is defined as $[\eta] = \Sigma_{xy}^p/(\mu \dot{\gamma}\phi )$, where $ \Sigma_{xy}^p$
is the particle contribution to shear stress, while $\mu$ is the suspending fluid viscosity.
The particle contribution to the stress tensor is given by \citep{kennedy94}
\begin{equation}
\Sigma_{ij}^p = \frac{1}{V} \sum_{m=1}^{N_p} \int_{S_m} [ \Delta f_i x_j + \mu(\lambda-1)(u_i n_j + u_j n_i) ] dS,
\end{equation}
where the sum in the right hand side is over all the particles in the system.
As can be seen in Fig. (\ref{fig:visc}c), $[\eta]$ is nearly identical for simulations
run with $\alpha \Delta y_m =0.5$ and $\alpha \Delta y_m =0.417$, thereby
demonstrating the convergence of the simulation with respect to GGEM parameters.
All the remaining simulations were performed with $\alpha \Delta y_m = 0.5$.
The effect of viscosity ratio on the apparent intrinsic viscosity is shown in Fig. (\ref{fig:visc}d).
These results represent an average over the last $10$ time units in the $N_p=120$ particle systems,
which have a confinement ratio of $2a/H=0.134$. For a direct comparison, the plot also shows $[\eta]$
for a dilute suspension of capsules obtained from single particle simulations in the same
geometry. In dilute suspensions, a non-monotonic variation of the viscosity with $\lambda$
is obvious -- this behavior has been demonstrated before in the literature \citep{bagchi10}. However, the 
non-monotonicity vanishes at the non-dilute volume fraction of $\phi=0.15$. This indicates that the
contribution to the overall stress from particle-particle interactions is a monotonically increasing 
function of $\lambda$, and, in non-dilute suspensions, easily compensates the non-monotonic variation of the 
isolated particle contribution. Hence caution is warranted before extrapolating
trends from dilute systems to non-dilute systems like blood flow.

\subsection{Computational complexity}\label{sec:algo_com}
We devote the remainder of this section to analyzing the overall computational
complexity of our algorithm. Before presenting the timing results from the 
detailed numerical simulations presented above, it will be useful to first discuss
the expected computational cost associated with various steps in the algorithm, and
consequently the overall implementation. The first step in the solution procedure,
as discussed in Sec. (\ref{sec:sol}), involves iteratively solving for the single
layer density $\mathbf{q}^b$ (step 1). This is followed by the computation of the fluid
velocity $\mathbf{u}^b$ at the element nodes (step 2) using the single layer density $\mathbf{q}^b$
computed in the previous step.  The computational cost associated with each iteration of step 1 and that
of step 2 has an identical optimal scaling with $N=N_\Delta \times N_p$, each of which
is essentially determined by the computational cost associated with the Stokes flow
solver (GGEM) described in Sec (\ref{sec:greens_fn}).
Therefore, for a direct cost comparison with the multiparticle flow problem here,
we consider an auxiliary problem involving a collection of $N$ point forces; both problems
then have an identical computational cost scaling with $N$. At this stage, it will also be worth pointing
out that the computational complexity analysis presented here closely follows the corresponding analysis
in PME like methods \citep{tornberg10,tornberg11} as the underlying ideas are fairly similar.

The overall cost associated with the GGEM Stokes flow solver is the sum of costs associated with the local
problem and the global problem. The cost of the local solution scales as
the product of the number of point forces $N$ and the number of neighbors
within a distance $r_{cut} \sim \alpha^{-1}$ of each of the point forces.
If we require the computational cost of the local problem to scale as
$t_l \sim O(N)$, then the number of neighbors per point force must 
stay constant with changing system size (meaning $N$ here). The 
system size $N$ can be increased in two contrasting ways: 
(i) by increasing volume at constant density  (i.e., by increasing $V$ while maintaining $N/V$ constant, $V$ is the system volume),
and (ii) by increasing density at constant volume (i.e., by increasing $N/V$ while maintaining $V$ constant).
If the system size is increased at 
constant density, we require  that $\alpha$ (or $r_{cut}$) be held constant,
and if the system size is increased at constant volume, we require that 
$\alpha \sim N^{1/3}$ (or $r_{cut} \sim N^{-1/3}$). This scheme for
choosing $\alpha$ ensures that the average number of near 
neighbors per point force is independent of the system size. Hence,
irrespective of how $N$ is varied, we always obtain $t_l \sim O(N)$.
Next, we determine the computational cost of the global solution procedure. 
Before we proceed further, it is important to realize that the
error in the global solution is essentially controlled by the 
parameter $\alpha \Delta y_m$\footnote{the error in the local solution is set by the 
choice of the parameter $\alpha \, r_{cut}$, which is unchanged with changing $N$};
see Sec. (\ref{sec:global_sol}). Therefore, for the error in the global solution to remain of the same order with 
changing system size $N$, we require that $\alpha \Delta y_m$ 
be held constant; this implies $\Delta y_m \sim \alpha^{-1}$. 
Coupling this requirement with the choices of $\alpha$ discussed
above in different scenarios, we conclude that the total number
of mesh points involved in the calculation of the global 
solution $n=N_xN_yN_z$ must be varied proportionally to $N$, i.e. $n \sim N$.
Having determined the scaling of $n$, we next present the expression
for the computational cost of the global solution $t_g$ as follows:
\begin{equation}
t_g \sim  \frac{n}{N_x} O(N_x \log N_x ) + \frac{n}{N_z} O(N_z\log N_z) + \frac{n}{N_y}(O(N_y) + O(N_y\log N_y)),
\end{equation}
where the first two terms on the right hand side denote the cost associated with the
FFT operations in $x$ and $z$ directions respectively, while
the last term is associated with the cost of the Chebyshev-tau solver in the
wall normal $y$ direction. Note that the $O(N_y)$ cost in the expression for the Chebyshev-tau solver
is associated with the quasi-tridiagonal solve, while $O(N_y\log N_y)$
cost is associated with the computation of the Chebyshev transforms and its inverse
with FFTs.  Simplifying the above expression and noting that $n\sim N$, we obtain the following asymptotic scaling 
\begin{equation}
t_g \sim  n \log n \sim N \log N .
\end{equation}
The overall cost per iteration of step 1 or of step 2 is therefore,
\begin{equation}
t = t_l + t_g \sim O(N) + O(N\log N) \sim N \log N.
\end{equation}
The total cost of step 1 is the cost per iteration times the number
of iterations required for convergence.
Now, if the number of iterations in step 1 is independent of $N$,
then the computational cost of the overall algorithm will scale as $N\log N$.
On the other hand, if the number of iterations in step 1 is
dependent on the system size, say it scales as $N^e$, then the
computational cost of the overall algorithm will scale as:
\begin{equation}
t \sim N^{1+e}\log N.
\end{equation}
Note that the need for iterative solution of step 1 or a related second kind integral equation is not unique to the present formulation, but is a general feature of any accelerated boundary integral method with $\lambda\neq 1$ \citep{freund10}. 
\begin{figure}[!t]
\centering
\includegraphics[width=0.5\textwidth]{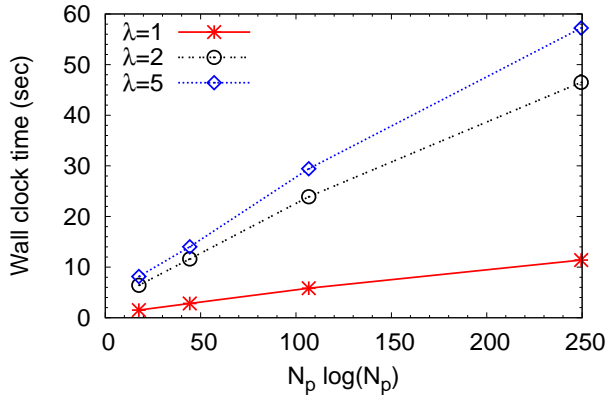}
\caption{Wall clock time per stage of a two stage midpoint method for various
viscosity ratios $\lambda$ and for various system sizes. The four data points in the plot
correspond to $N_p=15$, $30$, $60$, and $120$ respectively. Note that
the time is plotted against $N_p \log N_p$.}\label{fig:tscale}
\end{figure}

Having determined the expected scaling of our algorithm, we next report
timing results from the detailed numerical simulations presented in Sec. (\ref{sec:susp_visc}).
All runs were performed on a single core of a eight core machine with a 2 GHz Intel Xeon
processor running Linux. We plot the time required per stage of the two stage midpoint
time stepping algorithm in Fig. (\ref{fig:tscale}).
Note that the abscissa in the plot is $N_p\log N_p$; the four data points
in this plot are respectively for $N_p=15$, $N_p=30$, $N_p=60$ and $N_p=120$ as
discussed above.  One can conclude from this plot that in all cases the computational cost
scales approximately as $t \sim N_p \log N_p$ (in fact, in this case, the increase 
in the computational cost appears to be slower than the expected $N_p \log N_p$).
Another important feature to note in the plot is the jump in computational cost
as one moves from a matched viscosity problem to a non-matched one, which
is expected as no iterations are required in matched viscosity problems. 
The number of iterations required for convergence was found to be independent
of the system size for problems considered here, though it was found to increase
with increasing $\lambda$. The simulations with $\lambda=2$ capsules
required  approximately 2 iterations on an average, while the simulations with $\lambda=5$ capsules
required approximately 3 iterations. To summarize this section, we note that
a near perfect scaling of $N \log N$ is obtained for both matched viscosity
and non-matched viscosity problems. A few words of caution are necessary here, though, 
as at higher volume fractions and/or at much larger system sizes, the number of 
iterations for convergence is expected to become system size dependent. 
It must be emphasized here that this aspect is not specific to our implementation, but is intrinsic to 
the boundary integral equation for non-matched viscosity problems.
Future work on enhancements in the algorithm should address this by the
development of efficient preconditioners. Another obvious enhancement 
in the algorithm will be its parallelization  to take advantage of 
cheaply available multicore processors.

\section{Conclusions} 
A new accelerated boundary integral method for multiphase
Stokes flow in a confined geometry was presented. The  complexity of the method
scales as $O(N\log N)$ for the slit geometry discussed
in the present paper. The acceleration in the method
was provided by the use of General Geometry Ewald-like (GGEM)
method for the fast computation of the velocity and stress
fields driven by a set of point forces in the geometry of interest.
Due to non-periodic nature of the domain,
an alternative boundary integral formulation was employed, necessitated by the
requirements of the acceleration technique. An efficient methodology
was presented to compute the resulting double and single layer
integrals using the GGEM technique. The resulting implementation
was validated with several test problems. The computational
complexity of the algorithm was verified to be $O(N\log N)$ with
timing results from several large scale multiparticle simulations.

\section*{Acknowledgments}
The authors gratefully acknowledge helpful discussions with Pratik Pranay, Yu Zhang and Juan Hernandez-Ortiz
on the implementation of GGEM. This work was supported by NSF Grants CBET-0852976 and CBET-1132579.
\clearpage

\begin{figure}[!t]
\centering
\includegraphics[width=0.6\textwidth]{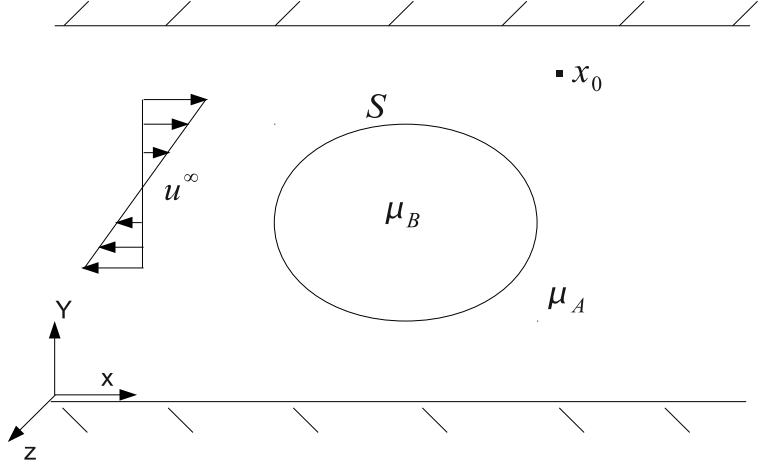}
\caption{Schematic of the problem described in \ref{sec:bi_derive}.
The figure shows the dispersed phase with viscosity $\mu_A$
and a particle with viscosity $\mu_B$. The surface of the
particle is denoted by $S$. Also shown is the
undisturbed flow  denoted by $\mathbf{u}^{\infty}$.}\label{fig:geom1}
\end{figure}

\appendix

\section{Derivation of Boundary Integral equation}\label{sec:bi_derive}
Consider a two phase flow in a specified geometry as shown in Fig. (\ref{fig:geom}).
The two fluids are respectively denoted by $A$ and $B$  in the
figure with viscosity $\mu_A$ and $\mu_B$. The fluid $B$ is assumed to be enclosed
by an impermeable interface denoted by $S$, which has its own characteristic properties
(e.g. drops, capsules, vesicles, etc). Now, consider a  point $\mathbf{x}_0$ in
fluid $A$ as shown in Fig. (\ref{fig:geom1}). Applying the reciprocal theorem to the disturbance
velocity in the region $A$ denoted by $\mathbf{u}^{DA}=\mathbf{u}^A-\mathbf{u}^{\infty}$,
and that due to a point force located at $\mathbf{x}_0$, we obtain
\begin{equation}\label{eq:ud1_a}
u_j^{DA}(\mathbf{x}_0) = \frac{-1}{8\pi\mu_A} \int_S f_i^{DA}(\mathbf{x}) \, G_{ij}(\mathbf{x},\mathbf{x}_0) \, dS(\mathbf{x}) + \frac{1}{8\pi} \int_S u_i^{DA} \,(\mathbf{x})T_{ijk}(\mathbf{x},\mathbf{x}_0) \,
n_k(\mathbf{x}) \, dS(\mathbf{x}),
\end{equation}
where $\mathbf{f}^{DA}$ is the hydrodynamic traction at the interface
on the side of fluid $A$ associated with the velocity field $\mathbf{u}^{DA}$,
while $\mathbf{G}$ is the Green's function for the specified geometry and $\mathbf{T}$
is the associated stress tensor. Next, employing the self-adjointness property of the Green's function $\mathbf{G}$,
we obtain the following form of the above equation 
\begin{equation}\label{eq:ud1}
u_j^{DA}(\mathbf{x}_0) = \frac{-1}{8\pi\mu_A} \int_S f_i^{DA}(\mathbf{x}) \, G_{ji}(\mathbf{x}_0,\mathbf{x}) \, dS(\mathbf{x}) + \frac{1}{8\pi} \int_S u_i^{DA} \,(\mathbf{x})T_{ijk}(\mathbf{x},\mathbf{x}_0) \,
n_k(\mathbf{x}) \, dS(\mathbf{x}).
\end{equation}
Recall that the self-adjointness of the Green's function $\mathbf{G}$ implies that
\begin{equation}\label{eq:sadjoint}
G_{ij}(\mathbf{x},\mathbf{x}_0) = G_{ji}(\mathbf{x}_0,\mathbf{x}).
\end{equation}
Next, we apply the reciprocal theorem to the undisturbed flow $\mathbf{u}^{\infty}$
in region $B$ and due to a point force located at $\mathbf{x}_0$ in region $A$. This
yields,
\begin{equation}\label{eq:finf}
0 = \int_S f^{B\infty}_i(\mathbf{x}) \, G_{ji}(\mathbf{x}_0,\mathbf{x}) \, dS(\mathbf{x}) - \mu_B \int_S
u_i^{\infty}(\mathbf{x})\, T_{ijk}(\mathbf{x},\mathbf{x}_0) \,
n_k(\mathbf{x}) \, dS(\mathbf{x})
\end{equation}
Note that in the above equation, the normal $\mathbf{n}$ is
pointing out of region $B$ into region $A$ (outward normal). We next note the following
relation between $\mathbf{f}^{A\infty}$ and $\mathbf{f}^{B\infty}$
\begin{equation}\label{eq:finf_r}
\frac{\mathbf{f}^{B\infty}}{\mu_B} = \frac{\mathbf{f}^{\infty}}{\mu_A}.
\end{equation}
This follows from the continuity of the undisturbed
flow across the interface, and that the internal and external stresses
are computed with viscosity $\mu_B$ and $\mu_A$ respectively for the same flow field.
Using this relationship, we write Eq. (\ref{eq:finf}) as
\begin{equation}\label{eq:finf1}
0 = -\frac{1}{8\pi\mu_A}\int_S f^{A\infty}_j(\mathbf{x}) \, G_{ji}(\mathbf{x}_0,\mathbf{x}) \, dS(\mathbf{x}) +  \frac{1}{8\pi} \int_S u_i^{\infty}(\mathbf{x}) \, T_{ijk}(\mathbf{x},\mathbf{x}_0) \, n_k(\mathbf{x}) \, dS(\mathbf{x}).
\end{equation}
Adding Eqs. (\ref{eq:ud1}) and (\ref{eq:finf1}), we obtain
\begin{equation}\label{eq:bi_ext}
8\pi\mu_A \, u_j^{DA}(\mathbf{x}_0) =  -\int_S f_i^{A}(\mathbf{x}) \, G_{ji}(\mathbf{x}_0,\mathbf{x})\,dS(\mathbf{x}) + \mu_A \int_S u_i^{A}(\mathbf{x}) \, T_{ijk}(\mathbf{x},\mathbf{x}_0) \, n_k(\mathbf{x})\, dS(\mathbf{x}).
\end{equation}

Next, we apply reciprocal theorem to the fluid velocity in region 2 ($\mathbf{u}^B$)
and due to a point force located at $\mathbf{x}_0$. This yields
\begin{equation}\label{eq:bi_int}
0 = \int_S f^{B}_j(\mathbf{x}) \, G_{ji}(\mathbf{x}_0,\mathbf{x}) \, dS(\mathbf{x}) - \mu_B \int_S
u_i^{B}(\mathbf{x}) \, T_{ijk}(\mathbf{x},\mathbf{x}_0) \, n_k(\mathbf{x})\, dS(\mathbf{x}).
\end{equation}

For interfacial flows neither the interfacial velocity nor the interface
tractions on either side is known, though we have the following boundary
conditions at any point on the interface
\begin{subequations}
\begin{equation}
\mathbf{u}^A = \mathbf{u}^B = \mathbf{u}^I
\end{equation}
\begin{equation}\label{eq:sjump}
\Delta f = \mathbf{f}^A - \mathbf{f}^B = -\mathbf{f}^{I},
\end{equation}
\end{subequations}
which essentially implies the continuity of the velocity
across the interface and that the net force on an element due to hydrodynamic stresses
is balanced by the net force due to interfacial stresses, see Sec. (\ref{sec:memb}).
We note that the interfacial contribution $\mathbf{f}^I$ is assumed to be known by the
known constitutive equation of the interface and its known configuration.
Given the unknowns and the boundary conditions, a  very common approach is to
eliminate the unknown hydrodynamic interfacial tractions
$\mathbf{f}^A$ and $\mathbf{f}^B$ with the known traction jump $\Delta f$
in Eqs. (\ref{eq:bi_ext}) and (\ref{eq:bi_int}). This
leads to the widely employed second kind integral
equation for the unknown interfacial velocity. Here,
we adopt an alternative approach. Using Eqs (\ref{eq:bi_ext}) and (\ref{eq:bi_int})
we instead eliminate the double layer integral to obtain
\begin{equation}
8\pi u_j^{DA}(\mathbf{x}_0) =  - \frac{1}{\mu_A} \int_S f_i^{A}(\mathbf{x}) \, G_{ji}(\mathbf{x}_0,\mathbf{x}) \, dS(\mathbf{x}) + \frac{1}{\mu_B} \int_S f_i^{B}(\mathbf{x}) \, G_{ji}(\mathbf{x}_0,\mathbf{x})\,dS(\mathbf{x}).
\end{equation}
The above can be written in the following form
\begin{equation}
u_j(\mathbf{x}_0) =  u_j^{\infty}(\mathbf{x}_0) + \frac{1}{8\pi} \int_S \left(\frac{f_i^B(\mathbf{x})}{\mu_B} - \frac{f_i^A(\mathbf{x})}{\mu_A}\right) \, G_{ji}(\mathbf{x}_0,\mathbf{x}) \, dS(\mathbf{x}).
\end{equation}
This equation for the velocity can be shown to be valid inside, outside, as well
as on the boundary $S$. The drawback of this equation is that both the
velocity (including interfacial velocity) and the surface tractions are
unknown. The main advantage for our purposes here is that
we have switched the pole and the field point of the Green's function
using its self-adjointness property (Eq. \ref{eq:sadjoint}). For simplifying the notation,
we next express the operand of the Green's function by $\mathbf{q}$, i.e.
we define $\mathbf{q}$ as
\begin{equation}
\mathbf{q}(\mathbf{x}) = \frac{1}{8\pi} \left( \frac{\mathbf{f}^B(\mathbf{x})}{\mu_2}-\frac{\mathbf{f}^A(\mathbf{x})}{\mu_1} \right).
\end{equation}
Using the above definition, we write the velocity as
\begin{equation}\label{eq:bi_u}
u_j(\mathbf{x}_0) =  u_j^{\infty}(\mathbf{x}_0) +  \int_S q_i(\mathbf{x}) \, G_{ji}(\mathbf{x}_0,\mathbf{x}) \, dS(\mathbf{x}).
\end{equation}
The pressure associated with the above velocity field in the region external to $S$ can be written as
\begin{equation}\label{eq:bi_p}
p(\mathbf{x}_0) =  p^{\infty}(\mathbf{x}_0) +  \mu_A \int_S q_i(\mathbf{x}) \, P_{i}(\mathbf{x}_0,\mathbf{x}) \, dS(\mathbf{x}).\end{equation}
Using Eqs. (\ref{eq:bi_u}) and (\ref{eq:bi_p}), we can write the stress $\sigma_{jk}^A(\mathbf{x}_0)$ as
\begin{equation}\label{eq:bi_t1}
\sigma^A_{jk}(\mathbf{x}_0) =  \sigma_{jk}^{A\infty}(\mathbf{x}_0) +  \mu_A \int_S q_i(\mathbf{x}) \, T_{jik}(\mathbf{x}_0,\mathbf{x}) \, dS(\mathbf{x})
\end{equation}
A similar expression can be written for $\sigma_{jk}^B(\mathbf{x}_0)$
as shown below:
\begin{equation}\label{eq:bi_t2}
\sigma^B_{jk}(\mathbf{x}_0) =  \sigma_{jk}^{B\infty}(\mathbf{x}_0) +  \mu_B \int_S q_i(\mathbf{x}) \, T_{jik}(\mathbf{x}_0,\mathbf{x}) \, dS(\mathbf{x})
\end{equation}
We will now take the limit of the above equations as we approach the interface
from  either side and then dot it with the normal to get the tractions on
either side of the interface $\mathbf{f}^A$ and $\mathbf{f}^B$. We can express both of them
using the principal value of the double layer integral along with
the jump condition to obtain \citep{pozrikidis92}
\begin{subequations}
\begin{equation}\label{eq:f1}
f^A_j(\mathbf{x}_0) =  f_j^{A\infty}(\mathbf{x}_0) - 4\pi\mu_A \, q_j(\mathbf{x}_0) +  \mu_A \, n_k(\mathbf{x}_0) \int_S^{PV} q_i(\mathbf{x}) \, T_{jik}(\mathbf{x}_0,\mathbf{x}) \, dS(\mathbf{x}),
\end{equation}
\begin{equation}\label{eq:f2}
f^B_j(\mathbf{x}_0) =  f_j^{B\infty}(\mathbf{x}_0) + 4\pi\mu_B \, q_j(\mathbf{x}_0) +  \mu_B \, n_k(\mathbf{x}_0) \int_S^{PV} q_i(\mathbf{x}) \, T_{jik}(\mathbf{x}_0,\mathbf{x}) \, dS(\mathbf{x}),
\end{equation}
\end{subequations}
where $PV$ implies the principal value of the improper integral when the
observation point lies on the domain of the integration. Note that
the sign of the jump condition ($4\pi q_j(\mathbf{x}_0)$) depends on
the direction from which we approach the interface relative to the outward normal
defined above, i.e. whether we approach the interface parallel to the normal
or anti to it.

Taking the difference of Eqs. (\ref{eq:f1}) from (\ref{eq:f2})
and using Eqs. (\ref{eq:sjump}) and (\ref{eq:finf_r}), we have that
\begin{equation}
-\Delta f_j(\mathbf{x}_0) = (\lambda-1) f_j^{A\infty}(\mathbf{x}_0) + 4\pi\mu_A(\lambda+1)\, q_j(\mathbf{x}_0) \, + \,  (\lambda-1)\mu_A \, n_k(\mathbf{x}_0) \int_S^{PV} q_i(\mathbf{x}) \, T_{jik}(\mathbf{x}_0,\mathbf{x}) \, dS(\mathbf{x}),
\end{equation}
where we have now introduced the viscosity ratio $\lambda = \mu_B/\mu_A$.
Rearranging the above equation, we obtain a second kind integral equation for the
density of the Green's function $\mathbf{q}$ as follows:
\begin{equation}\label{eq:q}
q_j(\mathbf{x}_0) \, + \,    \frac{\kappa}{4\pi} \, n_k(\mathbf{x}_0) \int_S^{PV} q_i(\mathbf{x}) \, T_{jik}(\mathbf{x}_0,\mathbf{x}) \, dS(\mathbf{x}) = -\frac{1}{4\pi\mu_A} \left(\frac{\Delta f_j(\mathbf{x}_0)}{\lambda+1} + \kappa f_j^{A\infty}(\mathbf{x}_0) \right),
\end{equation}
where we have defined $\kappa$ as
\begin{equation}
\kappa = \frac{\lambda-1}{\lambda+1}.
\end{equation}
The above equation is used to solve for the unknown density $\mathbf{q}$,
which upon substitution in Eq. (\ref{eq:bi_u}) gives the
velocity at any point in the domain, including the interface.

\section{Undisturbed flow stress}\label{sec:finf}
For pressure driven flows, the stress tensor is given by
\begin{equation}
\sigma^{\infty}_{ij} = \frac{8 \mu U_0}{H^2} x \delta_{ij} + \frac{4\mu U_0}{H}(1-\frac{2y}{H}) e_{ij},
\end{equation}
where $e_{ij}$ is
\begin{equation}
\mathbf{e} = \left( \begin{array}{ccc}
0 & 1 & 0 \\
1 & 0 & 0 \\
0 & 0 & 0 \\
\end{array}
\right).
\end{equation}
Note that the velocity and pressure field in pressure driven flows is given by the following expressions
\begin{subequations}
\begin{equation}
u = 4U_0 \frac{y}{H}\left(1-\frac{y}{H}\right),
\end{equation}
\begin{equation}
p = - \frac{8\mu U_0}{H^2} x.
\end{equation}
\end{subequations}
In the above equations, $U_0$ is the centerline velocity.
The surface traction $\mathbf{f}^{\infty}(\mathbf{x})$ can be
obtained at a point $\mathbf{x}$ on the surface with normal
vector $\mathbf{n}(\mathbf{x})$ as 
\begin{equation}\label{eq:straction}
f^{\infty}_i(\mathbf{x}) = \sigma^{\infty}_{ij}(\mathbf{x}) n_j (\mathbf{x}).
\end{equation}
For simple shear flows, the stress tensor is given by
\begin{equation}
\sigma^{\infty}_{ij} = \mu \dot{\gamma} e_{ij},
\end{equation}
where $\dot{\gamma}$ is the shear rate. The surface
traction for this case can be obtained by substituting the stress
tensor in the above equation in Eq. (\ref{eq:straction}).

\section{Fast Spectral Stokes Flow Solver}\label{sec:inf_mat}
Here we discuss the solution procedure for the global problem 
in Eq. (\ref{eq:stokes_gg}) for a slit geometry (Fig. \ref{fig:geom}). 
We simplify the notation in Eq. (\ref{eq:stokes_gg}) and
represent it by the following set of equations:

\begin{subequations}\label{eq:stokes}
\begin{equation} \label{eq:stokes_a}
-\bm{\nabla} p(\mathbf{x})  + \mu \nabla^2 \mathbf{u}(\mathbf{x}) = \mathbf{f}(\mathbf{x}),
\end{equation}
\begin{equation}
\bm{\nabla} \cdot \mathbf{u}(\mathbf{x}) = 0,
\end{equation}
\end{subequations}

where $p$ is the pressure, $\mathbf{u} = (u,v,w)$ is the velocity, while $\mathbf{f}(\mathbf{x})=(f_x,f_y,f_z)$ 
is a known function obtained from the known distribution of global force 
densities. The above set of equations are supplemented by the
periodic boundary conditions in $x$ and $z$ directions, while a Dirichlet
boundary condition for the velocity is specified in the y direction:

\begin{subequations}
\begin{equation}
\mathbf{u}(x,z) = \mathbf{g}_1(x,z) \;\; \textnormal{at}\; y=0,
\end{equation}
\begin{equation}
\mathbf{u}(x,z) = \mathbf{g}_2(x,z) \;\; \textnormal{at}\; y=H.
\end{equation}
\end{subequations}

The velocity and pressure variables are first expanded in truncated Fourier series
in $x$ and $z$ directions as:
\begin{equation}\label{eq:fourier}
u(\mathbf{x}) =  \sum_{l=-N_x/2}^{N_x/2-1} \; \sum_{m=-N_z/2}^{N_z/2-1} \; \hat{u}_{lm}(y) \, e^{i 2 \pi l x/L_x} \,e^{i 2\pi m z/L_z}.
\end{equation}

Similar expressions are written for $v(\mathbf{x})$, $w(\mathbf{x})$, $p(\mathbf{x})$,  $f_x(\mathbf{x})$, 
$f_y(\mathbf{x})$, and $f_z(\mathbf{x})$ by replacing $\hat{u}_{lm}$ by $\hat{v}_{lm}$, $\hat{w}_{lm}$, $\hat{p}_{lm}$,  $\hat{f}_{xlm}$, $\hat{f}_{ylm}$, and $\hat{f}_{zlm}$ respectively in the above equation.
Substituting this in equation (\ref{eq:stokes}) and employing the Galerkin method, we obtain the following owing to the orthogonality of Fourier modes:
\begin{subequations}\label{eq:stokes_t}
\begin{equation}
-il \,\hat{p} - \mu (k^2-\frac{\partial^2}{\partial^2 y}) \hat{u} = \hat{f_x}, 
\end{equation}
\begin{equation}
- \frac{\partial \hat{p}}{\partial y} - \mu (k^2-\frac{\partial^2}{\partial^2 y}) \hat{v} = \hat{f_y},
\end{equation}
\begin{equation}
-im \,\hat{p} -  \mu (k^2-\frac{\partial^2}{\partial^2 y}) \hat{w} = \hat{f_z} ,
\end{equation}
\begin{equation}
il\, \hat{u} + \frac{\partial v}{\partial y} + i m \hat{w} = 0,
\end{equation}
\end{subequations}

where we have dropped the subscript $lm$ in the above equation. Next, using the last 
equation, we eliminate $\hat{u}$, $\hat{v}$ and $\hat{w}$ in the first three equations
to obtain the following equation for the pressure:
\begin{equation}\label{eq:press_helm}
 \frac{\partial^2 \hat{p}}{\partial^2 y} -k^2 \hat{p} = il \hat{f_x} + \frac{\partial \hat{f_y}}{\partial y} + im \hat{f_z}.
\end{equation}
The continuity equation can be replaced by the above equation for pressure along
with the boundary condition requiring the velocity to be divergence free \citep{kleiser84,canuto07}, i.e.
\begin{equation}\label{eq:cont_bc}
i l \hat{u} + \frac{\partial \hat{v}}{\partial y} + i m \hat{w} = 0 \;\; at \;\; y = 0 \;\; \;\; \& \;\; y=H.
\end{equation}
Alternatively, one solves the Eq. \ref{eq:press_helm} with the following 
pressure boundary condition 
\begin{subequations}
\begin{equation}
\hat{p} = \hat{p}_1 \;\; \textnormal{at} \;\; y=0,
\end{equation}
\begin{equation}
\hat{p} = \hat{p}_2 \; \; \textnormal{at} \; \; y = H,
\end{equation}
\end{subequations}
though, the pressure boundary conditions above are unknown a priori, but  instead they
must take a value so that the condition in Eq. (\ref{eq:cont_bc}) is satisfied.
Grouping all the equations to be solved, we have the following
\begin{subequations}\label{eq:stokes_t_all}
\begin{equation}
\displaystyle \frac{\partial^2 \hat{p}}{\partial^2 y} -k^2 \hat{p} = il \hat{f_x} + \frac{\partial \hat{f_y}}{\partial y} + im \hat{f_z},
\end{equation}
\begin{equation}
\displaystyle   \mu \frac{\partial^2 \hat{u}}{\partial^2 y} -\mu k^2 \hat{u} -il \,\hat{p} = \hat{f_x},
\end{equation}
\begin{equation}
\displaystyle  \mu\frac{\partial^2 \hat{v}}{\partial^2 y} -\mu k^2 \hat{v}  - \frac{\partial \hat{p}}{\partial y} = \hat{f_y},
\end{equation}
\begin{equation}
\displaystyle \mu \frac{\partial^2 \hat{w}}{\partial^2 y} -\mu k^2 \hat{w}    -im \,\hat{p}  = \hat{f_z} ,
\end{equation}
\end{subequations}
with the following boundary conditions
\begin{equation}\label{eq:bc_vel}
\begin{array}{c}
\hat{p} = \hat{p}_1, \; \hat{u} = \hat{g}_{1x}, \; \hat{v} = \hat{g}_{1y}, \; \hat{w} = \hat{g}_{1z} \;\; \textnormal{at} \;\; y=0, \\
\hat{p} = \hat{p}_2, \; \hat{u} = \hat{g}_{2x}, \; \hat{v} = \hat{g}_{2y}, \; \hat{w} = \hat{g}_{2z} \;\; \textnormal{at} \;\; y=H, \\
\end{array}
\end{equation}
where $\mathbf{g}_1 = (g_{1x},g_{1y},g_{1z})$ and $\mathbf{g}_2 = (g_{2x},g_{2y},g_{2z})$ and, as with the other quantities, $~\hat{}~$ denotes discrete Fourier transform in $x$ and $z$. 

The Kleiser-Schumann influence matrix approach involves solving three set of equations 
as in Eq. (\ref{eq:stokes_t_all}) with different boundary conditions as discussed shortly.
In the first set, one solves the equations in \ref{eq:stokes_t_all} with the correct
boundary conditions for the velocity in Eq. (\ref{eq:bc_vel}), but with homogeneous
boundary conditions for the pressure, i.e.
\begin{equation}
\hat{p} = 0 \;\; \textnormal{at}\;\; y=0 \;\; \& \;\; y=H
\end{equation}
Each of the equations for pressure and velocity components above are solved
here by expanding them in discrete Chebyshev polynomials and then employing
the Galerkin method to obtain equations for each of Chebyshev modes.
The appropriate boundary conditions are satisfied by employing
the tau method \citep{canuto06,peyret02} in which the equations
for the highest two Chebyshev modes are replaced by the boundary
condition equations. The solution for pressure is 
first computed, whose value is then substituted in the equations for velocity.
Thus, in each step, one needs to solve a Helmholtz equation using Chebyshev
polynomial expansion. In this case, the  equations for the unknown Chebyshev
coefficients can be reduced to a quasi-tridiagonal matrix equation with 
the last full row being full, while the rest being in the standard tridiagonal
form. Also, note that the equations for the even and odd Chebyshev modes
are decoupled and solved separately. These quasi-tridiagonal systems of equations can be 
solved in $O(N_y)$ time with a direct algorithm \citep{peyret02}. Also, the use of 
Chebyshev Gauss-Lobatto quadrature points \citep{canuto06, peyret02} for transforming 
a variable between physical and transform space  ensures that
FFTs can be employed for its evaluation \citep{canuto07}. Therefore, for 
each of the Fourier modes $(l,m)$, the asymptotic cost of the Chebyshev-tau
solution procedure scales as $N_y \log N_y$.
We denote this first set of solution thus obtained
for Fourier mode $(l,m)$ by ($\hat{u}_a,\hat{v}_a,\hat{w}_a,\hat{p}_a$). 

The next two sets of equations involve solving the homogeneous version of the differential 
equations in \ref{eq:stokes_t_all}, i.e., the 
right hand side of the each of the equations in \ref{eq:stokes_t_all}
is set to zero. Moreover, the velocity boundary condition for these
two sets of problems are also homogeneous. The only non-homogeneous equation in these
two problems are the pressure boundary conditions. In the first of these, 
the pressure boundary condition is the following:
\begin{equation}
\hat{p}(0) = 1 \;\;  \& \;\; \hat{p}(H) = 0, 
\end{equation}
while in the second the pressure boundary condition is 
\begin{equation}
\hat{p}(0) = 0 \;\;  \& \;\; \hat{p}(H) = 1 .
\end{equation}
We denote these two solutions by ($\hat{u}_b,\hat{v}_b,\hat{w}_b,\hat{p}_b$),
and ($\hat{u}_c,\hat{v}_c,\hat{w}_c,\hat{p}_c$). It is important to note that
the latter two set of equations are to be solved just once at the beginning
of the simulation and the results are stored. The overall solution for the 
velocity and pressure is then obtained as:
\begin{equation}\label{eq:cheb_sol}
\begin{array}{c}
\hat{u} = \hat{u}_a + \hat{p}_1 \hat{u}_b + \hat{p}_2 \hat{u}_c, \\
\hat{v} = \hat{v}_a + \hat{p}_1 \hat{v}_b + \hat{p}_2 \hat{v}_c, \\
\hat{w} = \hat{w}_a + \hat{p}_1 \hat{w}_b + \hat{p}_2 \hat{w}_c, \\
\hat{p} = \hat{p}_a + \hat{p}_1 \hat{p}_b + \hat{p}_2 \hat{p}_c. \\
\end{array}
\end{equation}
It is easy to see that the above solution satisfies both the differential
equations as well as the boundary conditions. The only remaining task is 
therefore to determine the pressure boundary conditions $\hat{p}_1$ and
$\hat{p}_2$. This is accomplished by the requiring that the velocity be 
divergence free at the boundary (Eq. \ref{eq:cont_bc}). Thus, one 
obtains the following equations for the pressure boundary conditions
\begin{equation}\label{eq:inf_mat}
\left(
 \begin{array}{cc} 
C_b(0)   & C_c(0) \\
C_b(H) & C_c(H) 
\end{array}
\right)
\left( \begin{array}{c} \hat{p}_1  \\ \hat{p}_2 \end{array} \right)
=
-\left(
\begin{array}{c} 
C_a(0) \\
C_a(H) 
\end{array}
\right)
\end{equation}

In the above equations, $C_a$, $C_b$ and $C_c$ are the continuity expression (Eq. \ref{eq:cont_bc}) 
evaluated at the appropriate boundary point, i.e. at $y=0$ or $y=H$.
For example, $C_a(0)$ is given by 
\begin{equation}
C_a(0) = \displaystyle i l \hat{u}_a(0) + \frac{\partial \hat{v}_a}{\partial y}(0) + i m \hat{w}_a(0) 
\end{equation}
The $2\times2$ coefficient matrix in Eq. (\ref{eq:inf_mat}) is known as the influence matrix 
and the resulting matrix equation is trivially solved at a cost of $O(1)$.
The solution in Eq. (\ref{eq:cheb_sol}) gives the Fourier coefficients
of the pressure and velocity as a function of the wall normal coordinate $y$,
which is then employed to obtain the pressure and velocity in the physical space
using inverse FFTs.

\end{document}